\documentclass[aps,pra,twocolumn,superscriptaddress,amsmath,amssymb,eqsecnum,nofootinbib,notitlepage]{revtex4-2}

\usepackage[T1]{fontenc}

\usepackage{comment}
\usepackage{lipsum}
\usepackage[usenames,dvipsnames]{xcolor}
\usepackage[breaklinks=true]{hyperref}
\hypersetup{
  colorlinks   = true, 
  urlcolor     = blue, 
  linkcolor    = blue, 
  citecolor   =  red 
}
\usepackage{dingbat}
\usepackage{mathrsfs}
\usepackage{siunitx}
\usepackage{mathtools} 

\usepackage{tikz}

\makeatletter
\def\@bibdataout@aps{%
 \immediate\write\@bibdataout{%
  @CONTROL{%
   apsrev41Control,author="08",editor="1",pages="0",title="0",year="1",eprint="1"%
  }%
 }%
 \if@filesw
  \immediate\write\@auxout{\string\citation{apsrev41Control}}%
 \fi
}%
\makeatother 

\usepackage{graphicx}

\usepackage{cooltooltips} 
\usepackage{bm} 

\usepackage[caption=false,subrefformat=simple,labelformat=simple,listofformat=subsimple]{subfig}

\usepackage{bbold}
\usepackage{dsfont}

\usepackage[capitalise]{cleveref}

\input{shorthand.sty}

\begin{document}

\title{Master equations and quantum trajectories for squeezed wave packets}
\author{Jonathan A. Gross}\email{jonathan.gross@usherbrooke.ca}
\affiliation{Center for Quantum Information and Control, University of New Mexico,
Albuquerque NM 87131-0001, USA}
\affiliation{Institut quantique, Universit\'{e} de Sherbrooke,
Sherbrooke QC J1K 2R1, Canada}
\author{Ben Baragiola }\email{ben.baragiola@gmail.com}
\affiliation{Center for Quantum Information and Control, University of New Mexico,
Albuquerque NM 87131-0001, USA}
\affiliation{Centre for Quantum Computation \& Communication Technology,
School of Science, RMIT University, Melbourne, VIC 3000, Australia}
\author{T. M. Stace}\email{stace@physics.uq.edu.au}
\affiliation{ARC Centre for Engineered Quantum System, Department of Physics,
University of Queensland, Brisbane, QLD 4072, Australia}
\author{Joshua Combes}\email{joshua.combes@gmail.com}
\affiliation{Center for Quantum Information and Control, University of New Mexico,
Albuquerque NM 87131-0001, USA}
\affiliation{Department of Electrical, Computer and Energy Engineering, University of Colorado Boulder, Boulder, Colorado 80309, USA}

\date{\today}

\begin{abstract}
The interaction between matter and squeezed light has mostly been treated within the approximation that the field correlation time is small.
Methods for treating squeezed light with more general correlations currently involve explicitly modeling the systems producing the light.
We develop a general purpose input-output theory for a particular form of narrowband squeezed light---a squeezed wave-packet mode---that only cares about the statistics of the squeezed field and the shape of the wave packet.
This formalism allows us to derive the input-output relations and the master equation. We also consider detecting the scattered field using photon counting and homodyne measurements which necessitates the derivation of the stochastic master equation.
The non Markovian nature of the field manifests itself in the master equation as a coupled hierarchy of equations. We illustrate these with consequences for the decay and resonance fluorescence of two-level atoms in the presence of such fields. 
\end{abstract}
\maketitle

\section{Introduction}\label{sec:intro}
Squeezed light \cite{yuen_twophoton_1976, caves_twophoton_1985, schumaker_twophoton_1985} exhibits fundamentally nonclassical properties that can be leveraged for metrology~\cite{lawrie2019quantum} and quantum information~\cite{hamilton2017gaussian}. 
Injecting squeezed light into Advanced LIGO~\cite{aligo_squeezing_2019,virgo_squeezing_2019} as originally proposed by Caves~\cite{caves_prl_1980,caves_prd_1981} 
markedly improves the sensitivity for gravitational wave detection.
Squeezed light is routinely used to conditionally prepare single-photon states~\cite{couteau2018spontaneous} that can act as carriers of quantum information for quantum computing~\cite{knill2001scheme,nielsen2004optical,gimeno2015from}. 
Alternatively, many squeezed states can be directly stitched together into continuous-variable cluster states~\cite{asavanant2019generation,larsen2019deterministic} that form the foundation for measurement-based computing schemes~\cite{menicucci2006universal}, with the squeezing itself being the fundamental resource~\cite{braunstein2005squeezing}.

Studying the interaction between matter and squeezed light is of great interest 
in part because squeezed light is the simplest nonclassical light to produce and thus probe matter with. 
For the most part, theoretical analyses are focused on broadband squeezed light~\cite{caves1986broadband}, where the squeezing is constant over all relevant frequencies.
This situation is usually treated in the Markovian white-noise formalism~\cite{InputOutput85, Gardiner86} whose crowning achievements are the master equation~\cite{InputOutput85} and stochastic master equation~\cite[Sec. 4.4.1]{WisemanPhD94} that describe the reduced state of a quantum system in contact with a time-invariant delta-correlated Gaussian field.
The mathematical literature has addressed similar issues with increased rigor~\cite{honegger1997, HellHone02, Gough:2003aa,Barchielli_2009,parthasarathy2012introduction,accardi2013quantum,DabGou2016a,DabrGough2016b}.

The broadband approximation has serious drawbacks, however, as squeezed fields treated in the white-noise formalism exhibit nonphysical behavior when one attempts to describe photon counting.
When a photodetector with infinite temporal resolution---that is, infinite bandwidth---measures such a field, it sees an infinite photon flux because the delta time correlations of the white-noise fields correspond to a flat power spectral density as discussed in \cite[Sec. 4.3.3]{wiseman_quantum_2010}.
Relaxations of the strict white-noise formalism are therefore necessary, not only because they avoid this issue but also because squeezed light produced in a laboratory is, at best, only approximately delta-correlated in time as determined by the squeezing bandwidth of the source, leading to an observable difference in the response of a physical system interacting with finite-bandwidth squeezed fields compared to the white-noise prediction~\cite{Gard_io_III_1987,toyli_resonance_2016}.

\begin{figure}[t]
\begin{center}
\includegraphics[width=0.9\hsize]{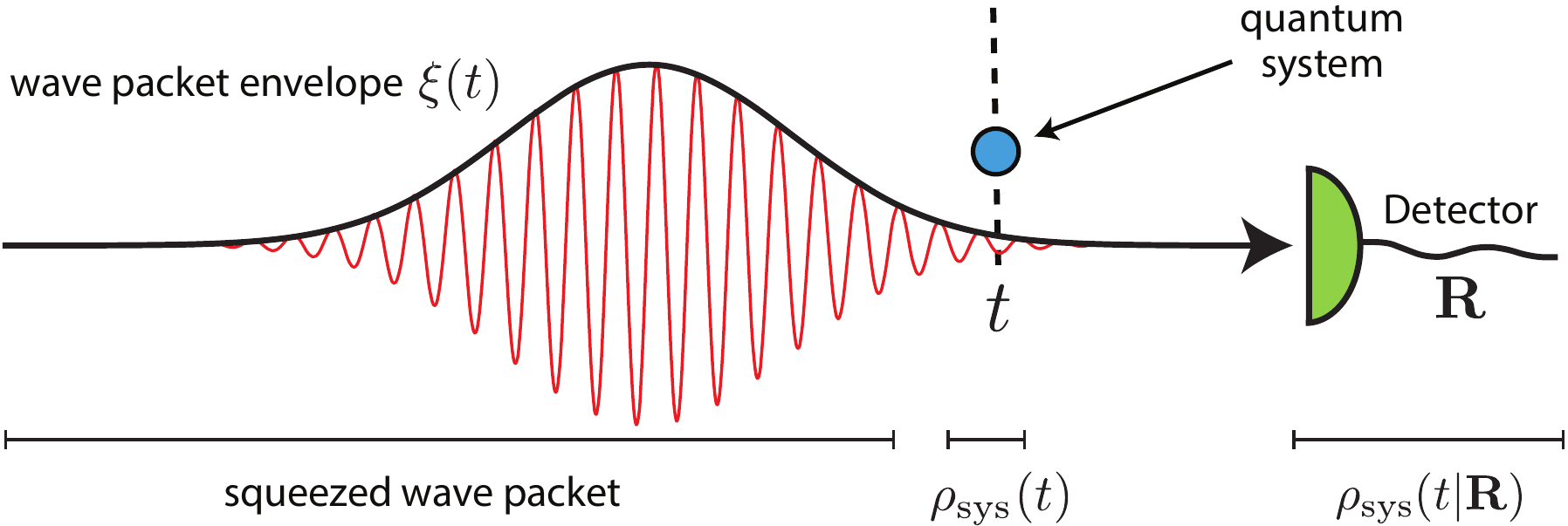}
\caption{
A squeezed wave packet propagating towards a quantum system followed by measurement of the scattered field. The temporal mode is given by a slowly-varying envelope $\xit$ (black) modulating fast oscillations at the carrier frequency (red).
At time $t$ a portion of the field has interacted with the system and is propagating away. Due to the nonlocal temporal correlations in the input field, the system is entangled with both the output portion of the field as well as the remaining input field. Performing a trace over the entire field gives the reduced state of the system $\rho_{\rm sys}(t)$ whose evolution is governed by a master equation.
Should one monitor the output field with a detector rather than let it escape, the state of the system conditioned on the measurement record $\Rec$ is a quantum trajectory denoted by the conditional state $\rho_{\rm sys}(t|\Rec)$ whose evolution is governed by a stochastic master equation.
} \label{fig1}
\end{center}
\end{figure}

The degree to which a squeezed field is ``broadband'' is best discussed with respect to
a reference system---such as a detector, a cavity, or an atom---with which the field interacts.
Consider a field characterized by carrier frequency $\omega_c$ and squeezing bandwidth $B_{\rm sq}$, such that the field exhibits appreciable squeezing at frequencies between $\omega_c \pm B_{\rm sq}/2$.
Given a coupling rate $\lw$ between the field and the reference, we adopt the threefold categorization from the master-equation literature and classify the squeezing as either broadband, finite bandwidth, or narrowband. 
Broadband squeezing has a bandwidth $B_{\rm sq} \gg \lw$, with infinitely broadband squeezing being the limit $\omega_c, B_\text{sq} \rightarrow \infty$, where the field exhibits appreciable squeezing at all frequencies.
Finite-bandwidth squeezing is characterized by $B_{\rm sq} \sim \lw$, and narrowband squeezing has a bandwidth $B_{\rm sq} \ll \lw$.
Early squeezing experiments were limited to narrow and finite-bandwidth squeezing \cite{30years_of_squeezing} though recent experiments have produced fields with high levels of squeezing~\cite{vahlbruch2016detection} with very wide bandwidths~\cite{wakui2014ultrabroadband, kashiwazaki2020continuous, vaidya2020broadband, pao-kang2021}.

A salient distinction between the white-noise formalism and finite- or narrow-bandwidth squeezing is the Markovian versus non-Markovian characteristic of the field.
Realistic squeezed fields have nonzero correlation times, which mean their effect on a system cannot be described by the inherently Markovian master equations enjoyed by broadband descriptions.
Non-Markovian evolution is quite difficult to deal with in full generality, and this difficulty is partly responsible for the attention direction at more tractable Markovian descriptions. 
Nevertheless, prior theoretical work has developed some methods to model the light-matter interaction for finite and narrow bandwidth squeezed fields.
The so called ``quasi Markoffian'' methods \cite{Carmichael_1973,Gard_io_III_1987, yeoman_influence_1996,Ficek97,Tanas1999,Kowalewska-kudlaszyk:2001aa} work in the finite-bandwidth regime, and the cascaded systems methods~\cite{parkins_rabi_1990,Cirac1992,GardPar94,KochCarm94,Tanas1999,SmythSwain1999,Messikh2000} and colored noise methods ~\cite{Ritsch:1988aa,Ritsch:1988ab} can work in both the finite and narrow-bandwidth regimes.

In this paper we place the state of the field front and center, without a mediating source. By considering squeezed wave packets, we are able to deal with arbitrarily large or small bandwidths in a controlled way.
Our model aims to provide a more efficient representation and deeper understanding of the unique non-Markovian properties of the field and the light matter interaction.
It additionally allows us to explicitly write down field-correlation functions, master equations, and even stochastic master equations, which have eluded previous treatments due to the previously mentioned bandwidth issues.
The wave packets we consider, and the subsequent results,
are in the style of the number-state and cat-state wave packets considered in \cite{gough_quantum_2012,baragiola_n-photon_2012,baragiola_quantum_2017,dabrowska_quantum_2017,dabrowska_quantum_2019,Dabrowska_2019_nphoton,kiilerich_input-output_2019,kiilerich_quantum_2020} and intimately related to the collisional models literature \cite{Cakmak2017,Daryanoosh:2018aa,Whalen:2019aa,Taranto:2019aa,Rodrigues:2019aa,cilluffo_collisional_2020,carollo_mechanism_2020,Cattaneo_collisional_2020,ciccarello_CM_review_2021}.

Our paper begins with, in \cref{sec:formalism}, summarizing the white noise formalism used which we use to derive the standard Markovian master equation for squeezed baths in \cref{sec:comp}.
Then we have three major contributions which are graphically summarized in \cref{fig1}. The first part of the figure is simply a squeezed wave packet which is covered in \cref{sec:the_field} where we talk about the unusual temporal correlations present in a finite bandwidth squeezed field and calculate photon-detection probabilities. The second part of the figure depicts an interaction where  we trace over the scattered field which allows us to derive the master equation which is discussed in \cref{sec:master_eqn}. We discuss in \cref{sec:SMEs} the case when the scattered field is intercepted by a detector, deriving the promised stochastic master equations. We illustrate our formalism with two numerical examples. First we perform a qualitative comparison between broadband and wave-packet squeezing of the effect on a two level atom whose decay rate is modified by the squeezing in \cref{sec:qual_comp} and \cref{sec:res_fluor}. As the equations we derive are very complicated  we discuss the issues associated with their Numerical implementation in \cref{sec:NumericalAnalysis}.

\section{Formalism}\label{sec:formalism}
We outline the formalism and notation that have historically been used in quantum optics, and upon which we build in this work.
We make no attempt at a thorough review of input-output theory, collisional models, or quantum stochastic differential equations (QSDEs) as well as their relation to quantum optical problems.
Interested readers are directed to recent pedagogical treatments  \cite{q_noise_gard2004,wiseman_quantum_2010,gross_qubit_2017,combes_slh_2017, TrivFisc2019}, which themselves are based on foundational work \cite{InputOutput85,WisemanPhD94,HudsonPartha1984,YurkeDenker84}.

Input-output theory, including its formulation using QSDEs, is used to describe the general quantum optical situation where an arbitrary quantum system is coupled to a propagating quantum field via a dipole-type interaction in the rotating-wave approximation. 
We follow the approach of continuous-mode quantum optics~\cite{Blow1990}, which takes the coupling, laid out in $\mathbf{k}$-space, and transforms it to frequency space (for each $\mathbf{k}$-direction) through the free-space dispersion relation, $\omega = c k$ where $c$ is the speed of light.\footnote{ In other settings, such as in dielectric waveguides, one still relies on this relation by linearizing the dispersion relation the around a central frequency of interest~\cite{ScatInputOutput2010}---typically that of the bare system, $\omega_0$.} 
The continuous-mode field is described by mode operators $b\dg(\omega)$ and $b(\omega)$ that create and annihilate photons at frequency $\omega$ and obey the canonical commutation relation $[ b(\omega), b\dg(\omega')] = \delta(\omega -\omega')$.  
The interaction Hamiltonian describing the system-field coupling is given by 
\begin{equation}
  \label{eq:interaction-hamiltonian}
  H_{\rm int}
  =
  \,i \hbar  \int d\omega \, \kappa(\omega) \big[c-c^\dag\big]\big[b(\omega)+b^\dag(\omega)\big] 
  ,
\end{equation}
where $\kappa(\omega)$ is the coupling strength at frequency $\omega$ (which in agreement with~\cite{q_noise_gard2004} we will approximate as $\sqrt{\lw/2\pi}$ within the relevant frequency range), and $c^\dagger$ (and its Hermitian conjugate $c$) are operators on the system that give its response to absorbing/emitting a photon.

To simplify the description of the system and field we employ a set of widely applicable approximations that together form the \emph{white-noise} approximation \cite{WisemanPhD94,accardi2013quantum, Handel_2005,gross_qubit_2017}. 
We begin by splitting the field into a sequence of temporal modes whose durations $\Delta t$ sit between several timescales of interest.
The spatial extent of each mode $c \Delta t$ is taken to be large enough 
so that the system (of size $\Delta x$) is pointlike.
At the same time, take $\Delta t$ to be small compared to the field-correlation time $\tau_c$, such that field correlations cannot be detected across different temporal modes.
The fields we consider are quasimonochromatic (QMC), meaning that they have a well defined carrier frequency $\carrierfreq$ with respect to which the spectral bandwidth is small.
The system-field coupling strength $\lw$ is weak, and system-field interactions are treated in the rotating-wave approximation (RWA).
This set of approximations is summarized by:
\begin{subequations}
\begin{align}
  \Delta x&\ll c\Delta t & & \text{First Markov,} \\
  \tau_c&\ll\Delta t & & \text{Second Markov,} \\
  \label{eq:quasi-mono-rwa}
  \carrierfreq^{-1}&\ll\Delta t\ll\lw^{-1} & & \text{RWA \& QMC.}
\end{align}
\label{eq:approximations}
\end{subequations}

The quasimonochromatic approximations and the limit $\Delta t \rightarrow dt$ allow us to write down mode Fourier-transformed space-time mode operators,
    \begin{align}
        b(z,t) \coloneqq \frac{1}{\sqrt{2\pi}} \int d \omega \,  b(\omega) e^{-i (\omega - \omega_c) (t- z/c)}
        \, ,
    \end{align}
where $z$ is the spatial coordinate associated with the $\mathbf{k}$-direction.
Input-output theory concerns itself specifically with the fields at the position of the quantum system, chosen to be $z=0$,
\begin{align}\label{eq:neg-freq-approx}
 b_t \coloneqq b(z=0, t) = \frac{1}{\sqrt{2\pi}} \int d \omega \,  b(\omega) e^{-i (\omega - \omega_c) t}
\end{align}
and the Hermitian conjugate $b^\dagger_t = b^\dagger(z=0,t)$.
These mode operators describe the fields incident upon the system at time $t$---\ie{} at space-time point $(z=0,t)$---where $t$ is treated a mode label rather than a time dependence. 

With the approximations \cref{eq:approximations} we simplify \cref{eq:interaction-hamiltonian} in the interaction picture to
\begin{align}
    \label{Eq::H_dp_RWA_Int}
    H_{\rm int}(t)
    &=
    i\hbar\sqrt{\lw}\big[b_t^\dagger c-b_tc^\dagger\big]
    \,.
\end{align}

For the sake of compact notation, we adopt the practice of writing arguments to these functions as subscripts rather than enclosed in parentheses. 
In this notation, the commutator is
\begin{align}\label{eq:neg-freq-approx-compact}
  [b_t^{},b^\dagger_s]=\delta(t-s),
\end{align}
which shows the field is delta-correlated. This delta-correlated field is reminiscent of classical white noise which is why the operators are called \emph{white noise operators}.

Integrating the field operators over time intervals that are effectively infinitesimal with respect to system dynamics  gives the \emph{quantum noise increments} $\dB$ and $\dLam$:
\begin{subequations}\label{noiseincrements}
\begin{align} 
  \dB
  &\defined\int_t^{t+dt}\!ds\,b_s
  \\
  d\Lambda_t
  &\defined \label{lambdanoiseincrements}
  \int_t^{t+dt}\!ds\,b^\dagger_s b_s\, ,
\end{align}
\end{subequations}
where $[\dB, \dBdag] = dt$. The integral of operator $d\Lambda_t$ is the photon-number operator.
One can also define quadrature increments
\begin{align}\label{eq:quadrature-increment}
    dQ_t(\varphi)
    &=
    ( e^{-i\varphi}dB_t+e^{i\varphi}dB_t^\dagger)/\sqrt{2}
\end{align}
which yield quadratures when integrated.
As special cases we obtain the $x$ quadrature with $\varphi=0$ and the $p$ quadrature with $\varphi=\pi/2$.
These differential increments obey the following rules of vacuum quantum  It\={o} calculus: $\dB \dBdag = dt $, $\dB d\Lambda_t = \dB $, $d\Lambda_t \dBdag = \dBdag $, $d\Lambda_t d\Lambda_t = d\Lambda_t$, and all other products are zero~\cite{InputOutput85,wiseman_quantum_2010,HudsonPartha1984}.
These non commutative operator relations correspond to the It\={o} rules for the classical noise increment $dW_t$ (a scalar random variable), so just as differential equations involving $dW_t$ are called stochastic differential equations, differential equations involving the quantum noise increments are called \emph{quantum stochastic differential equations} which is abbreviated to QSDEs.
For readers unfamiliar with  It\={o} calculus we recommend Ref.~\cite{wiseman_quantum_2010}.

The system and field evolve in time according to the interaction Hamiltonian in \cref{Eq::H_dp_RWA_Int}. 
In the white-noise limit, \cref{eq:approximations}, a slightly generalized Hamiltonian~\cite{combes_slh_2017} generates a unitary time evolution operator $U_t$ describing the joint dynamics from time $t=0$ to time $t$ that obeys the differential equation~\cite{HudsonPartha1984,InputOutput85}:
\begin{align}\label{eq:dU_propagator}
\begin{split}
  dU_t
  &=
  \big[ {-}dt(iH\system+\tfrac{1}{2}L^\dagger L) \otimes \Id_t
  -L^\dagger S\otimes\dB
  \\
  &\phantrel{=}{}
  +L\otimes\dBdag+(\tight{S-\Id\system})\otimes d\Lambda_t\big ]U_t
\end{split}
\end{align}
with initial condition 
\begin{align}\label{eq:dU_propagatorIC}
  U_0
  &=
  \Id\system\otimes \Id\field\,.
\end{align}
The Hamiltonian $H\system$ includes the bare Hamiltonian of the system as well as any others that act on the system alone (such as a control Hamiltonian). The system's interactions with the field are described by the jump operator $L=\sqrt{\lw}c$---determined by the underlying system-field Hamiltonian, \cref{Eq::H_dp_RWA_Int}---and the unitary scattering operator $S$.\footnote{
The scattering operator $S$ is used to describe phase shifts~\cite{XuFan_fano2016} or effective system-field interactions arising from adiabatic elimination~\cite{combes_slh_2017} of internal states of the system imparted by a mirror on the input field, for example.} 
Specifying these three operators fully determines the evolution a single system's interaction with the field, and multiple systems can be tied together in a quantum input-output network using the SLH formalism~\cite{combes_slh_2017}.
The differential equation \cref{eq:dU_propagator} can be solved formally to find the infinitesimal time-evolution operator 
\begin{align}
  \label{eq:inf-time-evol-op}
  \begin{split}
  \Uprop{t+dt}{t}
  &=
  \Id\system\otimes \Id_t + dU_tU_t^\dagger \\
  &=
  \Id\system\otimes\Id_t
  -dt\,(iH\system+\tfrac{1}{2}L^\dagger L)\otimes\Id_t
  \\
  &\phantrel{=}{}
  -L^\dagger S\otimes\dB+L\otimes\dBdag
  \\
  &\phantrel{=}{}   +(\tight{S-\Id\system})\otimes d\Lambda_t
  \,.
  \end{split}
\end{align}
that describes dynamics over an infinitesimal time step, from $t$ to $t+dt$. The famous {\em input-output} relations can be obtained from this time evolution operator by conjugating the input operator
\begin{align}
    \dB^{\rm out} &=  \Uprop{t+dt}{t}^\dagger \dB  \Uprop{t+dt}{t} \\
         & = \dB + L dt  \, .
\end{align}
Many authors will denote the input operator  $\dB^{\rm in}$; we do not to reduce notational clutter. Again we reiterate, the subscript $t$ on $\dB$ denotes the temporal mode label, not a time dependence, as explained in~\cite{gross_qubit_2017}.

\section{Markovian squeezing treatments}\label{sec:comp}
We now remind the reader of Markovian treatments of squeezed light interacting with a system, \ie{} the master equation.
This affords us the opportunity to introduce familiar objects that appear in unfamiliar contexts in our non Markovian treatment.
We discuss the Markovian treatment of broadband squeezing and related techniques that relax the broadband approximation while retaining a Markovian description.

In a single mode (with annihilation and creation operators satisfying $[b,b\dg]=1$), a pure squeezed vacuum state $\ket{\psi_{\text{sq}}}\defined S(r,\phi)\ket{\text{vac}}$
is generated from the vacuum state by application of the \emph{squeeze operator},
    \begin{align}
        \label{eq:squeeze-operator}
        S(r,\phi)\defined\exp\Big[\half r\big(e^{-2i\phi}b^2-e^{2i\phi}{b\dg}^2\big)\Big]\,.
    \end{align}
The parameter $\phi$ sets the squeezing angle, and the amount of squeezing in decibels is determined from the squeezing parameter $r$ by the relation, $r_{\rm dB} = 10 \log_{10}(e^{2r})$.

We define a squeezed mode annihilation operator $b_{\text{sq}}$ via a squeezing transformation,
\begin{align}\label{eq:squeezing-transformation}
    b_{\text{sq}} \defined S\dg(r,\phi)\,b\,S(r,\phi) = \cosh r b-e^{2i\phi} \sinh r b\dg
\end{align}
with $b^\dagger_{\text{sq}}$ following directly from Hermitian conjugation.
This transformation does not modify the vacuum expectation values of the mode operators, $\expt{b}_{\text{sq}}=\expt{b_{\text{sq}}}_{\text{vac}}=0$, but it does affect vacuum expectation values of their products,
\begin{subequations}\label{eq:Npuresqueezed}
\begin{align}
  N& = \expt{b\dg b}_{\text{sq}}=\expt{b\dg_{\text{sq}}b_{\text{sq}}}_{\text{vac}}=\sinh^2\!r\,, \\
  M& = \expt{b^2}_{\text{sq}}=\expt{b^2_{\text{sq}}}_{\text{vac}}=-e^{2i\phi}\sinh r\cosh r\,.
  \label{eq:Mpuresqueezed}
\end{align}
\end{subequations}

\noindent {\em White-noise regime:} The stationary statistics of a broadband (white-noise) squeezed field can be found by applying the single-mode squeeze transformation in \cref{eq:squeezing-transformation} (using fixed values of $\phi$ and $r$) to each infinitesimal temporal mode in \cref{eq:neg-freq-approx} of size $\Delta t$. An infinitely broadband field has no temporal correlations, since each infinitesimal temporal mode is transformed independently.
For a more details, see Ref.~\cite{gross_qubit_2017}.

Squeezing transformations on the noise increments in \cref{noiseincrements}, yield vacuum expectation values that follow directly from \cref{eq:Npuresqueezed} and \cref{eq:Mpuresqueezed} \cite{InputOutput85,Gard92,Hellmich02,Gough:2003aa},
\begin{subequations}\label{eq:bath-stats}
\begin{align}
  \nexpt{\dB} &= 0 \\
  \nexpt{\dBdag \dB} &= N\,\df t\,, \\
  \nexpt{\dB^2} &= M\,\df t\,, \\
  \nexpt{[\dB,\dB\dg]} &= \df t\,.
\end{align}
\end{subequations}
These relations have been used to derive the Markovian master equation for a quantum system interacting with stationary, broadband squeezed vacuum \cite{InputOutput85,Gardiner86},
\begin{align}\label{eq:pure-squeezed}
\begin{split}
  \df\rho&=\df t \Big ( -i[H\system,\rho]+ (N+1)\D{L}\rho+N\mathcal{D}[L\dg]\rho \\
  &\phantrel{=}{}
  +\frac{1}{2}M^*\comm{L}{\comm{L}{\rho}}+
  \frac12M[L\dg,[L\dg,\rho]]\Big ).
\end{split}
\end{align}
with the Lindblad superoperator on $\rho$ defined as
    \begin{equation} \label{Lindblad}
        \mathcal{D}[L] \rho \defined L \rho L^\dagger - \tfrac{1}{2}(L^\dagger L \rho + \rho L^\dagger L).    
    \end{equation}
A coherent driving field can be added to the squeezed vacuum bath by displacing to the input mode operators $ b \rightarrow b + \alpha(t)$, which effectively transforms the Hamiltonian $H_\text{sys}$ \cite{Mollow1969}. 

Many experiments do not conform to the approximations that led to the \cref{eq:pure-squeezed}, and thus are not modeled by this equation.
Some work has been done to break free of the Markovian regime to model the dynamics of quantum systems interacting with finite-bandwidth squeezed fields~\cite{parkins_rabi_1990,Cirac1992}. 

\noindent{\em Quasi-Markoffian regime:} By considering a driven two-level atom as done in Refs.~\cite{Carmichael_1973, yeoman_influence_1996,Ficek97,Tanas1999,Kowalewska-kudlaszyk:2001aa} and relaxing the broadband requirement such that the squeezing bandwidth can be narrow with respect to the Rabi frequency while still staying broad with respect to the atomic linewidth one can still describe the atomic dynamics with a Markovian master equation.
In this regime interesting modifications to resonance fluorescence appear that are impossible in the broadband setting.

\noindent {\em Cascaded systems and related methods:} Yet another variation on broadband squeezing arises by considering the squeezing to come from a source whose output is then cascaded to serve as the input to the system of interest~\cite{parkins_rabi_1990,GardPar94,KochCarm94,Tanas1999,SmythSwain1999,Messikh2000}. Generally the cascaded approach should let you model arbitrary input fields provided your cascaded model is realistic. 
Earlier work, see Refs.~\cite{Ritsch:1988aa,Ritsch:1988ab}, did a colored noise analysis that is intimately related the to the cascaded model.

Our hope in pursuing a non cascaded model is to attempt to get a more efficient representation and deeper understanding of the unique non-Markovian properties of the field and the light matter interaction.

\section{Fock and Squeezed wave packets}\label{sec:the_field}
The wave packets we consider are collective modes of the field defined as superpositions over temporal modes.
Fields with wave-packet modes in arbitrary states are not stationary in time and may have nontrivial temporal correlations.
Tracking the temporal correlations is the critical feature of any master equation, and these correlations are what drive non-Markovian reduced-state dynamics.

Thus, working with wave packets generally requires abandoning the second Markov approximation.
We begin by introducing temporal wave-packet modes followed by a brief summary of continuous-mode Fock states in~\cref{sec:fieldfock}.
Then, we introduce squeezed Fock wave packets in~\cref{sec:fieldsqfock}, which form the basis for the squeezed master-equation in \cref{sec:master_eqn}.

A temporal wave packet is described by a complex-valued square-normalized function $\xi$ evaluating to $\xi_t$ at time $t$:
    \begin{equation}
        \int dt \abs{\xit}^2 = 1.
    \end{equation}
In order to preserve the rotating-wave approximation (as explained in \cite[Sec. 3.3.1]{garrison_quantum_2008}), we require in our derivation that the wave packet has a \emph{slowly varying envelope}, expressed by the time-domain relations
\begin{align}
  \left|\frac{\partial^2\xi}{\partial t^2}\right|
  &\ll\carrierfreq\left|\frac{\partial\xi}{\partial t}\right|
  \ll\carrierfreq^2|\xi|\,.
\end{align}
This corresponds to the wave-packet bandwidth being much smaller than the carrier frequency $\carrierfreq$, in accord with the quasimonochromatic approximation.

The temporal-mode creation operator that destroys photons in the wave-packet mode defined by $\xi$ is
\begin{align} \label{tempmodeop}
  B^\dagger[\xi]& \defined \int dt\,\xit b^\dagger(t) = \int d\omega\, \tilde\xi_\omega b^\dagger(\omega)   \,,
\end{align}
the second equality gives the equivalent frequency domain description which requires $\int d\omega |\tilde\xi_\omega|^2=1$.
The wave-packet creation operator and its conjugate $B[\xi]$, which creates photons in $\xi$, satisfy the canonical commutation relation
\begin{align}
  \big[B[\xi],B^\dagger[\xi]\big]&=1\,.
\end{align}

\subsection{Fock states}\label{sec:fieldfock}
A continuous-mode Fock state contains a definite number of photons in a wave packet $\xi$. However, these photons do not have well defined arrival times (at a detector or system) but rather are in a temporal superposition weighted according to $\xi$.
A continuous-mode Fock state with $n$ photons, $\ket{n_\xi}$, is generated from vacuum by repeated application of the wave-packet creation operator $B^\dagger[\xi]$ (followed by normalization):
\begin{align} \label{nphotonstate}
  \nket{n_\xi}&= \frac{1}{\sqrt{n!}} \big( B^\dagger[\xi] \big)^n \nket{\text{vac}}\,.
\end{align}
These Fock states are orthogonal $\niprod{m_\xi}{n_\xi} = \delta_{m,n} \,$, and it is evident that the wave-packet creation and annihilation operators behave as single-mode operators for the temporally nonlocal mode defined by $\xit$.

For master-equation derivations, it can be useful to know the action of the infinitesimal annihilation number operators, \cref{noiseincrements} and \cref{lambdanoiseincrements}, on basis states. For Fock states, the relations are given by~\cite{baragiola_n-photon_2012},
\begin{subequations}  \label{Fockincaction}
\begin{align}
  \dB\nket{n_\xi}
  &=
  dt\,\xit \sqrt{n}\,\tightket{(n-1)_\xi}
  \\
  \dLam\nket{n_\xi}
  &=
  \dBdag\,\xit \sqrt{n}\,\tightket{(n-1)_\xi}\,.
\end{align}
\end{subequations}

These relations suggest an interesting temporal structure that we now examine further by performing a relative state decomposition of the field.
The field is decomposed with respect to an arbitrary, infinitesimal time interval $[t, t + dt)$.
It turns out that a complete basis for field states in the interval $[t,t+dt)$ can be constructed from the infinitesimal states $ \ket{0_t}$ and $\ket{1_t}$ where these kets denote vacuum and a single photon  \cite[Sec. V A]{JackColl00}.
A normalized single-photon state in the current time interval is \cite{wiseman_quantum_2010}
\footnote{Technically here we want to consider discrete and finite-duration operators and then take the limit $\lim_{\Delta t\rightarrow0}\Delta B/\sqrt{\Delta t}$. When we play fast and loose by writing $\sqrt{dt}$, this is what we mean.}
\begin{align} \label{Eq::InfSinglePhoton}
  \ket{1_t} \defined \frac{\dB}{\sqrt{dt}} \ket{0_t}.
\end{align}
Using $ \ket{0_t}$ and $\ket{1_t}$ as basis for a relative-state decomposition of an $n$-photon Fock state yields~\cite{baragiola_quantum_2017}
\begin{align}\label{Eq::Focktempdecomp}
  \begin{split}
    \ket{n_\xi}
    &=
    \big (\noprod{0_t}{0_t}+\noprod{1_t}{1_t} \big) \nket{n_\xi}
    \\
    &=
    \ket{0_t} \otimes \ket{n_{\overline{t}}}
    + \sqrt{n \, dt} \, \xit \ket{1_t} \otimes \ket{(n-1)_{\overline{t}}},
  \end{split}
\end{align}
where $\ket{n_{\overline{t}}} \defined \niprod{0_t}{n_\xi}$ is the (unnormalized) partial projection of Fock state $\ket{n_\xi}$ onto infinitesimal vacuum.
Further details about the temporal decomposition of Fock states can be found in~\cite{baragiola_quantum_2017}.
This expression is interpreted by imagining a detector that can measure in this instantaneous basis. If you get a vacuum detection then you would ensure that the remainder of the field is in an $n$-photon Fock state, if you detect one photon then it is in an $n-1$ photon Fock state.
In  either  case,  the  remaining  wave packet  mode  has  had  the  portion  at  the time interval removed.
If however you could detect in the basis $\ket{\pm}\defined (\ket{0_t} \pm \ket{1_t})/\sqrt{2}$ and got the result $\ket{+}$ you would have projected onto an infinitesimal superposition of $n$ and $n-1$.

\subsection{Squeezed Fock states}\label{sec:fieldsqfock}
With respect to the wave packet $\xi$, the wave-packet-mode operators, \cref{tempmodeop}, function just as single-mode operators do. Here, we use this fact to define a wave-packet squeezing operator  
\begin{subequations}
\begin{align} \label{tempsqop}
  \Sgx & \defined \exp\left[\frac{r}{2}\left(e^{-2i\phi}B[\xi]^2
  -e^{2i\phi}B^\dagger[\xi]^2\right)\right] \\
  \gamma&=re^{2i\phi},\label{eq:gam_param}
\end{align}
\end{subequations}
with squeezing factor $\gamma$ that relates the squeezing parameter $r$ and squeezing angle $\phi$.\footnote{It is worth noting \cref{tempsqop} is not the most general type of squeezing operator. For example one could consider squeezing operators such as $\exp(\gamma^* \Psi -\gamma \Psi ^\dagger)$, where $\Psi = \int d s_1 d s_2 \, \psi(s_1,s_2) b(s_1)b(s_2)$ and $\psi(s_1,s_2)$ is a temporal distribution function.}
Applying the squeezing operator to Fock states generates squeezed Fock states,
\begin{equation}
        \label{eq:squeezed-number-state-def}
        \nket{n_{\gamma,\xi}} \defined \Sgx  \ket{n_\xi}
        \, .
\end{equation}
with the squeezed vacuum state
\begin{align}
  \nket{0_{\gamma,\xi}}& \defined \Sgx \nket{\text{vac}}
  \, ,
\end{align}
being of particular interest in this work.
Since squeezed wave-packet Fock states are fixed unitary transformations on Fock states in \cref{nphotonstate}, they are similarly orthogonal,$ \niprod{m_{\gamma,\xi}}{n_{\gamma,\xi}} = \delta_{m,n}$,
and form a complete basis for states in the temporal mode $\xit$.

The wave-packet-mode quadratures are defined with angle $\varphi$:
\begin{align}\label{eq:quad_def}
  Q_\varphi[\xi]=( e^{-i\varphi}B[\xi]+e^{i\varphi}B^\dagger[\xi])/\sqrt{2}.
\end{align}
Thus for the squeezing angle $\phi=0$, the $x$ quadrature fluctuations (\ie{} the standard deviation of $Q_{\varphi=0}[\xi]$)  are scaled by a factor of $e^{-r}$ and $p$ quadrature fluctuations are scaled by a factor of $e^r$. When $\phi = \pi/2$ then $p$ is squeezed.

It is interesting to consider how wave-packet annihilation operator $B[\nu]$ transforms under wave-packet squeezing $\Sgx$ when the temporal modes are mismatched,
\begin{align}\label{eq:squeezed-wavepacket-op-mismatch}
  &\Sgxdag\,B[\nu]\,\Sgx =\\
  & B[\nu]+
 \Big[(\cosh r-1)B[\xi]-e^{2i\phi} \sinh r B\dg[\xi]\Big ]\int\! dt \,\nu_t^* \xit ,\nonumber
\end{align}
which can be derived from the results in \cref{app:lie_algebra}. The dissonance between this expression and single-mode case in \cref{eq:squeezing-transformation} arises from temporal mode mismatch.
When the modes are temporally matched, \ie{} $\nu_t=\xit$, then \cref{eq:squeezed-wavepacket-op-mismatch} becomes
\begin{subequations}\label{eq:squeezed-wavepacket-op}
\begin{align}
  \Sgxdag\,B[\xi]\,\Sgx 
  &=   \coshr B[\xi]-e^{2i\phi} \sinhr B\dg[\xi]\\
  \coshr&\defined\cosh r
  \\
  \sinhr&\defined\sinh r,
\end{align}
\end{subequations}
as expected from the single-mode case in \cref{eq:squeezing-transformation}.

Our next task is to determine the action of the noise increments on squeezed vacuum.
The easiest way to figure out the action of the increments is to perform the following trick
\begin{align}\label{eq:trick2}
  \begin{split}
  \dB\nket{0_{\gamma,\xi}}
  &= 
  \Sgx \Sgxdag \dB \Sgx \ket{\rm vac}
  \\
  &=
  \Sgx \dBsq \ket{\rm vac}\, ,
  \end{split}
\end{align}
where we define a squeezed white-noise increment
\begin{align}\label{eq:squeezed_noise_inc_dB}
  \begin{split}
  \dBsq
  &\defined
  \Sgxdag\,\dB\,\Sgx
  \\
  &=
  \dB+dt\,\xi_t\big[(\tight{\coshr-1})B[\xi]-e^{2i\phi} \sinhr B\dg[\xi]\big]\,.
  \end{split}
\end{align}
This right-hand side of the above equation has a mix of $\dB$ and $B[\xi]$ in it, on account of squeezing the $\dB$ operator with a squeezer defined using the $B[\xi]$ operators.
This can be intuitively explained by the temporal mode mismatch between $\dB$ and $\Sgx$. That is the mix occurs because we're squeezing the infinitesimal-mode operator $\dB$, effectively a square infinitesimal wave packet, with the finite wave-packet squeeze operator $\Sgx$.

By analogy we also define an infinitesimal squeezed number operator
\begin{align}\label{eq:squeezed_noise_inc}
  \begin{split}
  \dLamsq
  &\defined
  \Sgxdag\,\dLam\,\Sgx
  \\
  &= 
  \dLam+\xit\dBdag\big[(\tight{ \coshr-1})B[\xi]-e^{2i\phi}   \sinhr B\dg[\xi]\big]
  \\
  &\phantrel{=}{}
  +\xi^*_t\big[(\tight{ \coshr-1})B\dg[\xi]-e^{-2i\phi} \sinhr B[\xi]\big]\dB
  \\
  &\phantrel{=}{}
  +dt\,\abs{\xit}^2\big[(\tight{ \coshr-1})B\dg[\xi]-e^{-2i\phi} \sinhr B[\xi]\big]
  \\
  &\phantrel{=}{}
  \times \big[(\tight{ \coshr -1})B[\xi]-e^{2i\phi} \sinhr B\dg[\xi]\big]\,.
  \end{split}
\end{align}
The appearance of terms like $ \coshr -1$ again indicates the temporal modes are mismatched which can be seen by conjugating the wave-packet photon-number operator by the squeeze operator $  \Sgxdag\,B\dg[\xi]B[\xi]\,\Sgx $.

The expressions for $\dBsq$ and $  \dLamsq$ may look unwieldy, but they are very convenient for the derivations performed in \cref{sec:master_eqn}. 

In the derivation of the master equation, 
we will require the action of the squeezed increments $\dBsq$ and $\dLamsq$ on Fock states:
\begin{subequations}\label{eq:sq_quantum_noise_on_fock}
\begin{align}
  \begin{split}
  \dBsq\nket{n_\xi}
  &=
  dt\,\xit\big(\coshr\sqrt{n}\,\tightket{(n-1)_\xi}
  \\
  &\phantrel{=}\hphantom{dt\,\xit\big(}{}
  -e^{2i\phi}\sinhr\tightsqrt{n+1}\,\tightket{(n+1)_\xi}\big)
  \end{split}\label{eq:dB_on_ketN}
  \\
  \begin{split}
  \dLamsq\nket{n_\xi}
  &=
  \dBsqdag\,\xit\big(\coshr\sqrt{n}\,\tightket{(n-1)_\xi}
  \\
  &\phantrel{=}\hphantom{\dBsqdag\,\xit\big(}{}
  -e^{2i\phi}\sinhr\tightsqrt{n+1}\,\tightket{(n+1)_\xi}\big).
  \end{split}
\end{align}
\end{subequations}
Notice that \cref{eq:dB_on_ketN} is analogous to the action of $b_{\text{sq}}$, see \cref{eq:squeezing-transformation}, on single mode Fock states $\ket{n}$. We can also use the action of the squeezed increments to work out two-time correlation functions, such as
\begin{subequations}
\begin{align}
    \nbra{n_{\gamma,\xi}}b_{t+\tau}b_t\nket{n_{\gamma,\xi}}
    &=
    \frac{1}{dt}\nbra{n}dB_{t+\tau,\text{sq}}dB_{t,\text{sq}}\nket{n}
    \\
    &=
    -dt\,\xi_{t+\tau}\xi_t \coshr \sinhr e^{2i\phi}(2n+1)
    \, .
\end{align}
\end{subequations}
Because of the wave packet these correlation functions necessarily depend on $t$ and $\tau$ rather than $\tau$ alone as is the case for stationary fields.

The first and second moments of $\dBsq$ and $\dBsqdag$ in the wave-packet Fock state $\ket{n_\xi}$ (equivalently, the moments of $\dB$ and $\dBdag$ in the squeezed-wave-packet Fock states $\ket{n_{\gamma,\xi}}$) are
\begin{subequations}\label{eq:wavepkt-bath-stats}
\begin{align}
  \nexpt{  \dBsq} &= 0 \\
  \nexpt{ \dBsqdag  \dBsq}
  &= \df t^2\, |\xit|^2 [\coshr^2 n + \sinhr^2 (n+1) ]=0, \\
  \nexpt{ \dBsq^2} &= -\df t^2\, \xit^2 \coshr \sinhr e^{2i\phi} ( 2n+1) =0, \\
  \nexpt{[\dBsq,\dBsqdag]} &= \df t
  \,,
\end{align}
\end{subequations}
with terms proportional to $dt^2$ vanishing.
Additionally, the first moment of the squeezed infinitesimal number operator is
\begin{align} \label{eq:dlamexp}
  \nexpt{\dLamsq} &= \df t\, |\xit|^2 [\coshr^2 n + \sinhr^2 (n+1) ], 
\end{align}
which can be can be understood intuitively as $\nexpt{ \dBsqdag  \dBsq} /dt$.
These expressions can be useful calculating output field quantities.

We make an important comparison between the wave-packet moments in \cref{eq:wavepkt-bath-stats} and those for stationary broadband squeezed fields in \cref{eq:bath-stats}.
First, notice that for $n=0$, we recover expressions {\em close} to the broadband case 
for stationary squeezed vacuum. 

The broadband case is recovered when the wave packet is chosen to be $\xit \sim 1/\sqrt{\df t}$, giving 
\begin{subequations}
\begin{align}
  \nexpt{\dBsqdag \dBsq}
  & \longrightarrow
  \sinh^2r\,\df t
  \\
  \nexpt{\dBsq^2}
    & \longrightarrow
  - \, e^{2i\phi} \cosh r \sinh r\,\df t ,
\end{align}
\end{subequations}
which, noting that $\nexpt{\dBsqdag \dBsq}$ and $\nexpt{\dBsq^2}$ are no longer zero, gives the explicit expressions for the $M$ and $N$ parameters in \cref{eq:Npuresqueezed}.
For clarity in this comparison, we have used the notation $\cosh r$ and $\sinh r$ instead of $\coshr$ and $\sinhr$ here.
A striking difference is that the infinitesimal photon number, \cref{eq:dlamexp}, is no longer infinitesimal for stationary squeezed vacuum: 
\begin{equation} \label{brokendLam}
  \nexpt{\dLamsq}  \longrightarrow  \sinh^2 r
  \, .
\end{equation}
This expression leads to unphysical consequences, most notably that the photon flux for stationary squeezed fields is infinite---a problem that has been identified previously and remained unresolved until now, see \cite{wiseman_quantum_2010}.
The resolution is the wave-packet description we present here, which for any square-normalizable wave packet describing a physical state and gives a finite photon flux and finite total average photon number (in the wave packet).
Indeed, the choice of wave packet required to recover the stationary squeezed field relations is $\xit = 1/\sqrt{\Delta t}$.
To arrive at the continuous-time description we must take the limit $\Delta t\to 0$ which gives the problematic expression $\xit \sim 1/\sqrt{\df t}$. 
In \cref{sec:shrt-wp}, we will show how this same choice of wave packet also yields that standard broadband squeezing master equation from the squeezed wave-packet master equations we present below.

The temporal structure of squeezed Fock states can be analyzed using a temporal decomposition, just as was done above for Fock states in \cref{sec:fieldfock}.
This procedure, whose details are presented in \cref{app:temp_decomp}, gives the resulting temporal decomposition:
\begin{align}
  \nket{n_{\gamma,\xi}} \nonumber
  &= \nket{0_t}\otimes\Big \{\nket{n_{\gamma,\overline{t}}}
  -\tfrac{1}{2}\abs{\xi_t}^2dt\, 
  \Big([n(2\sinhr^2+1)+\sinhr^2]\nket{n_{\gamma,\overline{t}}}
  \vphantom{\tightsqrt{(n+1)}} \nonumber
  \\
  &\phantrel{=}\hphantom{\nket{0_t}\otimes\Big[}
  {}-\coshr\sinhr\Big[e^{-2i\phi}\tightsqrt{n(n-1)}\tightket{(n-2)_{\gamma,\overline{t}}}+\nonumber
  \\
  &\phantrel{=}\hphantom{\nket{0_t}\otimes\Big[{}-\coshr \sinhr \Big[}{}
  e^{2i\phi}\tightsqrt{(n+1)(n+2)}\tightket{(n+2)_{\gamma,\overline{t}}}\Big]\Big)
  \Big\} \nonumber
  \\
  &\phantrel{=}{}+\nket{1_t}\otimes\Big[ \xi_t\sqrt{dt}\,
  \big(\coshr\sqrt{n}\tightket{(n-1)_{\gamma,\overline{t}}} \nonumber\\
  &
 \phantrel{=+\nket{1_t}\otimes\Big[ }{} -e^{2i\phi}\sinhr\tightsqrt{n+1}\tightket{(n+1)_{\gamma,\overline{t}}}\big) \Big ]
  \label{eq:squeezed-temp-decomp}
\end{align}
Like in the Fock-state case this expression is interpreted by imagining a detector that can measure in this instantaneous basis. 
This decomposition can be interpreted in a similar way as was the Fock-state case, \cref{Eq::Focktempdecomp}, with very different conclusions. Consider the conditional state after an ideal, infinite resolution photon detector performs a measurement in the $\{\ket{0_t}, \ket{1_t} \}$ basis over the infinitesimal interval $[t, t + dt)$. Vacuum detection projects the remainder of the field into a squeezed $n$-photon state whose squeezing strength is gently decreased from $r$ to $r-|\xi_t|^2dt\,\cosh r\sinh r$ by the order-$dt$ corrections (see \cref{app:temp_decomp} for details).
Detection of a photon projects the remaining field into a non-Gaussian state---a superposition of squeezed $(n-1)$- and $(n+1)$-photon Fock states with relative amplitudes that depend on the squeezing $r$ in the initial state.

We can also read off the probability of detecting a single photon in the time interval $[t,t+dt)$ from \cref{eq:squeezed-temp-decomp}
\begin{subequations}
\begin{align}
    \Pr(1_t)&=|\xi_t|^2dt\,[n(2\sinhr^2+1)+\sinhr^2]\label{eq:pr_click}\\
    \Pr(0_t)&=1- \Pr(1_t),
\end{align}
\end{subequations}
or equivalently from \cref{eq:dlamexp}.
The fact that the broadband case has infinite photon flux, as noted earlier, means that the temporal decomposition is a product state of squeezed states at each time $t$, giving an order unity probability of photon detection for each infinitesimal time interval $[t,t+dt)$. Determining the probabilities for subsequent detection events requires the machinery we develop in \cref{sec:SMEs}.

\section{Squeezed-wave-packet Master equation}\label{sec:master_eqn}

In this section, we present the dynamical equations for a quantum system $\rho\system$ interacting with a wave packet $\xit$  prepared in a the state $\ket{0_{\gamma,\xi}}$. Initially, the system and field are initially are described by the product state  $\rho_0 \otimes \oprod{0_{\gamma,\xi}}{0_{\gamma,\xi}}$, and at later times they have interacted and are, in general, entangled. The squeezed-wave-packet master equations below are a set of dynamical equations whose solution gives the time evolution of the reduced system state
\begin{align} \label{reducedstate}
  \rho(t)
  & \defined
  \tr\field\big[\Ut(\rho_0\otimes\noprod{0_{\gamma,\xi}}{0_{\gamma,\xi}})\Utdag\big] \, ,
\end{align}    
to save space we denote $\rho(t)$ and $d \rho(t)$ as $\rho_t$ and $d\rho_t$.
The details of the derivation are given in \cref{app:mastereqnderivation}.
Here we highlight the crucial points in the derivation and state the main result. At the end of this section we explain how to generalize to the case where the initial field is expressed as a superposition or mixture of squeezed Fock states.

At the nucleus of the derivation is the QSDE for the time-evolution operator, $\dU$, given in \cref{eq:dU_propagator}.
To find squeezed wave packet dynamics, it is useful to define a \emph{squeezed time-evolution operator} by conjugating $\Ut$ with the wave packet squeezing operator in \cref{tempsqop}, $\Utsq \defined \Sgxdag \Ut \Sgx$. In this way, the effects of wave-packet squeezing are separated from the input field state and are entirely contained mathematically in $\Utsq$, effectively moving to a squeezed frame of reference. 
The QSDE for $\Utsq$ is similarly found by conjugating $\dU$ by the squeeze operator $\Sgx$:
\begin{align} 
  \begin{split} 
  \dUsq
  &\defined
  \Sgxdag\,\dU\,\Sgx 
  \\
  &=
  \big[ {-}dt(iH\system+\tfrac{1}{2}L^\dagger L)-L^\dagger S\otimes\dBsq 
  \\
  &\phantrel{=}{}
  +L\otimes\dBsqdag
  +(\tight{S-\Id\system})\otimes \dLamsq\big ]\Utsq.\label{squeezeddU}
  \end{split}
\end{align}
This differs from \cref{eq:dU_propagator} by the appearance of the squeezed noise increments derived in \cref{eq:squeezed_noise_inc_dB,eq:squeezed_noise_inc}.

Because the time evolution operator can add and subtract quanta, 
it is useful to define the generalized state matrices,
\begin{align}\label{eq:rhomn}
  \begin{split}
  \rhonorm{t}{m,n}
  & \defined
  \tr\field\big[\Ut(\rho_0\otimes\noprod{m_{\gamma,\xi}}{n_{\gamma,\xi}})\Utdag\big]
  \\
  &=
  \tr\field\big[\Utsq(\rho_0\otimes \noprod{m_{\xi}}{n_{\xi}})\Utsqdag \big]\,,
  \end{split}
\end{align}
where $m$ and $n$ denote the number of squeezed photons in the initial field, $\xi$ is the wave-packet envelope, and $\gamma$ is the squeeze parameter. The definition in \cref{eq:rhomn} implies that $\rhonorm{}{m,n}= [\rhonorm{}{n,m}]^\dagger$.
These generalized state matrices encode information about correlations between the future field and the state of the system. Further, note that the reduced state in \cref{reducedstate} is $\rho\system(t) = \rhonorm{t}{0,0}$.

Our derivation proceeds by finding the QSDE for the joint-state evolution and then taking a partial trace over the field~\cite{baragiola_n-photon_2012}. Thus the physical state of the system, given squeezed vacuum input, is always described by $\rhonorm{t}{0,0}$. As the interaction between the environment and system can add or subtract photons from the field we end up having to compute equations of motion for  $\rhonorm{t}{m,n}$. 
Thus the broad structure of the derivation corresponds to computing the stochastic differentials of the time-evolved generalized state matrices, \cref{eq:rhomn},
\begin{align}\label{eq:gen-state-matrix-trace}
\begin{split}
  d\rhonorm{t}{m,n}
  &=
  \tr\field\big[\dUsq(\rho_0\otimes\noprod{m_\xi}{n_\xi}\dg)\Utsqdag
  \\
  &\phantrel{=}{}
  \hphantom{\tr\field\big[}
  +\Utsq(\rho_0\otimes\noprod{m_\xi}{n_\xi}\dg)\dUsqdag
  \\
  &\phantrel{=}{}
  \hphantom{\tr\field\big[}
  +\dUsq(\rho_0\otimes\noprod{m_\xi}{n_\xi})\dUsqdag\big].
\end{split}
\end{align}
Again, the derivation is facilitated by working in a picture where the time-evolution operator itself carries the squeezing. Inserting the QSDE for $\dUsq$, \cref{squeezeddU}, and going through the machinations (detailed in \cref{app:mastereqnderivation}),
we arrive at the coupled master equations,
\begin{widetext}
\begin{align}\label{eq:mainresult}
  \begin{split}
    d\rhonorm{t}{m,n}
    &=
    \Big\{
    -i[H\system,\rhonorm{t}{m,n}] + \mathcal{D}[L]\rhonorm{t}{m,n}
    \\
    &\phantrel{=}\hphantom{\Big[}{}
    +\xi_t\coshr\sqrt{m}[S
    \rhonorm{t}{m-1,n},L^\dagger]
    +\xi^*_t\coshr\sqrt{n}\,
    [L,\rhonorm{t}{m,n-1}S^\dagger]
    \\
    &\phantrel{=}\hphantom{\Big[}{}
    +\xi_t\sinhr\,e^{2i\phi}\tightsqrt{m+1}[L^\dagger,S
    \rhonorm{t}{m+1,n}]
    +\xi^*_t\sinhr\,e^{-2i\phi}\sqrt{n+1}\,
    [\rhonorm{t}{m,n+1}S^\dagger,L]
    \\
    &\phantrel{=}\hphantom{\Big[}{}
    +\abs{\xi_t}^2\mathcal{D}[S]\Big(
    \coshr^2\sqrt{mn}\,\rhonorm{t}{m-1,n-1}
    -\coshr\sinhr\,e^{-2i\phi}\tightsqrt{m(n+1)}\,
    \rhonorm{t}{m-1,n+1}\\
    &\phantrel{=}\hphantom{\Big[{}
    +\abs{\xi_t}^2\mathcal{D}[S]\big(}{}
    -\coshr\sinhr\,e^{2i\phi}\tightsqrt{(m+1)n}\,
    \rhonorm{t}{m+1,n-1}
    +\sinhr^2\tightsqrt{(m+1)(n+1)}\,
    \rhonorm{t}{m+1,n+1}\Big)
    \Big\} \,dt\,,
  \end{split}
\end{align}
\end{widetext}
where $\coshr$ and $\sinhr$ are the abbreviations defined in \cref{eq:squeezed-wavepacket-op}.
From \cref{eq:rhomn}, it is clear that the generalized state matrices $\rhonorm{t}{m,n}$ are not defined for negative indices $m,n < 0$; in the master equations above, these terms should be set to zero.
This set of coupled differential equations, one for each $\rhonorm{t}{m,n}$, comprise the \emph{squeezed wave-packet master equations} and is the main result of this section. We also refer to the set of equations as the \emph{squeezed hierarchy}.
The terms on the first line of \cref{eq:mainresult} describe both unitary evolution due to a Hamiltonian and dissipation given by the Lindblad superoperator in \cref{Lindblad}.
The terms on the second and third lines describe driving of the system, or exchange of energy, from quanta in the input field.
The terms on the fourth and fifth lines describe evolution of the system due to direct coupling, via the system operator $S$, to the number of squeezed photons in the field at time $t$.
In the absence of squeezing, \ie{} when $\sinhr=0$ and $\coshr=1$, we recover the hierarchy of Fock-state master equations~\cite{baragiola_n-photon_2012}. 
Finally, with the cyclic property of the trace it is straightforward, albeit tedious, to find the equivalent Heisenberg-picture equations of motion as was done in \cite{baragiola_n-photon_2012}.

The initial conditions for the coupled equations in \cref{eq:mainresult} follow from the definitions of the generalized state matrices in \cref{eq:rhomn} at $t=0$ when $U_0=\Id$.
The diagonal equations $ \rhonorm{0}{n,n}$ are initialized with the initial system state $\rho_{\rm sys}$, while the off-diagonal equations, $ \rhonorm{0}{m,n}$ for $m\neq n$, should be initialized to zero:
\begin{subequations}\label{eq:hierarchy-init-cond}
\begin{align}
  \rhonorm{0}{n,n}
  &=
  \rho_{\rm sys}
  \\
  \rhonorm{0}{m,n}
  &=
  \Zero .
\end{align}
\end{subequations}

More generally, since squeezed Fock states comprise a basis for states in the wave packet $\xit$---see the text below \cref{eq:squeezed-number-state-def}---we can use them to describe an \emph{arbitrary} input field state given by the state matrix 
\begin{equation} \label{fieldstate}
        \Upsilon_\xi = \sum_{m,n} c_{m,n} \ket{m_{\gamma,\xi}}\bra{n_{\gamma,\xi}}
        \, ,
\end{equation}
with $\tr[\Upsilon_\xi] = 1$. The reduced system state at time $t$, $\rho{\system}(t)$, is found by taking superpositions of the hierarchy solutions, $\rhonorm{t}{m,n}$, using the expansion coefficients in \cref{fieldstate}; 
    \begin{equation} \label{rhosysdecomp}
        \rho\system(t) = \sum_{m,n} c_{m,n} \, \rhonorm{t}{m,n}
        \,
    \end{equation}
see Refs.~\cite{baragiola_n-photon_2012,baragiola_quantum_2017} for further details. This means that a formal solution to the squeezed hierarchy (\ie{} for all $\rhonorm{t}{m,n}$) can be used to construct reduced system states for \emph{all possible} input field states in the wave packet $\xit$ given initial system state $\rho\system$.

In practice, however, there are two issues that arise. First, such a solution consists of an infinite number of $\rhonorm{t}{m,n}$. Second, in general it is not clear that a formal solution exists for an infinite set of coupled differential equations.
To implement numerical simulations we need to enforce approximations on the squeezed hierarchy.
There are two obvious strategies to do so which have been called time-non-local and time local in the ``hierarchy equations of motion'' literature see \eg{}~\cite{Strumpfer2012}.
The simplest strategy is to set all equations above a chosen truncation, $\nmax$, to zero: $ \rhonorm{t}{\nmax+m',\nmax+n'}= \Zero$ for some integers $m',n'>0$. This kind of time-non-local truncation does not seem well justified as the scalar coefficients of the hierarchy grow with $m,n$. That is, even for $r \ll 1$, there exist $m,n$ for which products that appear in the upward-coupling terms such as $\sinhr \sqrt{m}$ become non-negligible.

The second strategy is to assert that all equations above the cutoff $\nmax$ get replaced with some approximate equation or solution.
There are good reasons to believe this is a more appropriate way to approximate the squeezed hierarchy. Consider the case where the operator norms of $||H\system||$ and $||L||$ are bounded by the scalar $|k|$ and $S=I$, the latter condition implies the last term in \cref{eq:mainresult} is zero. Then for large $(m,n)$ the term on the first line is negligible while the terms on the second and third lines of \cref{eq:mainresult} dominate. This suggests the solutions of these equations may be derived using tools from singular perturbation theory~\cite{neu2015singular} such as boundary layers and matched asymptotic expansions. Moreover it indicates that numerical solutions will have to ensure that $\max_t(\xit) \nmax\coshr \Delta t \ll 1 $.

\subsection{Connection to the Fock hierarchy}
As noted above \cref{eq:mainresult} limits to the \emph{Fock hierarchy} of Ref.~\cite{baragiola_quantum_2017} when $r=0$ \ie{} $\sinhr=0$ and $\coshr=1$, and \cref{eq:mainresult} becomes exactly the Fock hierarchy~\cite{baragiola_n-photon_2012}. Note that either hierarchy, including the squeezed hierarchy with any squeezing parameter $r$, can describe the reduced-state dynamics given \emph{any} input field state. One must simply represent that input field state in the appropriate squeezed Fock-state basis (or unsqueezed for the Fock hierarchy) as in \cref{fieldstate}. 
 
\begin{figure}[t]
\centering 
\includegraphics[width=0.9\columnwidth]{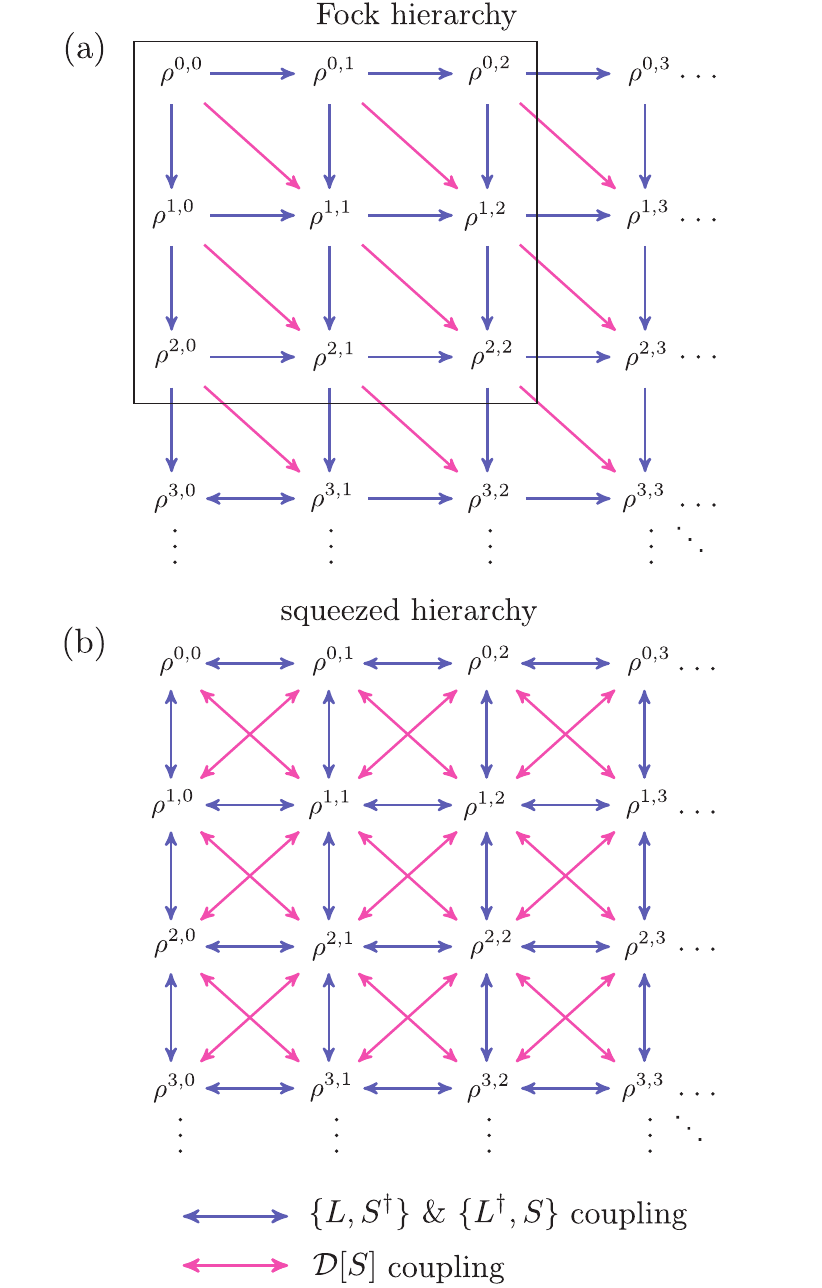}
\caption{Graphical depictions of the couplings between equations in the (a) Fock hierarchy and (b) squeezed hierarchy of master equations. An arrow from $\rhonorm{t}{m,n}$ to $\rhonorm{t}{m',n'}$ indicates that the differential equation for $\rhonorm{t}{m',n'}$ depends on $\rhonorm{t}{m,n}$. In the absence of squeezing $(r=0)$, the squeezed hierarchy becomes the Fock hierarchy. All coupling terms proportional to $\sinhr$ ($\sinh r$) vanish, and the structure of the hierarchy changes drastically such that the equations for all $\rhonorm{t}{m,n}$ do not depend on any $\rhonorm{t}{m',n'}$ with larger labels. This feature allows closed subsets of the Fock hierarchy to be solved---an example for $n_\text{max} = 2$ is indicated by the black box in (a).
} \label{fig:dynamicalmapcartoons}
\end{figure}

For no squeezing, the Fock hierarchy emerging from \cref{eq:mainresult} features master equations that only couple \emph{downward}---that is, the equation for each $\rhonorm{t}{m,n}$ couples to itself and to generalized state matrices with smaller labels---$\rhonorm{t}{m-1,n}$, $\rhonorm{t}{m,n-1}$, and  $\rhonorm{t}{m-1,n-1}$---which are visualized in  \cref{fig:dynamicalmapcartoons}(a).\footnote{The cat-state hierarchy in Refs.~\cite{gough_quantum_2012, dabrowska_quantum_2019} also has this feature, but that hierarchy is not directly related to the squeezed hierarchy here.} Alternatively one can think of the information flowing from the generalized state matrices $\rhonorm{t}{m-1,n}$ upwards to $\rhonorm{t}{m,n}$.
A consequence of downward coupling is that subsets of the hierarchy couple only amongst themselves and can be solved. Specifically, for a fixed, chosen Fock cutoff $n_\text{max}$, there is a closed set of $(n_\text{max}+1)^2$ coupled equations whose solutions can be used to construct the reduced system state given any input field state that can be represented in the basis of Fock states $\{0,1,\dots n_\text{max}\}$~\cite{baragiola_n-photon_2012, Baragiola:2014aa, baragiola_quantum_2017}, including truncated approximations to squeezed states.

For nonvanishing squeezing, $r > 0$, the equations in the squeezed hierarchy, \cref{eq:mainresult}, not only contain downward coupling (modified by squeezing-parameter dependencies $\cosh r$ and $\sinh r$), but they also contains \emph{upward} coupling to generalized state matrices with larger labels: $\rhonorm{t}{m+1,n}$, $\rhonorm{t}{m,n+1}$, $\rhonorm{t}{m-1,n+1}$, $\rhonorm{t}{m+1,n-1}$, and $\rhonorm{t}{m+1,n+1}$. These couplings are visualized in \cref{fig:dynamicalmapcartoons}(b). Unlike the Fock hierarchy, there is no closed subset of equations, so the entire hierarchy must be solved at once. Since there is an infinite number of coupled equations, the existence of formal solutions is not guaranteed and it remains to be shown that solutions exist.

\subsection{Output field quantities}
In addition to system observables, we may also be interested in features of the output field.
Of the many possible field operators we will focus on two field quantities: photon flux (photon number) and field quadratures.

Following the same prescription as in \cref{eq:trick2}, we may derive the equation of motion for an arbitrary field quadrature  $Q(\varphi)$ defined in \cref{eq:quad_def}. The key result needed is the field operator after the interaction
\begin{align}\label{eq:dbsq_out}
\dBsq^{\rm out} 
&=  \Sgxdag \Upropdag{t+dt}{t} \dB  \Uprop{t+dt}{t}   \Sgx  \nonumber \\
& =  L dt  + S \dBsq.
\end{align}
Expectation values are found by taking the trace of the output field operators using the physical state $\rho\system(t) = \rhonorm{t}{n,n}$ (for a squeezed Fock input, squeezed vacuum is the case where $n=0$) along with the expectation in the field using \cref{eq:wavepkt-bath-stats}, 
\begin{align}
        \expt{\dBsq^{\rm out} }  = dt \tr [ L \rhonorm{t}{n,n} ] .
\end{align}
This equation looks like an awkward mix of the Heisenberg and \sch\ pictures.
One can understand this as the temporal field mode $\dB$ scattering off a system initially in the time-evolved \sch\ state $\rhonorm{t}{n,n}$.
To arrive at a quadrature $Q(\varphi)$ 
we combine $\dB^{\rm in}$ with its dagger as in \cref{eq:quadrature-increment}.
Then we just take expectation with respect to vacuum or a Fock state.

The output photon-flux operator follows similarly,
    \begin{align}
        \dLamsq^{\rm out} &=  \Sgxdag \Upropdag{t+dt}{t} \dLam  \Uprop{t+dt}{t}   \Sgx  \nonumber \\
        & =  L\dg L dt  + L\dg S \dBsq + S\dg L \dBsqdag + \dLamsq \, .
    \end{align}
Its expectation value with respect to a squeezed Fock state is given by \cref{eq:wavepkt-bath-stats}
\begin{equation}
  \expt{\dLamsq^{\rm out}}
  =
   dt\,\tr [L\dg L \rhonorm{t}{n,n} ]
   +\df t\, |\xit|^2 [\coshr^2 n + \sinhr^2 (n+1) ].
\end{equation}

To compute higher order moments, such as a quadrature variance, one must derive the QSDE for operators like $ [dQ_t^{\rm out}(\varphi)]^2$, and then take expectation values of the resulting expression.
Moreover, if one wants to compute first or higher-order moments of output-field quantities for states other than a squeezed Fock state one can use the method described in \cite{baragiola_n-photon_2012}.

\subsection{System correlation functions: quantum regression theorem}\label{sec:qrt-wavepacket}

The squeezed-Fock master equation hierarchy can be used to calculate two-time correlation functions. For \sch-picture system operators $A \otimes I$ and $B \otimes I$, the two-time correlation $\langle A_t B_{t+\tau}\rangle$ is calculated using the quantum regression theorem \cite{Lax:1963aa} as follows. Given an initial joint system-field state, $\rho_\text{sys,field}(0)$, we have
\begin{align}
    \langle A_t B_{t+\tau}\rangle
    &=\tr\system\big(B\Lambda_{t+\tau,t}\big) \, ,
\end{align}
where we have defined the two-time system operator
\begin{align}  
  \label{eq:qrt-op-def}
  \Lambda_{t+\tau,t}&\defined\tr_{\text{field}}\big[\Uprop{t+\tau}{t}
  \rho_{\text{sys},\text{field}}(t)A\,\Upropdag{t+\tau}{t}\big]\,.
\end{align}
The operator $\Lambda_{t+\tau,t}$ evolves as a function of $\tau$ according to the open-system evolution given by the master equation, with initial condition ($\tau = 0$)
\begin{align}
  \Lambda_{t,t}&=\tr_{\text{field}}\big[\rho_{\text{sys},\text{field}}(t)A\big]
  =\rho_tA\,.
\end{align}
where $\rho_t$ is the reduced system state at time $t$.

When the open-system evolution of the reduced state is governed by a hierarchy of equations as in \cref{eq:mainresult}, there an associated hierarchy of two-time operators \cite{sathyamoorthy_quantum_2014, Baragiola:2014aa}
\begin{align}
  &\Lambda^{\hieridx{m,n}}_{t+\tau,t} \defined \nonumber \\
  &\tr_{\text{field}}\,\Big[\Uprop{t+\tau}{t}\big[\Uprop{t}{0} \rho\system(0)\otimes \noprod{m_{\gamma,\xi}}{n_{\gamma,\xi}}) \Upropdag{t}{0}\big]A\,\Upropdag{t+\tau}{t}\Big]\,.
\end{align}
These operators evolve from time $t$ to $t+\tau$ according to the same Schr\"{o}dinger-picture hierarchy of master equations as the generalized state matrices $\rhonorm{t}{m,n}$ do, \cref{eq:mainresult}, with initial conditions ($\tau = 0$),
\begin{align} \label{eq:QRT_boundaryconds}
  \Lambda^{\hieridx{m,n}}_{t,t}
  =\rhonorm{t}{m,n}A\,.
\end{align}
Calculations of two-time correlations, $\langle A_t B_{t+\tau}\rangle$, are performed in the following way:  first, find all of the generalized state matrices at time $t$, $\rhonorm{t}{m,n}$, and use them to calculate the boundary conditions in \cref{eq:QRT_boundaryconds}.
Second, evolve the hierarchy of two-time operators $\Lambda^{\hieridx{m,n}}_{t+\tau,t}$ from time $t$ to $t+\tau$ using the squeezed master equation hierarchy, \cref{eq:mainresult}.
Finally, take the trace of the two-time operator corresponding to the physical input state with system operator $B$.
For example, when the initial wave-packet is prepared in a squeezed vacuum state, the two-time system correlation function is 
\begin{align}
  \langle A_t B_{t+\tau}\rangle
  &=
  \tr\system\big(B\Lambda_{t+\tau,t}^{\hieridx{0,0}}\big)\,.
\end{align}
Since the input field states we consider are defined within a temporal envelope, in general there is no system steady state, and the correlation functions and related spectra are time dependent.  
\\

\subsection{Compact notation}\label{sec:compact-notation}
The previous section illustrates how equations describing the hierarchy of generalized state matrices, which serve  operators quickly become unwieldy.
Since the coupled equations have a structure that suggests a matrix of matrices, we employ the following notation to tame these expressions:
\begin{equation}
  \rhomat_t\defined\begin{pmatrix}
    \rhonorm{t }{0,0} & \rhonorm{t }{0,1} & \cdots \\
    \rhonorm{t }{1,0} & \rhonorm{t }{1,1} & \cdots \\
    \vdots & \vdots & \ddots
  \end{pmatrix},
 \end{equation}
 which we call the state tensor in analogy to the state vector and state matrix.
 Because each individual matrix obeys $\rhonormdag{t }{m,n} =  \rhonorm{t }{n,m}$ we have
 \begin{equation}
  \rhomat_t^\matdag\defined\begin{pmatrix}
    {\rhonormdag{t }{0,0}} & {\rhonormdag{t }{1,0}} & \cdots \\
    {\rhonormdag{t }{0,1}} & {\rhonormdag{t }{1,1}} & \cdots \\
    \vdots & \vdots & \ddots
  \end{pmatrix}=\rhomat_t
  \end{equation}
where ${}^\matdag$ defines a matrix-wise transpose conjugation of an operator tensor analogous to the Hermitian conjugation ${}^\dagger$.
Next we define a left ``scalar'' multiplication by a matrix:
  \begin{align}
  X\rhomat_t&\defined\begin{pmatrix}
    X\rhonorm{t }{0,0} & X\rhonorm{t }{0,1} & \cdots \\
    X\rhonorm{t }{1,0} & X\rhonorm{t }{1,1} & \cdots \\
    \vdots & \vdots & \ddots
  \end{pmatrix} \, .
\end{align}
Right ``scalar'' multiplication $ \rhomat_tX$ works analogously.
These operations are like standard scalar multiplication of a matrix except for the fact that the ``scalars'' don't commute.
Finally we introduce the tensor \vspace{-2pt}
\begin{equation}
\minusmat\defined\begin{pmatrix}
    0 & \sqrt{1}\Id\system & 0 & \cdots \\
    0 & 0 & \sqrt{2}\Id\system & \ddots \\
    0 & 0 & 0 & \ddots \\
    \vdots & \vdots & \ddots & \ddots \\
  \end{pmatrix}\,.
\end{equation}
Note the striking resemblance of our operator $\minusmat$---introduced here to express the couplings between various levels of the hierarchy that appear in our equations---to the annihilation operator for a harmonic oscillator.
The similarity continues as the structure of the couplings between levels suggests defining a squeezed generalized lowering operator
\begin{align}
  \minusmat_{\text{sq}}&\defined \coshr \minusmat-\sinhr e^{-2i\phi}\minusmat^\matdag\, ,
\end{align}
that resembles the action of a squeeze operator on a harmonic-oscillator annihilation operator given by \cref{eq:squeezing-transformation}.

Now we can express the unconditional master equation \cref{eq:mainresult}, using this notation:
\begin{widetext}
\begin{align}\label{eq:drho_tensor}
    d\rhomat_t&=
    dt\,\bigg[-i[H\system,\rhomat_t] + \mathcal{D}[L]\rhomat_t
    +\xi_t[S\minusmat^\matdag_{\text{sq}}\rhomat_t,L^\dagger]
    +\xi^*_t[L,\rhomat_t\minusmat_{\text{sq}}S^\dagger]
    +\abs{\xi_t}^2\mathcal{D}[S]\big(
    \minusmat^\matdag_{\text{sq}}\rhomat_t
    \minusmat_{\text{sq}}\big)
    \bigg]\,.
\end{align}
\end{widetext}
From these expressions it is easy to see how the vacuum equations are recovered by setting $\xi_t=0$ and how the number wave-packet equations are recovered by setting $\minusmat_{\text{sq}}=\minusmat$. For the case of a field initially in squeezed vacuum, in order to extract the reduced state of the system alone we must extract the matrix $\rhonorm{t }{0,0}$ from the state tensor $\rhomat_t$. Other cases follow by combining appropriate elements of the state tensor.

The initial condition for the  the state tensor  $\rhomat_t$ in \cref{eq:drho_tensor} is implied by \cref{eq:hierarchy-init-cond}.
Specifically, the diagonal matrices in the state tensor $\rhomat_t$ should be initialized with the initial system state $\rho_{\rm sys}$, while the off-diagonal matrices in the state tensor should be initialized to zero.
For squeezed vacuum input, calculating expectation values requires only the density operator $ \rhonorm{t}{0,0}$. If the initial state of the field can be represented as a mixture or a superposition of squeezed Fock states, as in \cref{fieldstate}, then the correct elements of the state tensor need to be extracted and combined; see Ref.~\cite{baragiola_n-photon_2012} for further details.

\subsection{Convergence to the Markovian squeezed master equation for short wave packets}\label{sec:shrt-wp}

The squeezed wave-packet master equation, \cref{eq:mainresult}, converges to the standard master equation for stationary broadband squeezing, \cref{eq:pure-squeezed}, in the limit where the field is described by a pulse train of very short squeezed-vacuum wave packets, which we show here.
For simplicity, we set $S=\Id$ and $H=\Zero$ and work with a rectangular wave packet
\begin{align}
  \xi_t&=\begin{cases}
    1/\sqrt{\dt} & 0\leq t\leq\dt
    \\
    0 & \text{otherwise}
  \end{cases}.
\end{align}
The restriction to $H=\Zero$ is not essential; however, it simplifies the calculation, because the Hamiltonian-driven dynamics are of the same order in $\dt$ as those from the $\D{\cdot}$ Lindblad dissipator---a fact that the arguments below rely on.

The squeezed wave-packet hierarchy gives us the equation of motion for the physical state
$\rhonorm{t}{0,0}$,
\begin{equation} \label{eq:yodawg}
    \rhonormdot{0,0}
    =
    \D{L}\rhonorm{t}{0,0}
    -\xi_t \sinhr e^{2i\phi}[\rhonorm{t}{1,0},L\dg]
    -\xi_t^* \sinhr e^{-2i\phi}[L,\rhonorm{t}{0,1}].
\end{equation}
We assume that at every time slice $\dt$ the system interacts with a wave packet $\xi_t\sim1/\sqrt{\dt}$ much shorter than the any system lifetime \ie{} $\dt\ll1/\lw$ where $L\sim\sqrt{\lw}$.
The strategy is to perform a second-order Taylor expansion of the state,
\begin{align} \label{taylorexpansion}
  \rho_{\dt}
  &\defined \rhonorm{t}{0,0}|_{t=0}+\dt\rhonormdot{0,0}|_{t=0}
   +\tfrac{1}{2}\dt^2\rhonormddot{0,0}|_{t=0} \, ,
\end{align}
in terms of the initial system state, $\rho_0$,
the first-order equation $\rhonormdot{0,0}$ in \cref{eq:yodawg}, and the second-order equation $\rhonormddot{0,0}$, which is found by taking the time derivative of \cref{eq:yodawg}:
\begin{align} \label{rhodoubledot}
\begin{split}
  \rhonormddot{0,0}
  =&    \D{L}\rhonormdot{0,0}
  -\sinhr e^{2i\phi}[\rhonormdot{1,0},L\dg]/\sqrt{\dt} \\
  &
  -\sinhr e^{-2i\phi}[L,\rhonormdot{0,1}]/\sqrt{\dt}.
  \end{split}
\end{align}
Note that the term proportional to $ \D{L}$ is $O(\sqrt{\dt})$. 
Also, the objects $\rhonormdot{1,0}$ and $\rhonormdot{0,1}$, appear in this expression; their equations of motion also follow from the squeezed hierarchy,
\begin{align}
  \begin{split}
    \rhonormdot{1,0}
    &=
    \D{L}\rhonorm{t}{1,0}
    -\frac{\sqrt{2}\sinhr e^{2i\phi}}{\sqrt{\Delta t}}[\rhonorm{t}{2,0},L\dg]
     \\
    &\phantrel{=}{}
    +\frac{\coshr}{\sqrt{\Delta t}}[\rhonorm{t}{0,0},L\dg]
    -\frac{\sinhr e^{-2i\phi}}{\sqrt{\Delta t}}[L,\rhonorm{t}{1,1}],
  \end{split}
\end{align}
Substituting these equations into \cref{rhodoubledot} gives the expression,
\begin{widetext}
\begin{align} \label{rhodoubledot2}
  \begin{split}
    \rhonormddot{0,0}&=
    -\frac{\sinhr\coshr e^{2i\phi}}{\dt}\big[ [\rhonorm{t}{0,0},L\dg],L\dg\big]
    +\sqrt{2}\frac{\sinhr^2e^{4i\phi}}{\dt}\big[ [\rhonorm{t}{2,0},L\dg],L\dg\big]
    +\frac{\sinhr^2}{\dt}\big[ [L,\rhonorm{t}{1,1}],L\dg\big]
    \\
    &\phantrel{=}{}
    +\frac{\sinhr^2}{\dt}\big[L,[\rhonorm{t}{1,1},L\dg]\big]
    -\frac{\sinhr\coshr e^{-2i\phi}}{\dt}\big[L,[L,\rhonorm{t}{0,0}]\big]
    +\sqrt{2}\frac{\sinhr^2e^{-4i\phi}}{\dt}\big[L,[L,\rhonorm{t}{0,2}]\big]\, .
  \end{split}
\end{align}
where we have only kept terms that do not vanish in the Taylor expansion, \cref{taylorexpansion}. In that expression, the second-order term is multiplied by $(\dt)^2$, so any term in $\rhonormddot{0,0}$ of higher order than $1/\dt$ vanishes.

We now substitute \cref{eq:yodawg} and \cref{rhodoubledot2} into the Taylor expansion for the state, \cref{taylorexpansion}, to get
\begin{align} \label{stateequationyowza}
  \begin{split}
    \rho_{\dt}
    &=\rho_0+\dt\big ( \D{L}\rho_0 \big ) + \tfrac{1}{2}\dt\Big(-M\big[ [\rho_0,L\dg],L\dg\big]
    +N\big[ [L,\rho_0],L\dg\big]+N\big[L,[\rho_0,L\dg]\big]
    -M^*\big[L,[L,\rho_0]\big]\Big) \, ,
  \end{split}
\end{align}
\end{widetext}
where we have used the initial conditions for the generalized state matrices, \cref{eq:hierarchy-init-cond}, specifically
$\rhonorm{t}{0,0}|_{t=0}=\rhonorm{t}{1,1}|_{t=0}=\rho_0$ and
$\rhonorm{t}{1,0}|_{t=0}=\rhonorm{t}{0,1}|_{t=0}=\rhonorm{t}{2,0}|_{t=0}
=\rhonorm{t}{0,2}|_{t=0}=\Zero$ 
and the fact that
$M=-\sinhr\coshr e^{2i\phi}$ and $N=\sinhr^2$.
Including the evolution from $H\system$ (ignored for convenience earlier on the grounds
that its corrections to the above derivation are of
negligible order in $\dt$) then \cref{stateequationyowza} yields the differential equation, 
\begin{align}
  \frac{\Delta\rho_0}{\dt}
  &=   -i[H\system,\rho_0]+
  (N+1)\D{L}\rho_0+N\D{L\dg}\rho_0 \nonumber
  \\
  &\phantrel{=}{}
  -\frac{M}{2}\big[L\dg,[L\dg,\rho_0]\big]
  -\frac{M^*}{2}\big[L,[L,\rho_0]\big]\, ,
\end{align}
which reproduces the broadband, white-noise squeezed master equation in \cref{eq:pure-squeezed}.
It is possible that this method could be extended to a 4th-order expansion, which may result in simple but nontrivial modifications to the Markovian evolution.

\section{driving a two level atom with wave-packet and broadband squeezing: a qualitative comparison} \label{sec:qual_comp}

To highlight the distinction between broadband and wave-packet squeezing we now compare the open-system dynamics of a two-level atom driven by squeezed vacuum in a square wave packet, \cref{eq:mainresult}, to that of the atom driven by broadband squeezed vacuum, \cref{eq:pure-squeezed}. 
In both situations the coupling of the atom to the environment is specified by the SLH parameters 
\begin{align}
S=\Id, \quad H\system= \Zero, \quad L= \sqrt{\lw} \sigma_-,
\label{eq:relaxing-atom-slh}
\end{align}
where $\lw$ is the coupling strength of the atom to the continuum field and $\sigma_- \defined \oprod{g}{e}$ lowers the atomic excited state to the ground state. For reasons we explain below we take the initial state of the atom to be
\begin{equation}\label{eq:inital_state_atom}
    \rho_0
    =
    \half (I +\tfrac{1}{\sqrt{2}}\sigma_x +\tfrac{1}{\sqrt{2}}\sigma_y)
    \,,
\end{equation}
which is a state halfway between the positive $\sigma_x$ and $\sigma_y$ eigenstates. We use the square wave packet
\begin{equation}\label{eq:square_xi_t}
    \xi_T(t) = 
\left\{\begin{array}{rl}
 \frac{1}{\sqrt{T}}, & \text{if } 0<t<T\\
 0, & \text{otherwise }
\end{array}\right.
\end{equation}
to remove spurious time dependence due to the wave-packet shape.

Comparing the squeezed wave packet and broadband squeezed evolutions is difficult, in part because of the fundamental difference between the expectation values of the quantum noises \cref{eq:bath-stats} and \cref{eq:wavepkt-bath-stats}.
In the broadband squeezed field the photon flux in an infinitesimal time step is not infinitesimal---that is, $\expt{\dLam}= N$ and not $\expt{\dLam}=dt N$.
Nevertheless, in this section we adopt two approaches to comparing these disparate cases by matching either
\begin{enumerate}
\item initial decay rates of the Bloch components, or
\item the steady-state excitation probability of the atom.
\end{enumerate} 
The first condition reflects the fact that squeezing introduces unequal (dissipative) coupling of the field continuum to the atomic quadratures.
The second condition is akin to using the steady-state atomic population as a proxy for the number of photons absorbed by the atom.

To match the resulting evolutions under narowband and broadband squeezing we perform a numerical optimization with respect to conditions 1 and 2.
Because numerical simulations of the squeezed wave-packet evolution are numerically intensive, we fix the parameters for that simulation and numerically optimize the squeezing parameters for the broadband Markovian evolution.
In earlier sections we denoted both the wave-packet and broadband squeezing parameter and angle were denoted by $r$ and $\phi$.
Here we need to distinguish the two cases so we denote the broadband squeezing parameter with an ${\rm M}$ as it is Markovian evolution $r_{\rm M}$ and $\phi_{\rm M}$. The wave-packet parameters are subscripted with WP.

The fixed parameters we use for the squeezed-master-equation hierarchy [see  \cref{eq:gam_param}] are
\begin{align}\label{eq:r_params}
r_{\rm WP}\approx 0.5181\quad  {\rm and\ }\quad \phi_{\rm WP} = 0 \, ,
\end{align}
which correspond to driving the atom with a resonant squeezed wave packet with $\SI{4.5}{dB}$ of squeezing on resonance with zero phase.

\begin{figure}[t]
\centering 
\includegraphics[width=0.47\textwidth]{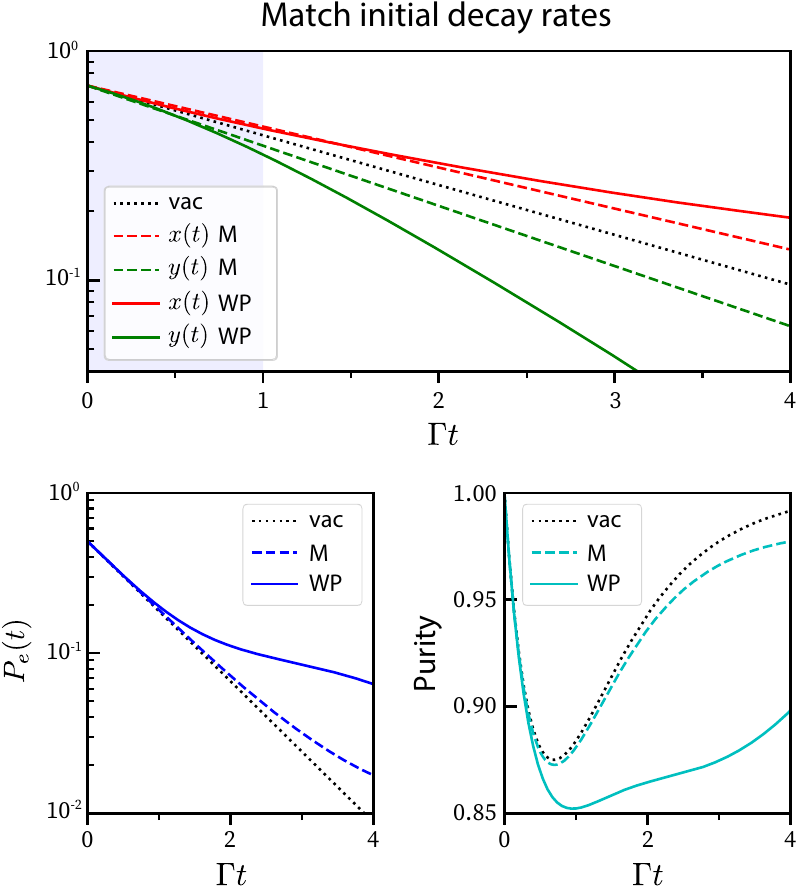}
\caption{
Comparison of broadband Markovian (M) and wave-packet (WP) dynamics using Approach 1---matching the inital decay rates of the Bloch-vector components. The shaded region indicates the time interval and the variables , \ie{} $x(t)$ and $y(t)$, optimization was perform on. The wave-packet is constant (square) on the interval $[0,4/\lw)$ and the squeezing simulation parameters are $r_{\rm WP}\approx 0.5181$ and $\phi_{\rm WP} = 0$. The broadband or Markovian evolution parameters that minimizes \cref{eq:short_time_min} are  is $r_{\rm M} = 0.0957$.
} \label{fig:blochcmp}
\end{figure}

 {\em Approach 1---}It is well known that broadband squeezed light
 can modify the coupling of the $L$ operator to the $x$ and $y$ Bloch components of the atom, see Ref.~\cite[sec. 10.3]{q_noise_gard2004}.
It is due to this asymmetry that we chose the initial state in \cref{eq:inital_state_atom} to be halfway between the positive $\sigma_x$ and $\sigma_y$ eigenstates.
We simulate the wave-packet dynamics with a square wave packet of length $T = 4/\lw$ using the parameters in \cref{eq:r_params}.
Then, we simulate the Markovian dynamics using the squeezed master equation, \cref{eq:pure-squeezed}. 
To compare the initial decay rates we construct a vector of values of the Bloch components, \eg{} $x(t)= \tr[\sigma_x \rho(t)]$, at discrete times $t_k$ from an initial time $t_i$ to a final time $t_f$:
$\mathbf{x}_{a} = [x_a(t_i), x_a(t_1), ..., x_a(t_f)]$, 
where $a\in \{ {\rm WP}, {\rm M} \}$ denotes wave-packet or Markovian evolution.
These are then used to solve the following optimization problem
\begin{equation}\label{eq:short_time_min}
\arg\min_{r_{\rm M}} \big [ \Vert\mathbf{x}_{\rm WP} - \mathbf{x}_{\rm M}(r_{\rm M})\Vert^2 +\Vert\mathbf{y}_{\rm WP} - \mathbf{y}_{\rm M}(r_{\rm M})\Vert^2\big ],
\end{equation}
from $t_i=0$ to $t_f=1/\lw$. The final time of $t_f=1/\lw$ is chosen to be short enough, relative to the wave-packet length, to ensure we are matching initial decay rates. With the chosen parameters, this optimization procedure gives a squeezing of $r_{\rm M} \approx 0.096$ corresponding to $\SI{0.83}{dB}$ of broadband squeezing. One way to understand the big discrepancy between the broadband squeezing parameter and the wave packet is to realize the total squeezing over the time interval is $r_{\rm M, tot} \approx  4\times 0.096 \approx 0.384$ which is of the same order of magnitude as $r_{\rm WP}\approx 0.5181$.

\begin{figure}[t!]
\centering 
\includegraphics[width=0.466\textwidth]{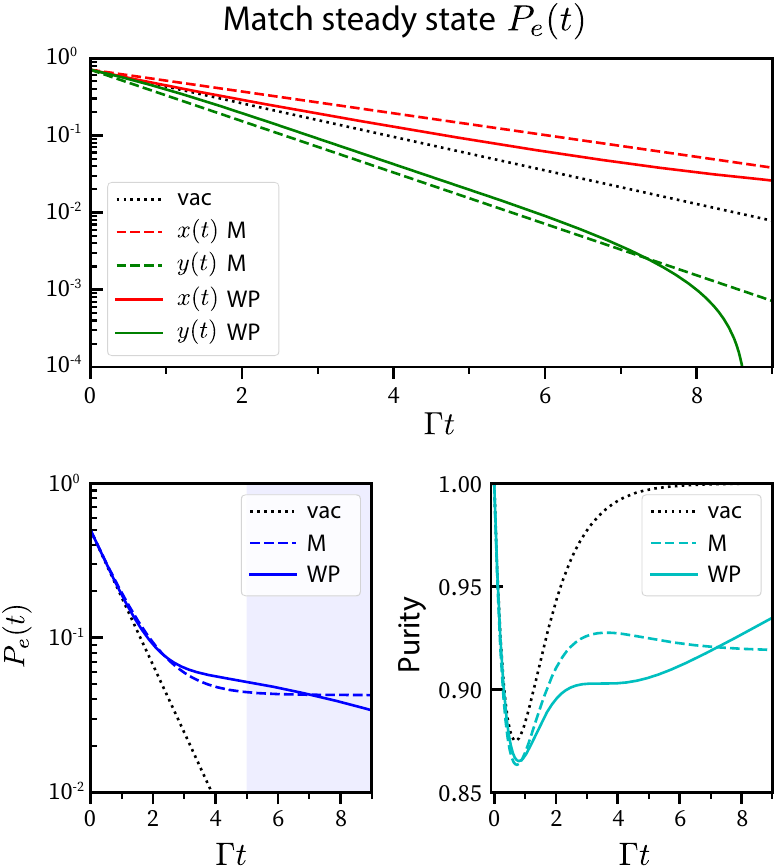}
\caption{
Comparison of Markovian and wave-packet and  dynamics using Approach 2---matching the steady-state excitation probability. The shaded region indicates the time interval and the variables, \ie{} $P_e(t)$, optimization was perform on.
The wave-packet is constant (square) on the interval $[0,9/\lw)$ and the squeezing simulation parameters are $r_{\rm WP}\approx 0.5181$ and $\phi_{\rm WP} = 0$. The squeezing parameter for Markovian evolution that minimizes \cref{eq:excite-prob-min} is  $r_{\rm M} = 0.2141$. The presence of two exponential decay rates is evident in the wave packet excitation probability.
} \label{fig:longtime}
\end{figure}

In the first row of \cref{fig:blochcmp} the Markovian evolution shows the $y$ component of the atom experiencing an increased decay rate relative to a vacuum environment.
This is because the atomic $\sigma_y$ couples to the optical $x$ quadrature, whose vacuum fluctuations are scaled by a factor of $e^{-r_{\rm M}}$ for $\phi_{\rm M}=0$.
These reduced vacuum fluctuations render the coupled atomic observable more visible to the environment.
Similarly the atomic $\sigma_x$ quadrature is coupled to the $y$ quadrature of the field which has fluctuations scaled by $e^{r_{\rm M}}$, resulting in a decreased decay rate for the $x$ component of the atom.
The resulting modified Bloch-component decay rates for arbitrary squeezing parameters are 
$\Gamma_x = (N+M+1/2)\lw \approx 0.413\lw $ and  $\Gamma_y = (N-M+1/2)\lw \approx 0.605\lw$~\cite[sec. 10.3]{q_noise_gard2004}.
To understand why the squeezing parameter reported by the optimization is so small consider naively using the wave-packet squeezing parameter in the Markovian evolution to compute the Bloch-component decay rates then the new rates would be
$\Gamma_x' \approx \Gamma_x/2.327$ and $\Gamma_y' \approx 2.327 \Gamma_y$.
Such a difference would be very noticeable in the second row of the figure.

In contrast, the wave-packet evolution exhibits nonexponential decay, with both quadratures initially decaying at the vacuum rate before diverging to suppressed and enhanced rates from the squeezing.
For comparison, we include the evolution under the (unsqueezed) vacuum master equation, which emphasizes the fact that in the squeezed case the quadratures decay at significantly different rates.
The second row shows a direct comparison of the $x$ and $y$ components to see how well the optimization performed under the eye-ball norm $\Vert\cdot\Vert_\text{\eye}$. Clearly the Bloch components are already diverging for $t=0.5/\lw$ and it gets worse for $t>1/\lw$.

In the second row we plot the long-time dynamics for the excitation probability $P_e(t) \defined \tr[ \oprod{e}{e}\rho\system(t)]$ and purity $\tr[\rho\system(t)^2]$.
These figures show good agreement between the two types of evolution for short times but significant divergence for times $t>0.5/\lw$.
The excitation probability in the broadband case reaches a steady state at $t> 4/\lw$ while the wave-packet evolution transitions to a second exponential decay rate at $t\approx 2/\lw$.
If we use purity as a proxy for system-field entanglement (as both are initially in pure states) the wave packet evolution is qualitatively more entangled.
The purity of the two evolutions seems to be more similar, under the qualitatively eye ball norm comparisons.

{\em Approach 2--} Our second strategy for comparison is motivated by the observation that, for long wave packets, the atom comes to a quasi-equilibrium with the input field.
By matching the excitation probability of the ``quasi-steady state'' of the Markovian evolution with that for the wave-packet evolution we obtain this additional point of comparison.
 
We simulate the wave-packet dynamics for a square wave packet of length $T = 9/\lw$ using the parameters in \cref{eq:r_params} and the Markovian dynamics for a range of $r_M$.
From the solutions, we construct vector of excitation-probability dynamics, $\mathbf{P}_{a} = [ P_{e,a}(t_i), P_{e,a}(t_1), ..., P_{e,a} (t_f)]$, again for $a\in \{ {\rm WP, M }\}$.
To ensure the atom is well away from its initial decay behavior we consider times from $t_i=5/\lw$ to $t_f=9/\lw$ and then optimize:
\begin{align}\label{eq:excite-prob-min}
\arg\min_{r_{\rm M}} \big [ [\mathbf{P}_{\rm WP} - \mathbf{P}_{\rm M}(r_{\rm M})]^2 \big ].
\end{align}
For the chosen parameters, the optimization procedure gives a squeezing of $r_{\rm M} \approx 0.214$ corresponding to $\SI{1.86}{dB}$ of broadband squeezing. This amount of squeezing is already comparable to the amount of wave-packet squeezing, unlike the previous approach. We will see that the broadband excitation probability is in approximate equilibrium over the optimization time interval $[5/\lw,  9/\lw)$ .

In \cref{fig:longtime} we examine the difference between the Markovian and wave packet evolution in approach 2. It is important to note that the time axis is much longer in these plots relative to \cref{fig:blochcmp}. We notice that, under the eye-ball norm $\Vert\cdot\Vert_\text{\eye}$, the evolution of the $x$ and $y$ Bloch components remains much closer; with the exception of the late time behaviour of the $y(t)$ Bloch component which diverges on a logarithmic scale.
Indeed the late time dynamics of most of the subplots all look a lot closer (compared to \cref{fig:blochcmp}).  Moreover it is surprising that the initial decays match so closely. 
The excitation probability in the broadband case reaches a steady state at $t\approx 5/\lw$ while the wave-packet evolution is transitions to a second exponential decay rate at $t\approx 3/\lw$.
The purity is much more similar using approach 2.
Combined these results suggest that using the ``long time'' or ``quasi-steady state'' atomic properties to compare the Markovian and wave-packet evolution is a fairer comparison than matching the initial decay, where the late time evolutions diverged.

Independent of whether we choose to fit the short or long time dynamics there are many discrepancies between the two approaches.
The most obvious is the two time scales present in the excitation probability $P_e$ plots. This leads to the inability to match short or long time dynamics.
Second, if we take (im)purity of the state as an indicator of system and field correlation then generally the wave packet evolution is (unsurprisingly) more correlated. Perhaps the final takeaway message is if the wave packet is long enough to induce a quasi steady steady state then it is possible to get somewhat similar behavior from a Markovian evolution, at least under the eyeball norm.

\section{Resonance fluorescence}\label{sec:res_fluor}

Resonance fluorescence describes the light emitted into the quantized electromagnetic field by a two-level atom driven by strong classical field at frequency $\omega_c$ near the atomic resonance frequency $\omega_a$.

This scenario is captured by an atomic Hamiltonian $H\system=\dr\sigma_x$ and field-coupling jump operator $L=\sqrt{\lw}\sigma_-$, where the resonant Rabi frequency $2\dr$ is determined by the classical drive amplitude.\footnote{We follow the convention of Carmichael~\cite{carmichael_resonance_1987}, where the Rabi frequency is $2\dr$ and the sidebands are at $\pm\dr$.}
Resonance fluorescence is characterized by the power spectral density of the scattered light---the rate of photon emission at each frequency---when the atom is in its steady state.

Typical investigations of resonance fluorescence consider the quantized field in the vacuum state~\cite{Mollow1969} or in a stationary squeezed state~\cite{carmichael_resonance_1987}.
For the case of vacuum, \citet{Mollow1969} predicted that the power spectral density for a two-level atom driven on resonance ($\omega_c=\omega_a$) would have three peaks---a central peak at the atomic resonance frequency $\omega_a$ and two peaks at $\omega_a \pm \dr$ known as the sidebands.
In his honor these three peaks are collectively referred to as the Mollow triplet.
Later, Gardiner~\cite{Gardiner86} and Carmichael~\cite{carmichael_resonance_1987,Carmichael:1987aa} examined the spectrum when the quantized field was squeezed and found that properties of the Mollow triplet, including the width of the central peak, depend on the relative phase the driving field and the squeezed quadrature.
We refer the interested reader to two reviews of broadband and narrowband squeezed light interacting with two-level atoms by \citet{Swain2004} and for three-level atoms by \citet{Ficek2004} for further information.

\begin{figure}[t]
  \centering
  \includegraphics[width=0.9\columnwidth]{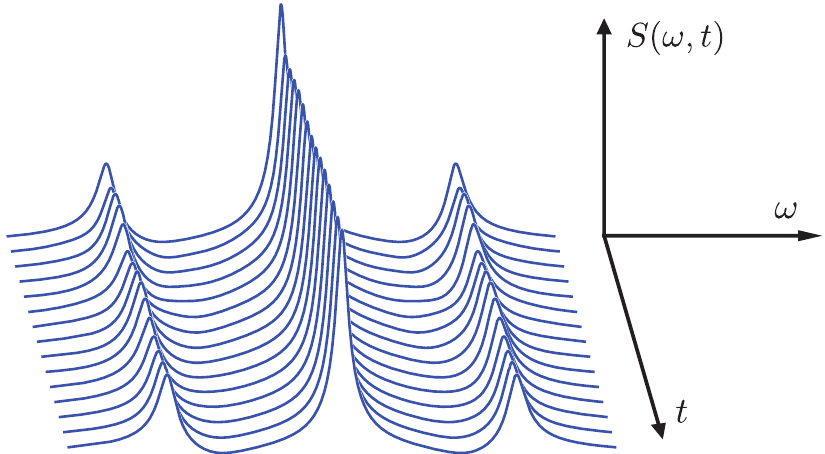}
  \caption{Illustration of nonstationary modifications to resonance fluorescence when a driven two-level atom interacts with a squeezed-vacuum wave packet.
  Because the correlation function $\expt{\sigma_+(t) \sigma_- (t+\tau)}$ depends on a reference time $t$ in addition to the relative time $\tau$, the power spectrum $S(\omega,t)$ is a function of the chosen reference time, see \cref{eq:spectrumform}.
  For long wave packets, the reference-time dependence is minimal, as shown above for various values of $t\in[T/16,15T/16]$ with a square wave packet at the red sideband $\carrierdetuning=-\Omega$ of duration $T=2/\lw$ squeezed with amplitude-squeezing factor $e^r=2$ and squeezing phase $\mu=0$.}
  \label{fig:joy-div}
\end{figure}

Here we use the squeezed hierarchy to perform a numerical investigation of the resonance fluorescence of two-level atoms in the presence of narrowband squeezed vacuum~\cite{JoshiPuri1994,Ficek97,SmythSwain1999,Tanas1999,Messikh2000}. 
Such investigation complements previous analyses of modifications that squeezing makes to the fluorescence spectrum, which have focused on broadband \cite{carmichael_resonance_1987,Carmichael:1987aa} (wider than the sideband separation given by the Rabi frequency) and finite-bandwidth \cite{parkins_rabi_1990,yeoman_influence_1996} (narrower than the sideband separation but wider than the atomic linewidth $\lw$) squeezed vacua.
Cascaded systems can model arbitrary bandwidth~\cite{parkins_rabi_1990,GardPar94,KochCarm94}, but this work has typically focused on CW fields with stationary statistics.
Such cascaded systems can model the wave packets we consider here, with the cost of complicated time dependence in the SLH parameters of the systems.

\begin{figure}[t!]
\centering 
\begin{tikzpicture}[scale = 8]
\draw[<->,thick] (0,0) -- (16/15,0);
\foreach \x in {1/15,2/15,1/5,4/15,5/15,2/5,7/15,8/15,3/5,10/15,11/15,4/5,13/15,14/15,1} \draw[shift={(\x,0)}] (0pt,2/8pt) -- (0pt,-2/8pt);
\draw[shift={(1/5,-0.01)}] node[below] {\footnotesize $\lw$};
\draw[shift={(2/5,-0.01)}] node[below] {\footnotesize $\dr$};
\draw[shift={(3/5,-0.01)}] node[below] {\footnotesize $\frac{1}{dt}$};
\draw[shift={(5/5,-0.01)}] node[below] {\footnotesize $\omega_c$};
\draw[shift={(2.5/5,-0.075)}] node[below] {\footnotesize Frequency  Bandwidth (Time$^{-1}$)};
\draw[<-,very thick, cyan] (0,36/8pt) -- (5.35/5,36/8pt);
\draw[shift={(3.125/5, 36/8 pt)}]  node[above]{\footnotesize{This work}} -- (3/5,1/2 pt);
\draw[|-,very thick, blue] (1.5/5,22/8pt) -- (5.35/5,22/8pt);
\draw[shift={(3.125/5, 21/8 pt)}]  node[above]{\footnotesize{Quasi-Markovian}} -- (3/5,1/2 pt);
\draw[|-,very thick, green] (2.5/5,8/8pt) -- (5.35/5,8/8pt);
\draw[shift={(3.125/5, 8/8 pt)}]  node[above]{\footnotesize{Markovian }} -- (3/5,1/2 pt);
\end{tikzpicture}
\caption{Applicability regimes of different squeezing formalisms based on squeezing bandwidths.
On the $x$ axis we denote relevant timescales such as the atomic linewidth $\lw$, the half Rabi frequency $\dr$, and the carrier frequency $\omega_c$.
The different bandwidth ranges represent frequency bandwidths of squeezing that can be modeled by various formalisms.
In particular these ranges cover the allowed widths of the Fourier transform of the two time correlation function $f(\omega) = \mathfrak F {\expt{b(t+\tau)b(t)}}$.
} \label{fig:freq_scales}
\end{figure}
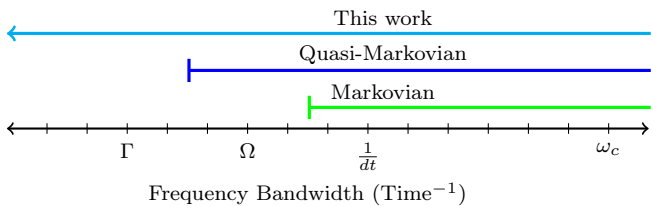

To model squeezed vacuum with a very narrow bandwidth, consider very long wave packets with a square envelope so that $\xi_t$ is constant like in \cref{eq:square_xi_t}.
The two-time correlation function of the input field $\nexpt{b_t\dg b_{t+\tau}}= \xi_t^* \xi _{t+\tau}\sinh^2r$
is then also virtually constant for the (long) duration of the wave packet, which corresponds to a squeezing power spectrum that approaches a delta function as the duration increases.
Since the derivation of our equations assumed the squeezing was at the carrier frequency we need to modulate the wave packet to have our narrow-band squeezing at other frequencies.
This is accomplished by the transformation
\begin{equation}\label{eq:detunewavepacket}
    \xi_t \mapsto \exp(-i \carrierdetuning t)\xi_t
    \, ,
\end{equation}
such that $\carrierdetuning=0$ corresponds squeezing at the carrier frequency and $\carrierdetuning=-\Omega$ corresponds to squeezing at the red sideband. As we have a square wave packet, the spectral content of the main peak gets shifted to the frequency range $\omega \in (\carrierdetuning-2\pi/T, \carrierdetuning +2\pi /T)$.\footnote{The spectral support of $\xi_t$ can be determined by the Fourier transform \cref{eq:detunewavepacket}. Assuming it is in a square wave packet $\xi(t) = e^{-i \carrierdetuning t} {\rm HeavisidePi} [(t - T/2)/T]$, the absolute value of the Fourier transform is $f(\omega) = \sqrt{\frac{2}{\pi }}\sin[ T (\omega -\carrierdetuning)/2] / (\omega -\carrierdetuning)$. Thus, the spectral content of the main peak of the ${\rm sinc}$ function is in frequency range $\omega \in (\carrierdetuning-2\pi/T, \carrierdetuning +2\pi /T)$.} For simulations will use $T=4/\lw$ giving a bandwidth $\pi\lw/2$ that is comparable to the atomic linewidth and therefore outside the quasi-Markoffian regime as illustrated in \cref{fig:freq_scales}.

The strategy for calculating the resonance fluorescence spectrum is to take the Fourier transform of the atomic correlation function~\cite[Sec.~5.7]{steck2007_QO_textbook}. Just as in Ref.~\cite{carmichael_resonance_1987}, this assumes we are studying the fluorescent scattering into a small set of unsqueezed modes, where the scattered light is entirely described by the atomic correlation function:
\begin{align} \label{eq:spectrumform}
    S(\omega, t)
    \propto \int d\tau \, e^{i \omega \tau}\expt{\sigma_+(t) \sigma_-(t+\tau)}
    \, .
\end{align}
Since the squeezed field is non-stationary, there is no atomic steady state, and the spectrum will vary with time $t$.
We calculate this correlation function using the quantum regression theorem for the squeezed hierarchy as described in \cref{sec:qrt-wavepacket}.
Because the statistics are not stationary, the spectrum will depend on our choice of $t$, as illustrated by \cref{fig:joy-div}.
However, because the square wave packet is long and constant, the dependence on the choice of $t$ is minimal.

\begin{figure}[b]
  \centering
  \includegraphics[width=\columnwidth]{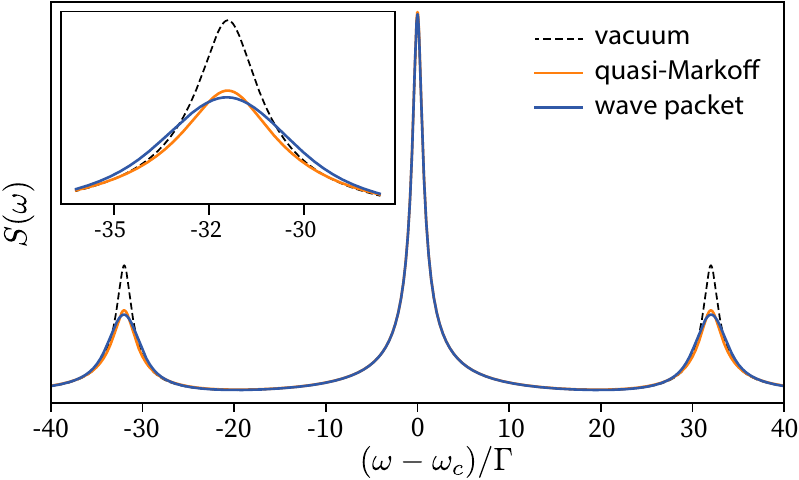}
  \caption{Comparison of the resonance fluorescence spectra $S(\omega)$ for vacuum,  quasi-Markoffian squeezed, and narrowband squeezed wave packet input fields.
    A rectangular wave packet with duration $T=4/\lw$ is prepared in squeezed vacuum with squeezing parameter $e^r=2$ ($\SI{6.02}{dB}$ of squeezing) and with classical drive $\Omega=32\Gamma$. The spectrum for $t = 1/(2\Gamma)$ is shown. 
    The quasi-Markoffian squeezing is optimized over cavity bandwidth $\gamma_c$ and squeezing-Hamiltonian strength $\epsilon$ for quasi-Markoffian squeezing coming from a degenerate parametric amplifier, so that the mean-squared-error between the resonance fluorescence spectra is minimized with respect to the wave packet.
    The optimized parameters for the degenerate parametric oscillator are linewidth $\gamma_c=0.48\lw$ and amplification constant $\epsilon=3.17$, translating to $\SI{2.65}{dB}$ of squeezing at the atomic transition frequency.
    The quasi-Markoffian spectrum that best matches is not as broad as the wave packet spectrum.
  }\label{fig:quasi-markoff-opf-mollow}
\end{figure}

\begin{figure*}[ht!]
  \centering
  \includegraphics[width=0.99\textwidth]{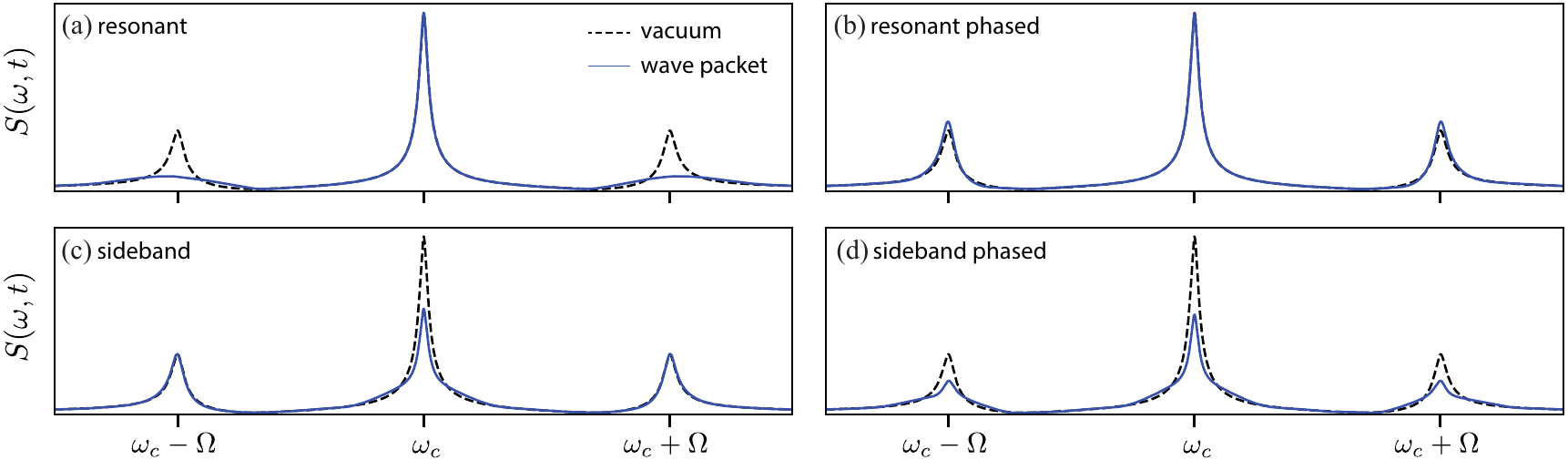}
  \caption{
    Resonance-fluorescence spectra for a two-level atom driven by a classical drive $\Omega=8\lw$ in the presence of long ($T=4/\lw$), strongly squeezed ($e^r=4$) wave packets. The spectra shown are at time $t = 1/(4\Gamma)$.
    (a) The ``resonant'' spectrum is on resonance ($\carrierdetuning=0$) with squeezing angle $\mu=0$ relative to the drive.
    (b) The ``resonant phased'' spectrum differs in that the squeezing angle relative to the drive is $\mu=\pi/2$.
    (c) The ``sideband'' spectrum is obtained by driving at the blue sideband $\carrierdetuning=\Omega$, and (c) the ``Sideband Phased'' spectrum again differs by setting $\mu=\pi/2$.
    The central peak is only affected by squeezing at the sidebands, so the resonant spectra have unmodified central peaks, although their sidebands are broadened or slightly narrowed depending on the relative phase between the squeezing and the drive.
    Squeezing at the sidebands dramatically alters the central peak to look similar to a superposition of two Lorentzians, suggesting two timescales similar to those observed in \cref{fig:blochcmp,fig:longtime}.
  }\label{fig:narrowband-squeezed-mollow-triplets}
\end{figure*}
To appreciate the difference between the extremely narrow bandwidth of these wave packets and previously studied spectra, we compare to the quasi-Markoffian regime studied by Yeoman and Barnett~\cite{yeoman_influence_1996}; in  \cref{app:yeoman_barnett} we summarize their formalism.
They identify that, in this regime where the squeezing bandwidth is allowed to be narrow with respect to the Rabi sideband splitting but still broad compared to the linewidth of the atom, modifications to the linewidth of the central peak of the Mollow triplet depend entirely on the squeezing at the sideband frequencies, which are offset from the central atomic frequency by the half Rabi frequency $\dr$.
Our equations allow us to investigate this phenomenon in the extreme regime where the squeezing bandwidth is narrower than the atomic linewidth, as illustrated in~\cref{fig:freq_scales}.

\Cref{fig:quasi-markoff-opf-mollow} illustrates the most comparable quasi-Markoffian spectrum to the spectrum of our long squeezed wave packet, illustrating some subtle features that apparently cannot be reproduced in the quasi-Markoffian regime such as the more broadening with less suppression of the sidebands. Because the quasi-Markoffian contains the Markovian regime as a special case we do no include a separate spectrum.

We also illustrate the difference between squeezing on resonance and squeezing at the sideband in \cref{fig:narrowband-squeezed-mollow-triplets}.
Because of the extreme narrowness of the squeezing bandwidth, the modified peaks exhibit qualitatively different features than seen previously in the literature.
Most strikingly, the modified central peak when squeezing is at the sideband appears as a sum of two Lorentzians of different widths, evocative of the two timescales present in the dynamics of \cref{fig:blochcmp,fig:longtime}.

In light of uncertainty regarding the convergence of the squeezed hierarchy---discussed in more detail in \cref{sec:NumericalAnalysis}---these spectra were simulated using the Fock hierarchy with $n_\text{max}=20$.
In cases where we observe the squeezed hierarchy converging to a solution as we increase $n_\text{max}$ the results of the two hierarchies agree with one another.

\section{Numerical analysis}\label{sec:NumericalAnalysis}
Because the squeezed hierarchy is formally infinite, any numerical implementation must truncate it at some point.
Here we investigate the effects of finite truncation for the squeezed hierarchy presented in this paper, and compare with the effects of finite truncation for the Fock hierarchy given Ref.~\cite{baragiola_n-photon_2012}. Formally the the Fock hierarchy is infinite when used to model squeezed wave packets.
Before turning to numerical experiments, we first explain the intuition behind why using the squeezed hierarchy should be advantageous over using the Fock hierarchy when dealing with wave packets initially in squeezed vacuum.

The simplest truncation of the squeezed hierarchy is the time-non-local truncation described in \cref{sec:master_eqn}, where $\rhonorm{t}{m,n}=0$ if either $m$ or $n$ are larger than some cutoff $\nmax$.
For the Fock hierarchy, truncation at a chosen $\nmax$
 forces the use of an approximate wave-packet state, since we must effectively approximate a squeezed-vacuum state as a superposition of a finite number of Fock states \ie{} $\ket{\psi_{\rm approx}} = \sum_{n=0}^{\nmax} \oprod{n}{n}\ket{\psi_{\rm sqz}}$.
In such an approximation, the amount of population thrown away is
\begin{align}
  \label{eq:num-disc-pop}
  \discardpop_{\text{Fock}}(\nmax,r)
  &=
  \sum_{n=\nmax+1}^\infty |\langle n | \psi_{\rm sqz}\rangle|^2
  \\
  &=\frac{1}{\cosh r}\sum_{k>\nmax/2}\left(\frac{\tanh^2r}{4}\right)^k
  \frac{(2k)!}{(k!)^2}\,.
\end{align}
This expression serves as a first guess for the amount of error incurred by truncating the Fock hierarchy at $\nmax$ as opposed to simulating the full hierarchy.

\begin{figure}[ht!]
  \centering
  \includegraphics{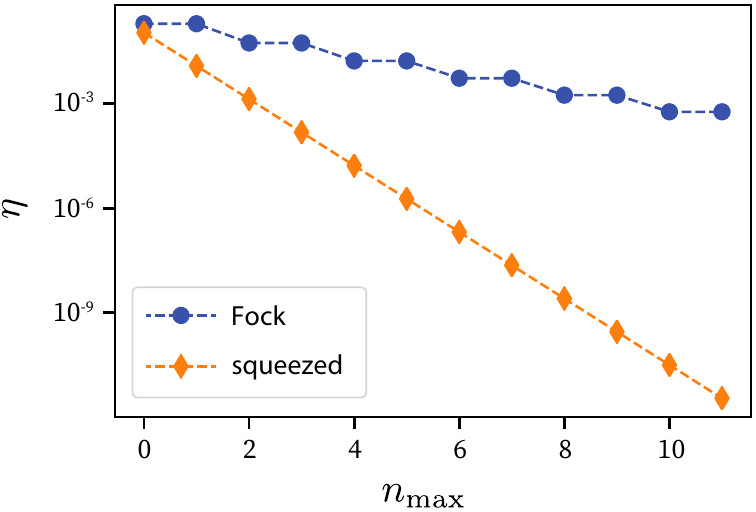}
  \caption{Discarded population for truncating the Fock and Squeezed hierarchies for squeezing strength $e^r=2$. As expected, by this measure the squeezed hierarchy is much better at modelling squeezed vacuum.}
  \label{fig:disc-pop}
\end{figure}

\begin{figure*}[ht!]
  \centering
  \includegraphics{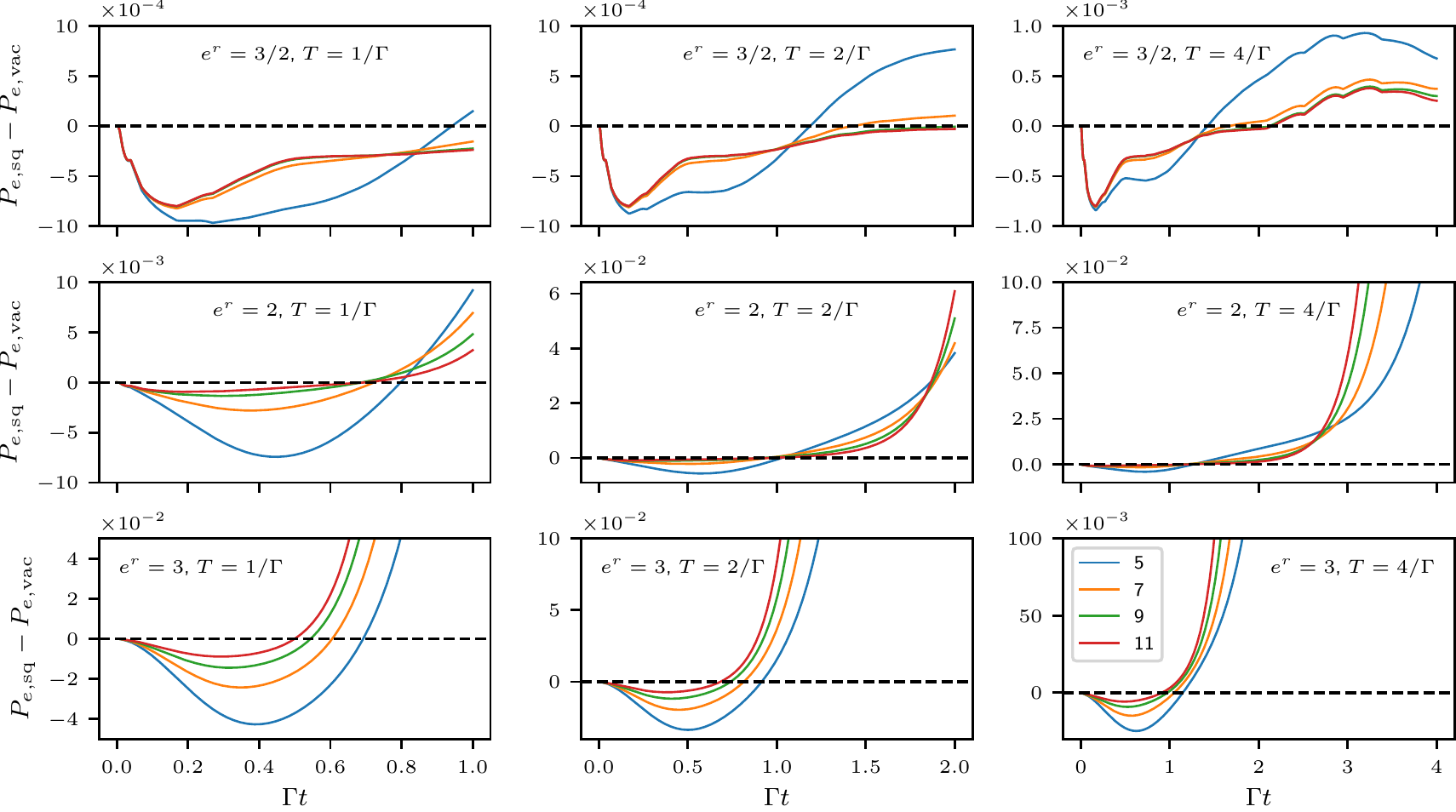}
  \caption{Differences between the excited-state probability given by the truncated squeezed hierarchy (with initial state chosen to approximate vacuum) and those given by the vacuum master equation.
  Titles of the subfigures indicate squeezing strength $r$ and the length $T$ of the rectangular wave packet.
  For small-enough squeezing and short-enough wave packets, increasing the number of levels in the hierarchy corresponds to decreasing difference with the vacuum master equation.
  However, for larger squeezing strength and longer wave-packet length, one sees divergent behavior as the difference with the vacuum master equation increases with increasing size of the hierarchy.}
  \label{fig:vacuum-diffs}
\end{figure*}

\begin{figure}[ht!]
  \centering
  \includegraphics{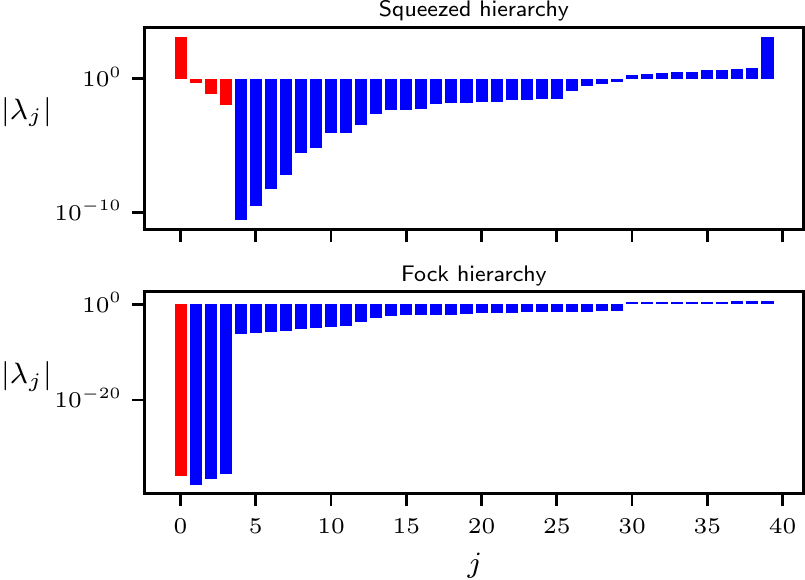}
  \caption{Numerical eigenvalues of the Choi matrices for the maps from system to system-plus-hierarchy states given by evolving under the hierarchy master equation for the wave-packet duration.
  The wave packets are both rectangular, with width $T=1/\lw$, and the cutoff is set at $\nmax=9$.
  The strength of the squeezing for the squeezed hierarchy is set such that $e^r=2$.
  The bars indicate the magnitude of the eigenvalues, which are sorted increasing from left to right.
  Blue bars indicate to positive eigenvalues and red bars indicate negative eigenvalues.
  Up to numerical precision, the eigenvalues of the Fock hierarchy are all positive, demonstrating that the map is physical, while there are substantial negative eigenvalues for the squeezed hierarchy, revealing that the evolution is unphysical.}
  \label{fig:choi-eigvals}
\end{figure}

The squeezed hierarchy introduces errors more subtly, since it represents the initial state of the wave packet exactly.
The errors introduced have to do with imprecise representation of the correlations within the wave packet, which the levels of the hierarchy are meant to keep track of.
To understand the effect of truncating the hierarchy, we must therefore understand the strength of correlations within the wave packet.
In \cref{sec:num-anal-det} we use a source model to quantify the loss of this correlation information in terms of the effective population thrown away in the reduced wave-packet state, arriving at the result
\begin{align}\label{eq:sq-disc-pop}
    \discardpop_{\text{sq}}(n_{\text{max}},r)
    &=
    \left(\frac{\cosh r-1}{\cosh r+1}\right)^{\nmax+1}\,.
\end{align}
As one can see in \cref{fig:disc-pop}, this analysis suggests that the error in both hierarchies ought to decrease exponentially in $\nmax$, with a larger constant in the exponent for the squeezed hierarchy.
As we shall see, however, this heuristic fails to capture many difficulties in using the squeezed hierarchy.

To get a more precise picture of how the squeezed hierarchy performs for various truncations, we make use of the fact that the squeezed Fock states form a basis for the Hilbert space and simulate the vacuum master equation in the squeezed-wave-packet hierarchy.
Since we know the exact solution for the vacuum master equation, this gives us an exact picture of the errors.
\Cref{fig:vacuum-diffs} illustrates the errors incurred in simulating the vacuum master equation with truncations of the squeezed hierarchy for various squeezing strengths and rectangular wave packets of various widths.
Interestingly, including more levels of the hierarchy is harmful once the strength of the squeezing or the width of the wave packet become large enough.
The culprit seems to be that truncating the squeezed hierarchy results in unphysical evolution, unlike truncating the Fock hierarchy, where the truncation does not affect the physicality of the evolution.

To explore this truncation issue further we consider the super-operator action of both the hierarchies.
Specifically, consider the map from the system state space to the system plus hierarchy state space defined by integrating the squeezed-hierarchy master equation for the duration of the wave packet.
This is a composition of the map initializing the hierarchy from the initial system state according to \cref{eq:hierarchy-init-cond} with the time ordered exponential $\mathcal{T}\exp(\int dt\mathcal{L}_t)$ of the generator $\mathcal{L}_t$ defined by the master equation \cref{eq:mainresult} and finally taking the linear combination of hierarchy solutions corresponding to the original wave packet state (for squeezed vacuum, this is just $\rhonorm{t}{0,0}$) in order to recover the final reduced system state.
The structure of both the Fock and the squeezed hierarchies ensures that this map is trace preserving, so the only physical condition to check is complete positivity.
If this map is not completely positive, the Choi matrix of the map will have at least one negative eigenvalue.
Numerically evaluating the eigenvalues of the Choi matrix as shown in \cref{fig:choi-eigvals}, one finds significant negative eigenvalues for the squeezed hierarchy (indicating that the evolution is not completely positive) in contrast to the purely positive (up to numerical precision) eigenvalues of the Choi matrix for the Fock hierarchy.

This analysis highlights a fundamental distinction between the way squeezed wave packets are approximated in the two hierarchies.
The Fock hierarchy takes the conservative approach of leaving the dynamics unchanged and approximating the initial state.
The squeezed hierarchy takes the more radical approach of representing the initial states exactly by virtue of a change of basis.
The approximation is then relegated to the dynamics.
While \cref{fig:disc-pop} gives us reason to hope this approximation might be more efficient that the approximation of the initial state by finitely many Fock states, the approximation to the infinitesimal dynamics breaks fundamental properties of the resulting map, taking us to non positive states whose errors can rapidly amplify.
We leave for future work the question of whether the promise of the squeezed hierarchy can be rescued from these pathologies, perhaps by using tools from singular perturbation theory~\cite{neu2015singular} such as boundary layers and matched asymptotic expansions.

\section{Stochastic Master Equation}\label{sec:SMEs}
The master equation in \cref{eq:mainresult,eq:drho_tensor} governs the evolution of systems coupled to fields described in the squeezed Fock basis when the field simply propagates away after the interaction.
The trace in \cref{eq:gen-state-matrix-trace} is the tool by which we ignore the departing field and whatever information it carries away about the system.
If, instead of ignoring the departing field, we measured it, then we would gain information about the system and be able to write down a refined state matrix conditioned on the measurement result.
The differential equation for this conditional evolution of the system is called a stochastic master equation (SME).
In this section we derive SMEs for photon-counting measurements and homodyne measurements of the field for a squeezed-vacuum input field.
The heterodyne case follows trivially from the homodyne~\cite{gross_qubit_2017,baragiola_quantum_2017} and one can use the more general results in \cref{app:sme_derivation} to derive the SMEs for more general input states such as squeezed Fock states.

Photon counting and homodyning are both time-continuous measurements of the field, which we model as a sequence of measurements on the infinitesimal time intervals introduced in \cref{eq:approximations}.
Individually, these sequential measurements are quite simple, since states in the infinitesimal time intervals are essentially restricted to the 0- and 1-photon subspace, as the relative state decompositions in  \cref{Eq::Focktempdecomp,eq:squeezed-temp-decomp} reveal.
The integral (or sum) of these infinitesimal measurements resolve the identity on the full Fock space for any finite time interval $[t_0,t)$~\cite[Sec. V]{JackColl00}.
We derive the SME by determining the conditional evolution via Kraus operators for all possible outcomes within an infinitesimal time interval and combining these random changes into a single differential equation that depends on a classical random variable $R$ representing the result of the measurement:
\begin{align}
  \rho_{t+dt|\Rec,R}
  &=
  \rho_{t|\Rec}+d\rho_{t|R}\,.
\end{align}
Note that the state of the system at time $t$ depends not only on the evolution time but also the record $\Rec$ of all previous measurement results.

Because of the length and complexity of the derivations of the stochastic master equations we simply state the resulting equations and point the interested reader to \cref{app:sme_derivation} for the details of the derivations.\vspace{2pt}

\subsection{Photon counting}
After interacting with the quantum system, the output fields are sent to a photodetector, which continuously measures the field in the infinitesimal basis $ \ket{0_t}$ and $\ket{1_t}$ defined in \cref{Eq::InfSinglePhoton}. At each infinitesimal time, the detector records outcome $R$ and over larger time intervals produces a photodetection record $\mathbf{R}$.
We adopt Carmichael's notation and give the infinitesimal measurement results the labels $R=\emptyset$ for vacuum detection and $R=J$ for single-photon or ``jump'' detection.
We also keep track of how many photons have been detected up to a certain time $t$, representing this quantity as $N_t$.
In each infinitesimal interval $[t,t + dt)$ change in the number of photons detected is a random variable $dN_t$, which takes a value of $0$ when $R_t=\emptyset$ and a value of $1$ when $R_t=J$.
We use this random variable to express the conditional evolution under continuous photon counting as a set of coupled SMEs
\begin{align}
  \begin{split}
  d\rhonorm{t}{m,n}&=(1-dN_t)d\rhonorm{t|\emptyset}{m,n}
  +dN_td\rhonorm{t|J}{m,n}
  \\
  &=
  d\rhonorm{t|\emptyset}{m,n}
  +dN_td\rhonorm{t|J}{m,n}
  \\
  &= d\rhonorm{t|\emptyset}{m,n}
  +(d\mathcal{J}^N_t
  +\mathbb{E}[dN_{t}])d\rhonorm{t|J}{m,n}\, ,
  \end{split}
  \label{eq:stochastic-differential-photon-counting}
\end{align}
The simplification in the second line follows from the observation that $\mathbb{E}[dN_t]\in\BigO(dt)$ and $d\rhonorm{t|\emptyset}{m,n}\in\BigO(dt)$, meaning their product vanishes to order $dt$. On the third line we have introduced the ``innovations'' and the expectation
\begin{align}
  d\mathcal{J}^N_t
  &\defined
  dN_t-\mathbb{E}[dN_t]
  \\
  \mathbb{E}[dN_t]
  &=
  0\times \Pr(\tight{R_t=\emptyset}) + 1\times\Pr(\tight{R_t=J})
  \nonumber\\
  &=
  \Pr(\tight{R_t=J})\,.
\end{align}
Given squeezed vaccum input, the probability of a photon detection is given by
\begin{align}\label{eq:pr_click_sme}
  \begin{split}
    \Pr(\tight{R_t=J})
    &=
    dt\,\tr\big(L\rhonorm{t}{0,0}L\dg
    -e^{-2i\phi}\xi_t^*\sinhr L\rhonorm{t}{0,1}S\dg\\
    &\phantrel{=}
    -e^{2i\phi}S\xi_t \sinhr \rhonorm{t}{1,0}L\dg
    +\abs{\xi_t}^2\sinhr^2\rhonorm{t}{1,1}\big)\,.
  \end{split}
\end{align}
as shown in \cref{sec:app_photon_sme}.
Since there are only two possible detection results for a particular time $t$, the probability of no jump is $\Pr(\tight{R_t=\emptyset})=1-\Pr(\tight{R_t=J})$.  \Cref{eq:pr_click_sme} is consistent with \cref{eq:pr_click} when we set $n=0$ (squeezed vacuum) and when the system is uncoupled (\ie{} $L=0$ and $S=I$) then  $ \Pr(\tight{R_t=J}) = |\xit|^2 dt \sinhr^2$.
The corresponding probabilities for more general input states than squeezed vacuum follow from taking the trace of \cref{eq:detection-unnorm-update} for diverse values of $m$ and $n$.

To simplify the statement of the SME, we define a ``Lindbladian/Liouvillian'' $\Lind{t}{m,n}{\mathbf{G},\xi}$ as shorthand notation for the master-equation map in \cref{eq:mainresult}:
\begin{align}
  d\rhonorm{t}{m,n}&=
  dt\,\Lind{t}{m,n}{\mathbf{G},\xi}\,.
\end{align}
Here, $\mathbf{G}$ denotes the SLH triple $(S,L,H)$ and $\xi$ is the wave packet.
This allows us to write the photon-counting SME in innovation form as
\begin{widetext}
\begin{align}\label{eq:sme_component_photon_count}
\begin{split}
  d\rhonorm{t}{m,n}
  &=
  dt\,\Lind{t|\Rec}{m,n}{\mathbf{G},\xi}
  +d\mathcal{J}_t^N
  \bigg(\Big[
  L\rhonorm{t|\Rec}{m,n}L\dg
  +\xi_t^*L\big(\coshr \rhonorm{t|\Rec}{m,n-1}-\sinhr e^{-2i\phi}\rhonorm{t|\Rec}{m,n+1}\big)S\dg
  +\xi_tS\big(\coshr \rhonorm{t|\Rec}{m-1,n}-\sinhr e^{2i\phi}\rhonorm{t|\Rec}{m+1,n}\big)L\dg
  \\
  &\phantrel{=}{}
  +|\xi_t|^2S\big(
  \coshr^2\rhonorm{t|\Rec}{m-1,n-1}
  -\sinhr \coshr e^{-2i\phi}\rhonorm{t|\Rec}{m-1,n+1}
  -\sinhr \coshr e^{2i\phi}\rhonorm{t|\Rec}{m+1,n-1}
  +\sinhr^2\rhonorm{t|\Rec}{m+1,n+1}
  \big)S\dg
  \Big]\big/\Pr(\tight{R_t=J})
  -\rhonorm{t|\Rec}{m,n}
  \bigg)
  \end{split}
\end{align}
\end{widetext}
If we average over the possible measurement outcomes, then because $\mathbb{E}[d\mathcal{J}_t^N]=0$ we recover the master equation as expected.
In the absence of squeezing ($\sinhr=0$) we recover the Fock SMEs in \cite[see Eq. 50]{baragiola_quantum_2017}. 

The expression in \cref{eq:sme_component_photon_count} is unwieldy, so we translate it into the compact tensor notation introduced in \cref{sec:compact-notation}.
We first introduce a conditional state tensor,
\begin{align}
  \rhomat_{t|\Rec}
  &\defined
  \begin{pmatrix}
  \rhonorm{t|\Rec}{0,0} & \rhonorm{t|\Rec}{0,1} & \cdots \\
  \rhonorm{t|\Rec}{1,0} & \rhonorm{t|\Rec}{1,1} & \cdots \\
  \vdots & \vdots & \ddots
  \end{pmatrix}
  \\
  \rhomat_{t|\Rec}^\matdag
  &=
  \rhomat_{t|\Rec}.
\end{align}
Moreover, we define nonlinear ``superoperators'' for these matrices of matrices that are analogous to those used in the Markovian equations:
\begin{align}
  \hiertr(\varmat)
  &\defined
  \tr\Big([\varmat]_{0,0}\Big)
  \\
  \mathcal{G}[\varmat]\,\rhomat_t
  &\defined
  \frac{\varmat\rhomat_t\varmat^\matdag}
  {\hiertr(\varmat\rhomat_t\varmat^\matdag)}
  -\rhomat_t \,.
  \label{eq:supop-defs}
\end{align}

Using all of this notation, we can now express the photon-counting stochastic master equation in a compact form:
\begin{align}
  \label{eq:simplified-photon-counting-sme}
  \begin{split}
    d\rhomat_t
    &=
    dt\,\bigg[\mathcal{D}[L]\rhomat_{t|\Rec}-i[H\system,\rhomat_{t|\Rec}]
    +\xi_t[S\minusmat^\matdag_{\text{sq}}\rhomat_{t|\Rec},L^\dagger]\\
    &\phantrel{=}\hphantom{dt\,\bigg[}{}
    +\xi^*_t[L,\rhomat_{t|\Rec}\minusmat_{\text{sq}}S^\dagger]
    +\abs{\xi_t}^2\mathcal{D}[S]\big(
    \minusmat^\matdag_{\text{sq}}\rhomat_{t|\Rec}
    \minusmat_{\text{sq}}\big)
    \bigg]\\
    &\phantrel{=}\hphantom{dt\,\bigg[}{}
    +d\mathcal{J}_t^N\mathcal{G}[L+\xi_tS\minusmat_{\text{sq}}^\matdag]
    \rhomat_{t|\Rec}\,.
  \end{split}
\end{align}
From these expressions, it is easy to see how the vacuum equations are recovered by setting $\xi_t=0$ and how the Fock hierarchy (for photon counting) is recovered by setting $\minusmat_{\text{sq}}=\minusmat$.

\subsection{Homodyne}
Homodyne detection of the arbitrary quadrature $dQ_t^\phi$~\cite{GoetGrah94,JackColl00,baragiola_quantum_2017}, see \cref{eq:quadrature-increment}, can be viewed as a measurement of $\sigma_x$ on the infinitesimal field modes~\cite{gross_qubit_2017}.
We denote the infinitesimal quadrature eigenstates and the eigenvalue equation as
\begin{subequations}
\begin{align}
  |\pm_t^\phi\rangle&\defined
  \frac{1}{\sqrt{2}}\Big(\!\ket{0_t}\pm e^{i\phi}\ket{1_t}\!\Big) \\
dQ_t^\phi|\pm_t^\phi\rangle&=
    \pm\sqrt{dt}\,|\pm_t^\phi\rangle\,.
  \label{eq:quadrature-eigenstates}
\end{align}
\end{subequations}
The derivation of this eigenvalue equation  requires that the action of $\dBdag$ on $\nket{1_t}$ may be discarded. In the context of the derivation of the homodyne stochastic master equation one uses Kraus operator calculated via partial-expectation sandwiches such as $\bra{\pm_t}U_t\nket{n_{\gamma,\xi}}$. Thus discarding the action of $\dBdag$ on $\nket{1_t}$ is justified since these terms are too high order in $dt$ and don't contribute to the state-update equations.

We calculate the outcome probabilities for squeezed vacuum as before by taking the trace when $m=n=0$:
\begin{widetext}
\begin{align}
  \begin{split}
    \Pr(\tight{R_t=\pm})&=\tfrac{1}{2}\Big[1
    \pm\sqrt{dt}\tr\big(
    e^{-i\phi}\xi_t \sinhr \tightsqrt{m+1}S\rhonorm{t|\Rec}{1,0}
    +e^{-i\phi}L\rhonorm{t|\Rec}{0,0}
    +e^{i\phi}\xi_t^* \sinhr\tightsqrt{n+1}\rhonorm{t|\Rec}{0,1}S\dg
    +e^{i\phi}\rhonorm{t|\Rec}{0,0}L\dg
    \big)\Big]
    \\
    &=\tfrac{1}{2}\Big(1\pm\sqrt{dt}\hiertr\big[
    e^{-i\phi}(\xi_tS\minusmat_{\text{sq}}+L)\rhomat_{t|\Rec}
    +e^{i\phi}\rhomat_{t|\Rec}(\xi_tS\minusmat_{\text{sq}}+L)^\matdag
    \big]
    \Big)\,.
  \end{split}
\end{align}
\end{widetext}
As before, one obtains outcome probabilities for more general input field states by taking the trace for additional values of $m$ and $n$.

Finally, we express everything as a differential update using the compact notation from \cref{sec:compact-notation}:
\begin{align}
    d\rhomat_{t|\pm}&=
    dt\,\bigg[\mathcal{D}[L]\rhomat_{t|\Rec}-i[H\system,\rhomat_{t|\Rec}]
    +\xi_t[S\minusmat^\matdag_{\text{sq}}\rhomat_{t|\Rec},L^\dagger]
    \nonumber\\
    &\phantrel{=}{}
    +\xi^*_t[L,\rhomat_{t|\Rec}\minusmat_{\text{sq}}S^\dagger]
    +\abs{\xi_t}^2\mathcal{D}[S]\big(
    \minusmat^\matdag_{\text{sq}}\rhomat_{t|\Rec}
    \minusmat_{\text{sq}}\big)
    \bigg]
    \nonumber\\
    &\phantrel{=}{}
    +dW_t\Hc{e^{-i\phi}(\xi_tS\minusmat_{\text{sq}}+L)}\rhomat_{t|\Rec}
    \, ,
\end{align}
where
\begin{align}
  \mathcal{H}[\varmat]\rhomat_t
  &\defined
  \varmat\rhomat_t+\rhomat_t\varmat^\matdag
  -\hiertr\big(\varmat\rhomat_t
  +\rhomat_t\varmat^\matdag\big)\rhomat_t\,.
\end{align}
The Wiener process $dW_t$ is the innovation obtained by subtracting the mean from the random variable $\pm\mapsto\pm\sqrt{dt}$. 

A heterodyne SME follows from this homodyne SME by considering measuring $dQ_t^0$  (like $\sigma_x$ in the qubit model of Ref.~\cite{gross_qubit_2017}) with probability $\tfrac{1}{2}$ and  $dQ_t^{\pi/2}$ (like $\sigma_y$) with probability $\tfrac{1}{2}$ in each time step. Other generalizations such as inefficient detection, superpositions, and mixtures of squeezed Fock states in the same wave packet, as well as additional decoherence, follow in the same way as for the Fock hierarchy presented in Ref.~\cite{baragiola_quantum_2017}.

\section{conclusion}

The formalism developed here has produced the master equation, input-output relations, and stochastic master equations for a system interacting with a squeezed wave packet.
The master equation encodes the system and wave packet correlations in a hierarchy of equations that both recovers the broadband squeezing master equation in a short wave-packet limit and also, away from this limit, allows the description of non-Markovian features that distinguish general wave-packet evolution.
Particularly interesting in this regard is the behavior of a two-level atom in the presence of such a wave packet, exhibiting two decay timescales when relaxing (see ~\cref{sec:qual_comp}) and unique fluorescence spectra when being driven (see \cref{sec:res_fluor}).

The equation we obtain is unusual when compared with other wave-packet master equations.
While the hierarchy for wave packets in Fock states~\cite{Gheri:1998aa,gough_quantum_2012, baragiola_n-photon_2012,baragiola_quantum_2017,Dabrowska_2019_nphoton} or cat states~\cite{gough_quantum_2012,dabrowska_quantum_2019} involves only a finite set of differential equations, the corresponding hierarchy for wave packets expressed in squeezed Fock states always involves an infinite set of differential equations, resulting in a subtle tradeoff between the efficiency of representing states in the squeezed Fock basis and the increased evolution complexity in this basis.

Continuous-wave broadband-squeezing treatments have the unfortunate side effect of delivering infinite photon flux (also interpreted as a photon-count rate)~\cite{wiseman_quantum_2010}.
The squeezed wave packet formalism we present gives sensible and physical expressions for the flux that can be used for broadband squeezed fields with bandwidths inversely proportional to the wave packet duration. 
The continuous-wave broadband squeezing limit is realized as a train of wave packets, each being exceedingly short.
In this limit, one expects unphysical consequences from the equations.

The perspective on finite-bandwidth squeezing developed in this work may be further developed in several ways.
A similar approach using wave-packet modes could provide a description for more general fields with multi modal squeezing.
Understanding the numerical difficulties in truncating the squeezed hierarchy could be important for more generally developing more robust numerical techniques for infinite dimensional systems.
In particular, much of the numerical difficulty comes from truncation of the hierarchy, suggesting that future researchers might be able to find ways to approximate the solution of the higher levels of the hierarchy rather than truncating it.
Preliminary results (not presented here) also show that time-convolutionless (TCL) and the convolutionless projection methods can capture some of the dynamics we describe. 
Moreover because squeezed vacuum is entirely described by second order correlation functions, approaches like Refs.~\cite{DiosiFerialdi2014, Ferialdi2016} might afford a more natural description than the hierarchical equations we derived.
Finally one might imagine generalizing the elegant work of~ Kiilerich and M\o lmer \cite{kiilerich_input-output_2019,kiilerich_quantum_2020} and Trivedi \emph{et al.}~\cite{trivedi_Malz_Cirac_convergence_2021} to the squeezed bath case.

{\em Acknowledgements.} The authors would like to thank Adrian Chapman, Carl Caves, Ivan Deutsch, and Gerard Milburn for many discussions over many years on these topics.
This research was undertaken thanks in part to funding from the Canada First Research Excellence Fund and from NSERC.
This work was supported by the Australian Research Council (ARC) via the
Centre of Excellence for Engineered Quantum Systems (EQUS, CE170100009) and the Centre of Excellence in Quantum Computation and Communication Technology (CQC$^2$T, CE170100012), and a Discovery Early Career Research Award (DECRA) project number DE160100356.

\appendix

\section{Squeeze operator action on temporal mode}\label{app:lie_algebra}
To work out the action of the wave-packet squeeze operator on a temporal-mode operator, it is helpful to recall that conjugating by the unitary $e^{iH}$ is equivalent to exponentiating the commutator map of $iH$:
\begin{subequations}\label{eq:Ad_ad}
\begin{align}
  e^{iH}Ae^{-iH}
  =
  \exp\big([iH,\cdot]\big)A\,.
\end{align}
\end{subequations}
With this tool in hand, we work out the action of $\Sgx $ on $b_t$.  The first step is to define
\begin{subequations}
\begin{align}
  \Sgx 
  &=
  e^{-iH_\gamma[\xi]}
  \\
  -iH_\gamma[\xi]
  &=
  \frac{1}{2}\left(\gamma^*B[\xi]^2
  -\gamma B^\dagger[\xi]^2\right)
\end{align}
\end{subequations}
explicitly work out \cref{eq:Ad_ad} 
\begin{subequations}
\begin{align}
  \begin{split}
    \Sgxdag b_t\Sgx
    &=
    \exp\big([iH_\gamma[\xi],\cdot]\big)b_t
  \end{split}
  \\
  \begin{split}
    \big[iH_\gamma[\xi],b_t\big]
    &=
    \tfrac{1}{2}[b_t,\gamma^*B[\xi]^2-\gamma B^\dagger[\xi]^2]
    \\
    &=
    -\gamma\xi_tB^\dagger[\xi]
  \end{split}
  \\
  \big[iH_\gamma[\xi],B^\dagger[\xi]\big]
  &=
  -\gamma^*B[\xi]
  \\
  \big[iH_\gamma[\xi],B[\xi]\big]
  &=
  -\gamma B^\dagger[\xi].
\end{align}
\end{subequations}
Since $[iH_\gamma[\xi],\cdot]$ is a linear operator that maps the subspace spanned by $\{b_t,B^\dagger[\xi],B[\xi]\}$ to itself, we can represent it as a $3\times3$ matrix in this basis.
The correspondence between the components in this representation and the operators we are representing is given by
\begin{align}
  xb_t+yB^\dagger[\xi]+zB[\xi]
  &\doteq
  \begin{bmatrix}
    x \\ y \\ z
  \end{bmatrix}\,.
\end{align}
The representation of $[iH,\cdot]$ in this basis is:
\begin{align}
  \begin{split}
    \big[iH_\gamma[\xi],\cdot\big]
    &\doteq
    \begin{bmatrix}
      0 & 0 & 0
      \\
      -\gamma\xi_t & 0 & -\gamma
      \\
      0 & -\gamma^* & 0
    \end{bmatrix}
    \\
    &=
    r\begin{bmatrix}
      0 & 0 & 0
      \\
      -e^{2i\phi}\xi_t & 0 & -e^{2i\phi}
      \\
      0 & -e^{-2i\phi} & 0
    \end{bmatrix}.
  \end{split}
\end{align}
We diagonalize this matrix to facilitate exponentiation:
\begin{subequations}
\begin{align}
  \big[iH_\gamma[\xi],\cdot\big]
  &\doteq
  PJP^{-1}
  \\
  P
  &=
  \begin{bmatrix}-\frac{1}{\xi_t} & 0 & 0
  \\
  0 & e^{2i\phi} & -e^{2i\phi} \\ 1 & 1 & 1\end{bmatrix}
  \\
  J
  &=
  \begin{bmatrix}0 & 0 & 0 \\ 0 & -r & 0 \\ 0 & 0 & r\end{bmatrix}.
\end{align}
\end{subequations}
Now we can easily work out the conjugate action of $\Sgxdag$ on $b_t$:
\begin{align}
  \begin{split}
    \Sgxdag b_t\Sgx
    &\doteq
    P\exp(J)P^{-1}
    \begin{bmatrix}1 \\ 0 \\ 0\end{bmatrix}
    \\
    &=
    \begin{bmatrix}1 \\- \xi_te^{2i\phi} \sinhr
    \\
    \xi_t(\coshr-1)\end{bmatrix}\\
    &\doteq
    b_t+\xi_t\big[(\coshr-1)B[\xi]-e^{2i\phi} \sinhr B^\dagger[\xi]\big],
  \end{split}
\end{align}

where we make use of the definitions for $\coshr$ and $\sinhr$ given in~\cref{eq:squeezed-wavepacket-op}.
By integrating the above result from $t$ to $t+dt$, we obtain \cref{eq:squeezed_noise_inc_dB}
\begin{align}
  \dBsq
  &=
  \dB+dt \xi_t\big[(\coshr-1)B[\xi]-e^{2i\phi} \sinhr B\dg[\xi]\big].
\end{align}
The above expression can be used to derive in~\cref{eq:squeezed_noise_inc}.

\section{Temporal decomposition}\label{app:temp_decomp}

From \cite{baragiola_quantum_2017} we have the following temporal decomposition for a
number state in the wave-packet mode:
\begin{align}
  \ket{n_\xi}&=\ket{0_t}\otimes\ket{n_{\overline{t}}}+\sqrt{n\,dt}\,\xi_t
  \ket{1_t}\otimes\tightket{(n-1)_{\overline{t}}}\,.
  \label{eq:number-temp-decomp}
\end{align}
We need to do this same decomposition for a squeezed number state in the wave-packet mode $\ket{n_{\gamma,\xi}}=\Sgx \ket{n_\xi}$.
A useful definition is
\begin{align}
  B[\xi]&=\int dt\,\xi^*_tb_t=(1-\tfrac{1}{2}|\xi_t|^2dt)\Btild+\xi^*_t\dB
\end{align}
where $\Btild$ is an annihilation operator for the mode $\xi^\prime$ obtained from $\xi$ by ignoring the segment $[t,t+dt)$.
The prefactor can be understood as necessary for ensuring the proper commutation relation.
To make progress we need to separate the squeeze operator into a part that acts on times other than $t$ and a part that couples time $t$ to the rest of the wave packet.
\begin{widetext}
\begin{align}
  \Sgx &=\exp\left[\frac{r}{2}\left(e^{-2i\phi}\big[(1-\tfrac{1}{2}|\xi_t|^2dt)\Btild+\xi^*_t\dB\big]^2
  -e^{2i\phi}\big[(1-\tfrac{1}{2}|\xi_t|^2dt)\Btilddag+\xi_t\dBdag\big]^2\right)\right]
  \nonumber\\
  &=
  \exp\left(\frac{r}{2}\left[(1-|\xi_t|^2dt)\big(e^{-2i\phi}\Btild^2-e^{2i\phi}\Btilddag^2\big)
  +2e^{-2i\phi}\xi^*_t\Btild\dB-2e^{2i\phi}\xi_t\Btilddag\dBdag\right]\right).
  \label{eq:squeeze-op-expansion}
\end{align}
\end{widetext}
We've made use of the fact that $\dB^2=\dB\,dt=0$ from \cref{tab:vac-ito} in the above expression.

We now exploit the structure in \cref{eq:squeeze-op-expansion} to write it as a squeeze operation on the mode $\xi^\prime$ conjugated by a beamsplitter between the mode $\xi^\prime$ and the infinitesimal mode $t$:
\begin{subequations}
\label{eq:squeeze-op-decomp}
\begin{align}
  \Sgx
  &=
  e^Xe^Ye^{-X}
  \\
  X
  &=
  -\xi^*_t\dB\Btilddag+\xi_t\dBdag\Btild,
  \\
  Y
  &=
  \frac{r}{2}(e^{-2i\phi}\Btild^2-e^{2i\phi}\Btilddag^2)\,.
\end{align}
\end{subequations}
We verify this expression by writing the conjugation as a series of commutators and working out the commutators:
\begin{subequations}
\begin{align}
  e^Xe^Ye^{-X}
  &=
  \exp(Y+[X,Y]+\tfrac{1}{2}[X,[X,Y]]+\cdots)
  \\
  [X,Y]
  &=
  r\left(\xi^*_te^{-2i\phi}\dB\Btild-\xi_te^{2i\phi}\dBdag\Btilddag\right)
  \\
  [X,[X,Y]]
  &=
  -2\abs{\xi_t}^2dt\,Y.
\end{align}
\end{subequations}
Since the triple commutator $[X,[X,[X,Y]]]$ vanishes this shows that \cref{eq:squeeze-op-expansion,eq:squeeze-op-decomp} are equivalent.

Since the beamsplitters are weak, we only need to keep a few terms in their series expansions.
Once we perform this expansion, we commute the squeezing operator to the left of all the field operators on mode $\overline{t}$
\begin{align}
  e^{-X}
  &=
  \Id+\xi^*_t\dB\Btilddag-\xi_t\dBdag\Btild \nonumber \\
 &\phantom{=} -\tfrac{1}{2}\abs{\xi_t}^2dt\,\Btilddag\Btild
  \\
  e^{\tilde{Y}}
  &=
  \tilde{S}
  \\
  \tilde{S}^\dagger\Btild\tilde{S}
  &=
 \coshr\Btild-e^{2i\phi}\sinhr\Btilddag
\end{align}\
to simplify 
\begin{widetext}
\begin{align}
  \begin{split}
    e^X\tilde{S}
    =
    \tilde{S}\tilde{S}^\dagger e^X\tilde{S}
    &=
    \tilde{S}\left(\Id-\xi^*_t\dB\tilde{S}^\dagger\Btilddag\tilde{S}
    +\xi_t\dBdag\tilde{S}^\dagger\Btild\tilde{S}
    -\tfrac{1}{2}\abs{\xi_t}^2dt\,\tilde{S}^\dagger\Btilddag\tilde{S}
    \tilde{S}^\dagger\Btild\tilde{S}\right)
    \\
    &=
    \tilde{S}\left\{\Id-\xi^*_t\dB(\coshr \Btilddag-e^{-2i\phi} \sinhr\Btild)
    +\xi_t\dBdag(\coshr\Btild-e^{2i\phi}\sinhr\Btilddag)\vphantom{\Btilddag^2}\right.
    \\
    &\phantrel{=}\left.{}
    -\tfrac{1}{2}\abs{\xi_t}^2dt\,
    \left[(2\sinhr^2+1)\Btilddag\Btild
    -\coshr \sinhr(e^{-2i\phi}\Btild^2+e^{2i\phi}\Btilddag^2)+\sinhr^2\right]\right\}\,.
  \end{split}
  \label{eq:beamsplitter-manipulations}
\end{align}
Combine these expressions recalling the vanishing products of $\dB$ that we've
already used along with the identities $\dBdag\dB=0$ and $\dB\dBdag=dt$.
\begin{align}
  \begin{split}
  e^X\tilde{S}e^{-X}
  &=
  \tilde{S}\left\{\Id-\xi^*_t\dB\left[
  (\coshr-1)\Btilddag-e^{-2i\phi}\sinhr \Btild\right]+\xi_t\dBdag\left[
  (\coshr-1)\Btild-e^{2i\phi}\sinhr\Btilddag\right]\vphantom{\Btilddag^2}\right.
  \\
  &\phantrel{=}\left.{}
  -\tfrac{1}{2}\abs{\xi_t}^2dt\,\left[
  2(\sinhr^2+1)\Btilddag\Btild-\coshr \sinhr(e^{-2i\phi}\Btild^2+e^{2i\phi}\Btilddag^2)
  +\sinhr^2-2(\coshr\Btilddag-e^{-2i\phi}\sinhr\Btild)\Btild\right]\right\}
  \\
  &=
  \tilde{S}\left\{\Id-\xi^*_t\dB\left[
  (\coshr-1)\Btilddag-e^{-2i\phi}\sinhr \Btild\right]+\xi_t\dBdag\left[
  (\coshr-1)\Btild-e^{2i\phi} \sinhr \Btilddag\right]\vphantom{\Btilddag^2}\right.
  \\
  &\phantrel{=}\left.{}
  -\tfrac{1}{2}\abs{\xi_t}^2dt\,\left[
  2\coshr(\coshr-1)\Btilddag\Btild-(\coshr-2)se^{-2i\phi}\Btild^2-\coshr \sinhr e^{2i\phi}\Btilddag^2
  +\sinhr^2\right]\right\}\,.
  \end{split}
  \label{eq:squeeze-op-simple-form}
\end{align}
Now that we have the wave-packet squeezing operator in this form, we can apply the finite polynomial of $\dB$, $\dBdag$, $\Btnorm$, and $\Btnormdag$ to the temporal decomposition of the number state, recalling that $\ket{1_t}=(\dBdag/\sqrt{dt})\ket{0_t}$ and
$\ket{n_{\overline{t}}}=(\Btilddag^n/\sqrt{n!})\ket{0_{\overline{t}}}$.
\begin{align}
  \begin{split}
  \Sgx \ket{n_\xi}&=\ket{0_t}\otimes\left[\ket{n_{\gamma,\overline{t}}}
  -\tfrac{1}{2}\abs{\xi_t}^2dt\,
  \left([n(2\sinhr^2+1)+\sinhr^2]\ket{n_{\gamma,\overline{t}}}
  \vphantom{\tightsqrt{(n+1)}}\right.\right.\\
  &\phantrel{=}\hphantom{\ket{0_t}\otimes{}}\left.\left.
  {}-\coshr \sinhr \left[e^{-2i\phi}\tightsqrt{n(n-1)}\tightket{(n-2)_{\gamma,\overline{t}}}
  +e^{2i\phi}\tightsqrt{(n+1)(n+2)}\tightket{(n+2)_{\gamma,\overline{t}}}\right]\right)
  \right] \\
  &\phantrel{=}{}+\ket{1_t}\otimes\xi_t\sqrt{dt}\,
  \left(\coshr \sqrt{n}\tightket{(n-1)_{\gamma,\overline{t}}}
  -e^{2i\phi} \sinhr\tightsqrt{n+1}\tightket{(n+1)_{\gamma,\overline{t}}}\right)\,.
  \end{split}
  \label{eq:squeezed-number-temp-decomp}
\end{align}
We will have need to use this decomposition to express the renormalized
wave-packet states in terms of the original wave-packet states:
\begin{align}
\begin{split}
  \ket{0_t}\otimes\ket{n_{\gamma,\overline{t}}}&=\ket{n_{\gamma,\xi}}
  -\xi_t\sqrt{dt}\,\ket{1_t}
  \otimes\left(\coshr\sqrt{n}\tightket{(n-1)_{\gamma,\overline{t}}}
  -e^{2i\phi}\sinhr\tightsqrt{n+1}\tightket{(n+1)_{\gamma,\overline{t}}}\right) \\
  &\phantrel{=}{}+\tfrac{1}{2}\abs{\xi_t}^2dt\,\ket{0_t}
  \otimes\left([n(2\sinhr^2+1)+\sinhr^2]\ket{n_{\gamma,\overline{t}}}
  \vphantom{\tightsqrt{(n+1)}}\right. \\
  &\phantrel{=}\left.{}-\coshr \sinhr \left[e^{-2i\phi}\tightsqrt{n(n-1)}
  \tightket{(n-2)_{\gamma,\overline{t}}}
  +e^{2i\phi}\tightsqrt{(n+1)(n+2)}\tightket{(n+2)_{\gamma,\overline{t}}}\right]\right)\,.
\end{split}
  \label{eq:renorm-wavepacket-state-rewrite}
\end{align}

We can understand the extra $\tightnket{(n-2)_{\gamma,\overline{t}}}$ and $\tightnket{(n+2)_{\gamma,\overline{t}}}$ terms by considering the squeezed Fock state $S(r-\delta r)\nket{n}$ expressed in the basis $\{S(r)\nket{m}\}_m$ of slightly-more-squeezed Fock states.
\begin{align}
    S(r-\delta r)\nket{n}
    &=
    \frac{1}{\sqrt{n!}}S(r)S(-\delta r)a^{\dagger n}S(\delta r)S(-\delta r)\nket{0}\,.
\end{align}
To first order in $\delta r$ we have
\begin{align}
\label{eq:squeeze-reduce-decomp}
    S(-\delta r)a^{\dagger n}S(\delta r)
    &=
    \big[a^\dagger\cosh(-\delta r)-a\sinh(-\delta r)\big]^n
    \nonumber
    \\
    &=
    a^{\dagger n}+\delta r\sum_{m=0}^{n-1}a^{\dagger n-1-m}aa^{\dagger m}
    \\
    S(-\delta r)\nket{0}
    &=
    \nket{0}+\frac{1}{\sqrt{2}}\delta r\nket{2}\,.
\end{align}
Combining the expressions gives
\begin{align}
    S(-\delta r)\nket{n}
    &=
    \frac{1}{\sqrt{n!}}\big(a^{\dagger n}
    +\delta r\sum_{m=0}^{n-1}a^{\dagger n-1-m}aa^{\dagger m}\big)
    \times\big(\nket{0}+\tfrac{1}{\sqrt{2}}\delta r\nket{2}\big)
    \\
    &=
    \nket{n}
    +\tfrac{1}{2}\delta r\big(\tightsqrt{n(n-1)}\tightnket{n-2}
    +\tightsqrt{(n+1)(n+2)}\tightnket{n+2}\big)
    \label{eq:infinitesimal-squeeze}
\end{align}
since
\begin{align}
    \sum_{m=0}^{n-1}a^{\dagger n-1-m}aa^{\dagger m}\nket{0}
    &=
    \sum_{m=0}^{n-1}m\sqrt{(n-2)!}\ket{n-2}
    =\frac{n(n-1)}{2}\sqrt{(n-2)!}\ket{n-2}\,.
\end{align}
The leftmost squeeze in \cref{eq:squeeze-reduce-decomp} simply takes us back to the squeezed Fock basis, and we now see that the $\tightnket{(n-2)_{\gamma,\overline{t}}}$ and $\tightnket{(n+2)_{\gamma,\overline{t}}}$ terms in \cref{eq:squeezed-number-temp-decomp} correspond to a reduction in squeezing by magnitude $\delta r=|\xi_t|^2dt\,\cosh r\sinh r$ conditioned on detecting no photon at time $t$.

\end{widetext}

\section{Master-equation derivation}\label{app:mastereqnderivation}
In this section we present the details for the derivation of \cref{eq:mainresult}.

The standard technique to derive a master equation is to start with a system and field/bath state that is a product state $\rho\system \otimes \sigma\field$.
Next we evolve this combined state under an interaction unitary and then traces out the bath state
\begin{align}
\rho_t = \tr\field\left[ \Ut\rho\system \otimes \sigma\field \Utdag\right].
\end{align}
Due to the oddities of quantum stochastic calculus, to derive the differential equation for $\rho\system$ we need to consider the following junk \cite{gross_qubit_2017}
\begin{align}\label{eq:drho_du}
\begin{split}
    d\rho_t  &=
  \tr\field\big[\dU(\rho\system\otimes\sigma\field )\Utdag
  \\
  &\phantrel{=}{}
  \hphantom{\tr\field\big[}
  +\Ut(\rho\system\otimes\sigma\field)\dUdag
  \\
  &\phantrel{=}{}
  \hphantom{\tr\field\big[}
  +\dU(\rho\system\otimes\sigma\field)\dUdag\big]\,.
  \end{split}
\end{align}
Because we're considering wave packets in squeezed vacuum, $\sigma\field=\oprod{0_{\gamma,\xi}}{0_{\gamma,\xi}}$.
We know that the infinitesimal unitary contains the lowering operator $\dBsq$ which can add photons to the field, see \cref{eq:dB_on_ketN}.
Anticipating these additional squeezed-Fock-state factors we introduce the more general state matrices
\begin{align}
  \rhonorm{t}{m,n}
  &=
  \tr\field\big[\Ut(\rho_0\otimes\noprod{m_{\gamma,\xi}}{n_{\gamma,\xi}})\Utdag\big]\,,
\end{align}
using outer products of squeezed Fock states \cref{eq:squeezed-number-state-def}.\\
In order to make the derivation simpler some additional algebraic contortions are required 
\begin{align}\label{eq:rhomn_sqeezed_def}
  \rhonorm{t}{m,n}
  &=
  \tr\field\big[\Ut(\rho_0\otimes \Sgx\noprod{m_{\xi}}{n_{\xi}}\Sgxdag)\Utdag\big]\,,\nonumber\\
  &= \tr\field\big[\Utsq(\rho_0\otimes \noprod{m_{\xi}}{n_{\xi}})\Utsqdag \big]\,.
\end{align}
In taking the derivative of \cref{eq:rhomn_sqeezed_def}, one might wonder whether terms proportional to $d\Sgx$ are necessary.
Careful inspection reveals that $\Sgx$ has no time dependence, and therefore $d\Sgx=0$, allowing us to write \cref{eq:drho_du} for our case of interest as
\begin{widetext}
\begin{align}
  d\rhonorm{t}{m,n}
  &=
  \tr\field\big[\dUsq(\rho_0\otimes\noprod{m_\xi}{n_\xi}\dg)\Utsqdag
  +\Utsq(\rho_0\otimes\noprod{m_\xi}{n_\xi}\dg)\dUsqdag
  +\dUsq(\rho_0\otimes\noprod{m_\xi}{n_\xi})\dUsqdag\big]\,.
  \label{eq:differential-mn-sq}
\end{align}

The first two terms are related by taking the Hermitian conjugate and swapping $m$ and $n$:
\begin{align}
  \tr\field\big[\Utsq(\rho_0\otimes\noprod{m_\xi}{n_\xi})\dUsqdag\big]
  &=
  \tr\field\big[\dUsq(\rho_0\otimes\noprod{n_\xi}{m_\xi})\Utsqdag\big]\dg\,.
\end{align}

Let us explicitly work out the first term
\begin{align}
  \begin{split}\label{eq:linear-dU-initial}
  \tr\field\big[\dUsq(\rho_0\otimes\noprod{m_\xi}{n_\xi})\Utsqdag\big]
  &=
  \tr\field\Big(\big[-dt(\tfrac{1}{2}L\dg L+iH\system)\otimes\Id\field
  -L\dg S\otimes\dBsq
  +L\otimes\dBsqdag
  \\
  &\phantrel{=}{}
  \hphantom{\tr\field\Big(\big[}
  +(\tight{S-\Id\system})\otimes\dLamsq\big]\Utsq(\rho_0\otimes\noprod{m_\xi}{n_\xi})\Utsqdag\Big)
  \end{split}
  \\
  \begin{split}\label{eq:linear-dU-final}
  &=-dt(\tfrac{1}{2}L\dg L\rhonorm{t}{m,n}+iH\system \rhonorm{t}{m,n})
  \\
  &\phantrel{=}{}
  -dt\,\xit(\coshr\sqrt{m}L\dg S\rhonorm{t}{m-1,n}-e^{2i\phi} \sinhr \tightsqrt{m+1}L\dg S\rhonorm{t}{m+1,n})
  \\
  &\phantrel{=}{}
  +dt\,\xit^*(\coshr\sqrt{n}L\rhonorm{t}{m,n-1}-e^{-2i\phi}\sinhr\tightsqrt{n+1}L\rhonorm{t}{m,n+1})
  \\
  &\phantrel{=}{}
  +dt\abs{\xit}^2\big[\coshr^2\sqrt{mn}(\tight{S-\Id\system})\rhonorm{t}{m-1,n-1}
  \\
  &\phantrel{=}{}
  \hphantom{+dt\abs{\xit}^2\big[}
  -e^{-2i\phi}\sinhr\coshr \tightsqrt{m(n+1)}(\tight{S-\Id\system})\rhonorm{t}{m-1,n+1}
  \\
  &\phantrel{=}{}
  \hphantom{+dt\abs{\xit}^2\big[}
  -e^{2i\phi} \sinhr\coshr \tightsqrt{(m+1)n}(\tight{S-\Id\system})\rhonorm{t}{m+1,n-1}
  \\
  &\phantrel{=}{}
  \hphantom{+dt\abs{\xit}^2\big[}
  +\sinhr^2\tightsqrt{(m+1)(n+1)}(\tight{S-\Id\system})\rhonorm{t}{m+1,n+1}\big]
  \end{split}
\end{align}
To go from \cref{eq:linear-dU-initial} to \cref{eq:linear-dU-final} we need to do two things. First we must act the quantum noises on the Fock states using the expressions in \cref{eq:sq_quantum_noise_on_fock}. Second we must then use the definition in \cref{eq:rhomn_sqeezed_def}.

We do the simplifications by commuting the squeezed noise increments at time $t$ past the squeezed time evolution operators $\Utsq$ (which only couple to system to times earlier that $t$), using the cyclic property of the trace as needed for the field operators, and pulling the system operators outside the partial trace, as illustrated for a single term below:
\begin{align}
  \begin{split}
  \tr\field\big[(L\otimes\dBsqdag)\Utsq(\rho_0\otimes\noprod{m_\xi}{n_\xi})\Utsqdag\big]
  &=
  L\tr\field\big[\Utsq(\rho_0\otimes\dBsqdag\noprod{m_\xi}{n_\xi})\Utsqdag\big]
  \\
  &=
  L\tr\field\big[\Utsq(\rho_0\otimes\noprod{m_\xi}{n_\xi}\dBsqdag)\Utsqdag\big]
  \\
  &=
  dt\,\xit^*(\coshr \sqrt{n}L\rhonorm{t}{m,n-1}-e^{-2i\phi} \sinhr \tightsqrt{n+1}\rhonorm{t}{m,n+1})\,.
  \end{split}
\end{align}

In the term quadratic in $\dUsq$ from \cref{eq:differential-mn-sq}, \ie{} $ \tr\field [\dUsq(\rho_0\otimes\noprod{m_\xi}{n_\xi})\dUsqdag]$, the only surviving terms correspond to $\dBsqdag\noprod{m_\xi}{n_\xi}\dBsq$, $\dBsqdag\noprod{m_\xi}{n_\xi}\dLamsq$, $\dLamsq\noprod{m_\xi}{n_\xi}\dBsq$, and $\dLamsq\noprod{m_\xi}{n_\xi}\dLamsq$.
Other terms are of too high order in $dt$.
These terms survive because $[\dBsq,\dBsqdag]=dt$.
The result is
\begin{align}
  \begin{split}
  \tr\field\big[\dUsq(\rho_0\otimes\noprod{m_\xi}{n_\xi})\dUsqdag\big]
  &=
  L\tr\field\big[\Utsq(\rho_0\otimes\dBsqdag\noprod{m_\xi}{n_\xi}\dBsq)\Utsqdag\big]L\dg
  \\
  &\phantrel{=}{}
  +L\tr\field\big[\Utsq(\rho_0\otimes\dBsqdag\noprod{m_\xi}{n_\xi}\dLamsq)\Utsqdag\big](\tight{S\dg-\Id\system})
  \\
  &\phantrel{=}{}
  +(\tight{S-\Id\system})\tr\field\big[\Utsq(\rho_0\otimes\dLamsq\noprod{m_\xi}{n_\xi}\dBsq)\Utsqdag\big]L\dg
  \\
  &\phantrel{=}{}
  +(\tight{S-\Id\system})\tr\field\big[\Utsq(\rho_0\otimes\dLamsq\noprod{m_\xi}{n_\xi}\dLamsq)\Utsqdag\big](\tight{S\dg-\Id\system})\,.
  \end{split}
  \\
  \begin{split}
  &=
  dt\,L\rhonorm{t}{m,n}L\dg
  \\
  &\phantrel{=}{}
  +dt\,\xit^*\big[\coshr \sqrt{n}L\rhonorm{t}{m,n-1}(\tight{S\dg-\Id\system})-e^{-2i\phi}\sinhr \tightsqrt{n+1}L\rhonorm{t}{m,n+1}(\tight{S\dg-\Id\system})\big]
  \\
  &\phantrel{=}{}
  +dt\,\xit\big[\coshr\sqrt{m}(\tight{S-\Id\system})\rhonorm{t}{m-1,n}L\dg-e^{2i\phi} \sinhr \tightsqrt{m+1}(\tight{S-\Id\system})\rhonorm{t}{m+1,n}L\dg\big]
  \\
  &\phantrel{=}{}
  +dt\abs{\xit}^2\big[\coshr^2\sqrt{mn}(\tight{S-\Id\system})\rhonorm{t}{m-1,n-1}(\tight{S\dg-\Id\system})
  \\
  &\phantrel{=}{}
  \hphantom{+dt\abs{\xit}^2\big[}
  -e^{-2i\phi}\sinhr\coshr \tightsqrt{m(n+1)}(\tight{S-\Id\system})\rhonorm{t}{m-1,n+1}(\tight{S\dg-\Id\system})
  \\
  &\phantrel{=}{}
  \hphantom{+dt\abs{\xit}^2\big[}
  -e^{2i\phi} \sinhr\coshr \tightsqrt{(m+1)n}(\tight{S-\Id\system})\rhonorm{t}{m+1,n-1}(\tight{S\dg-\Id\system})
  \\
  &\phantrel{=}{}
  \hphantom{+dt\abs{\xit}^2\big[}
  +\sinhr^2\tightsqrt{(m+1)(n+1)}(\tight{S-\Id\system})\rhonorm{t}{m+1,n+1}(\tight{S\dg-\Id\system})\big]\,.
  \end{split}
\end{align}

Combining all these expressions together gives us the master equation in \cref{eq:mainresult}.

\section{Quasi-Markovian master equation}\label{app:yeoman_barnett}
See Refs.~\cite{Carmichael_1973, yeoman_influence_1996,Ficek97,Tanas1999,Kowalewska-kudlaszyk:2001aa} for a derivation of the quasi-Markoffian master equation.
A key point in the derivation from \cref{eq:dU_propagator} is the assumption that
\begin{align}
\langle b\dg(\omega) b(\omega')\rangle= N(\omega) \delta(\omega-\omega') \quad \text{and} \quad \langle b(\omega) b(\omega')\rangle= M(\omega) \delta(2\omega_a - \omega-\omega') 
\end{align} 

In our code~\cite{gross2021pysme} we adapted the notation from~\citet{yeoman_influence_1996} (YB). 
Specifically, the quantity $\omega_L$ from YB is the laser frequency which we denote as $\omega_c$ in the main text, $\phi_L$ is the laser phase, $\Omega $ Rabi frequency, and the quantity $\omega_A$ from YB is the atomic transition frequency which we denote as $\omega_a$ in the main text.

Yeoman and Barnett make the additional assumption that the laser frequency is exactly resonant with the carrier frequency of the squeezed field.
This in turn leads to the $N(\omega)$ being evaluated at $\omega_A$ and $\omega_A \pm \Omega$. But because of the correlated nature of the squeezed field we find $N(\omega_A+\Omega) = N(\omega_a - \Omega)$, with a similar argument holding for $M(\omega)$.

We redefine the operators 
\begin{align}
\sigma_-  \mapsto \sigma_- e^{-i(\phi_L - \omega_L)t} \quad \text{and} \quad \sigma_+ \mapsto \sigma_+ e^{i(\phi_L - \omega_L)t}.
\end{align}
Then the master equation is
\begin{align}
d\rho_{\rm YB}
= -\quart \lw  \Big \{ (1+N_A) &(\sigma_+\sigma_- \rho + \rho \sigma_+ \sigma_-  - \sigma_+ \rho \sigma_+ - \sigma_- \rho \sigma_- -2 \sigma_- \rho \sigma_+)  \nonumber \\
+ N_A & (\sigma_-\sigma_+\rho + \rho  \sigma_- \sigma_+ - \sigma_+ \rho \sigma_+ -\sigma_-\rho \sigma_- -2 \sigma_+ \rho \sigma_- ) \nonumber \\
 +(1+N_{+\Omega}) &(\sigma_+\sigma_- \rho + \rho \sigma_+ \sigma_-  + \sigma_+ \rho \sigma_+ + \sigma_- \rho \sigma_- -2 \sigma_- \rho \sigma_+)  \nonumber \\
+N_{+\Omega} & (\sigma_-\sigma_+\rho + \rho  \sigma_- \sigma_+ + \sigma_+ \rho \sigma_+ +\sigma_-\rho \sigma_- -2 \sigma_+ \rho \sigma_- ) \nonumber \\
-M_{A} e^{-2i \phi_L} & (\sigma_-\sigma_+\rho + \rho  \sigma_+ \sigma_- -2 \sigma_+ \rho \sigma_+ -\sigma_+\rho \sigma_- -\sigma_- \rho \sigma_+ ) \nonumber \\
-M_{A}^* e^{2i \phi_L} & (\sigma_+\sigma_-\rho + \rho  \sigma_- \sigma_+ -2 \sigma_- \rho \sigma_- -\sigma_+\rho \sigma_- -\sigma_- \rho \sigma_+ ) \nonumber \\
-M_{+\Omega} e^{-2i \phi_L} & (-\sigma_-\sigma_+\rho - \rho  \sigma_+ \sigma_- -2 \sigma_+ \rho \sigma_+ +\sigma_+\rho \sigma_- +\sigma_- \rho \sigma_+ ) \nonumber \\
-M_{+\Omega}^* e^{2i \phi_L} & (-\sigma_+\sigma_-\rho - \rho  \sigma_- \sigma_+ -2 \sigma_- \rho \sigma_- +\sigma_+\rho \sigma_- +\sigma_- \rho \sigma_+ ) \Big \} \nonumber \\
-\half i \Omega [\sigma_x,\rho]& -\half i \omega_A [\sigma_z, \rho] \nonumber \\
+F(\omega) \{  [\sigma_x,\rho]& - \sigma_z \rho (\sigma_+ - \sigma_-) - (\sigma_+ - \sigma_-)\rho \sigma_z \}\nonumber \\
+G(\omega) [\sigma_z [\sigma_x,\rho]],
\end{align}
where $M_+ = M(\omega_a + \Omega)$ and $N_+ = N(\omega_a + \Omega)$.
Defining $\Delta = \omega - \omega_a$ the functions $F$ and $G$ are 
\begin{align}
F(\omega_a)  &=  -\frac{1}{4} i K^2(\omega_a) \mathscr{P} \int_{-\infty}^{\infty} \frac{d\Delta}{\Delta +\Omega} [M(\omega) e^{-2i\phi_L} +M^*(\omega) e^{2i\phi_L}  + 2N(\omega)]\\
G(\omega_a) &= -\frac{1}{4} i K^2(\omega_a) \mathscr{P} \int_{-\infty}^{\infty} \frac{d\Delta}{\Delta +\Omega} [M(\omega) e^{-2i\phi_L} +M^*(\omega) e^{2i\phi_L}],
\end{align}
where $\mathscr{P}$ denotes the principal value.

To recover the usual Markovian master equation 
\begin{equation}
N(\omega_a) = N(\omega_a \pm \Omega) = N \quad \text{and} \quad |M(\omega_a)| = |M(\omega_a\pm \Omega)| = |M|,
\end{equation}
which gives $F=0$ and $G=0$.
\end{widetext}

\section{Numerical-analysis details}
\label{sec:num-anal-det}
To understand the effect of truncating the squeezed hierarchy, consider a source model for the wave packet consisting of a leaky cavity with a mode prepared in a squeezed state, where the cavity decay rate is modulated as a function of time to produce the appropriate wave-packet shape.
In such a model, the cavity mode progresses from a pure squeezed state through thermal squeezed states to ultimately relax to vacuum.
To identify the path through state space taken by the cavity mode, consider the case of constant decay rate.
Diverse modulations of the decay rate will adjust the speed with which the mode traverses the path, but this detail may be ignored for our purposes.
The cavity mode starts in a squeezed state, and undergoes a Gaussian loss channel, so all states along the path will be Gaussian and characterized by their first and second moments.
Additionally, squeezed vacuum has no mean displacement, so the expectation of the annihilation operator, $\alpha(t)=\langle a(t)\rangle$, is identically zero.
This leaves two quadratic expectation values to calculate:
\begin{align}
  N(t)&=\langle a\dg(t)a(t)\rangle \\
  M(t)&=\langle a(t)a(t)\rangle\,.
\end{align}
For a lossy cavity with decay rate $\gamma$, the equation of motion for the
annihilation operator is
\begin{align}
  \dot{a}&=\gamma\big[a\dg aa-\tfrac{1}{2}(a\dg aa+aa\dg a)\big]
  =\tfrac{1}{2}\gamma[a\dg,a]a=-\tfrac{1}{2}\gamma a\,.
\end{align}
The solution $a(t)$ and the derived quantities $N(t)$ and $M(t)$ are
\begin{align}
  a(t)&=e^{-\gamma t/2}a(0) \\
  N(t)&=e^{-\gamma t}N(0) \\
  M(t)&=e^{-\gamma t}M(0)
\end{align}
In the Schr\"{o}dinger picture, we can think of $N$ and $M$ as coming from the squeezed thermal state of the cavity, parameterized by the expected number of photons of the unsqueezed thermal state $N_{\text{th}}$ and the strength $r$ and phase $\phi$ of the squeezing applied to it:
\begin{align}
  N
  &=
  (2N_{\text{th}}+1)\sinh^2\!r+N_{\text{th}}
  \\
  M
  &=
  -(2N_{\text{th}}+1)e^{2i\phi}\sinh r\cosh r\,.
\end{align}
Our expressions for $N(t)$ and $M(t)$, together with initial conditions $N(0)=\sinh^2 r$ and $M(0)=-e^{2i\phi}\sinh r\cosh r$, determine the quantity $N_{\text{th}}(t)$:
\begin{align}
  \begin{split}
    \big(N_{\text{th}}(t)+\tfrac{1}{2}\big)^2
    &=
    \big(N(t)+\tfrac{1}{2}\big)^2-|M(t)|^2
    \\
    &=
    \big(e^{-\gamma t}-e^{-2\gamma t}\big)\sinh^2r+\tfrac{1}{4}
  \end{split}
\end{align}
This ``thermal photon number'' is a kind of entanglement entropy, quantifying the amount of entanglement between the cavity field and the field modes into which it has been leaking.
Maximizing this quantity therefore ought to upper bound the amount of entanglement between the system and the future field that the equations need to keep track of:
\begin{align}
  \max_tN_{\text{th}}(t)
  &=
  \tfrac{1}{2}(\cosh r-1)\,.
\end{align}

We use this result to estimate how many levels of the hierarchy are needed to attain a given precision with the squeezed hierarchy.
The density operator for a squeezed thermal state has the same eigenvalue spectrum as a thermal state $\rho_{\text{th}}$ with mean photon number $N_{\text{th}}$.
We proceed under the assumption that including levels $\rhonorm{t}{m,n}$ in the squeezed hierarchy for $0\leq m,n\leq \nmax$ is analogous to including only number states for $0\leq n\leq \nmax$ in the eigendecomposition of $\rho_{\text{th}}$.
The amount of population neglected by stopping at $\nmax$ is
\begin{align}
  \label{eq:sq-disc-pop-app}
  \begin{split}
    \discardpop_{\text{sq}}(n_{\text{max}},r)
    &=
    \frac{1}{N_{\text{th}}(t_{\text{th}})+1}
    \sum_{n=n_{\text{max}}+1}^{\infty}
    \left(\frac{N_{\text{th}}(t_{\text{th}})}
    {N_{\text{th}}(t_{\text{th}})+1}\right)^n
    \\
    &=
    \left(\frac{\cosh r-1}{\cosh r+1}\right)^{n_{\text{max}}+1}\,.
  \end{split}
\end{align}
This expression serves as a first guess for the amount of error incurred by capping the squeezed hierarchy at $\nmax$ as opposed to simulating the full hierarchy.
To convert this into a hypothesis for the number of levels of the hierarchy one needs to account for all but $\discardpop$ of the population, simply invert:
\begin{align}
  n_{\text{max}}(\discardpop)&=\frac{-\ln\discardpop}{\ln(\cosh r+1)-\ln(\cosh r-1)}-1
\end{align}

\section{Details of the SME derivation}
\label{app:sme_derivation}
A foundational expression in the approach taken in \cite{baragiola_quantum_2017} to derive the SME is the relative state decomposition of the wave-packet state with respect to the temporal mode at time $t$.
Specifically, we aim to express the state in the form
\begin{align}
  \nket{0_{\gamma,\xi}}&=\nket{0_t}\otimes\nket{\unnorm{\psi}_0}
  +\nket{1_t}\otimes\nket{\unnorm{\psi}_1}\,,
\end{align}
where $\unnorm{\psi}_0$ and $\unnorm{\psi}_1$ are unnormalized states on the temporal-mode tensor-product space excluding the mode at time $t$.
We will refer to such a decomposition as a \emph{temporal decomposition}.
Our restriction to the subspace spanned by $\big\{\nket{0_t},\nket{1_t}\big\}$ is justified by the rules of quantum stochastic calculus that discard many-photon-creation terms like $\dBdag\dBdag$ as being of insignificant order in $dt$.
These product rules as given in \cref{tab:vac-ito} will prove indispensable for arriving at manageable expressions in the following derivation.

\begin{table}[h]
  \centering
  \begin{tabular}{c|c|c|c|c|}
                 & $\dB$ & $\dBdag$ & $d\Lambda_t$ & $\df t$ \\
    \hline
    $\dB$        & $0$   & $\df t$  & $\dB$        & $0$     \\
    \hline
    $\dBdag$     & $0$   & $0$      & $0$          & $0$     \\
    \hline
    $d\Lambda_t$ & $0$   & $\dBdag$ & $d\Lambda_t$ & $0$     \\
    \hline
    $\df t$      & $0$   & $0$      & $0$          & $0$     \\
    \hline
  \end{tabular}
  \caption{Vacuum It\={o} table for pairwise products of noise increments. The
  rows are labeled by the left factor in the product and the columns by the
  right factor in the product.}
  \label{tab:vac-ito}
\end{table}

It is useful to split the wave-packet operators into terms that act trivially at time $t$ and terms that act non trivially at time $t$:
\begin{align}
  \label{eq:wavepacket-annihil-terms}
  B[\xi]&=\Bt+\xi^*_t\dB\,,
\end{align}
$\Btdag$ is an sub normalized creation operator for the $t$-excluded mode $\overline{t}$.
This sub normalization follows from the commutation relations of $B[\xi]$ and $\dB$.
\begin{align}
  \big[B[\xi],\dB\dg\big]&=\xi_t^*[\dB,\dB\dg]=\xi_t^*dt
  \\
  \big[\Bt,\Bt\dg\big]&=\big[B[\xi],B[\xi]\dg\big]
  -\xi_t\big[B[\xi],\dB\dg\big]
  \nonumber\\
  &\phantrel{=}{}
  -\xi_t^*\big[\dB,B[\xi]\dg\big]+\abs{\xi_t}^2[\dB,\dB\dg]
  \nonumber\\
  &=1-\abs{\xi_t}^2dt\,.
\end{align}
The normalized creation operator for the mode $\overline{t}$ is therefore
\begin{align}
  \Btnorm&\defined\Bt\bigg/\sqrt{1-\abs{\xi_t}^2\!dt}
  =\sqrt{1+\abs{\xi_t}^2\!dt}\,\Bt\,.
\end{align}

The final squeezing operator on mode $\overline{t}$ then simply transforms number states in mode $\overline{t}$---such as those showing up in \cref{Eq::Focktempdecomp}---into squeezed number states in mode $\overline{t}$:
\begin{align}
  \begin{split}
  \nket{n_{\gamma,\xi}}&=\Sgx \nket{n_\xi} \\
  &=\nket{0_t}\otimes\big(\nket{n_{\gamma,\overline{t}}}
  +dt\,\nket{\unnorm{\psi}^n_{dt}}\big)
  +\sqrt{dt}\,\nket{1_t}\otimes\nket{\unnorm{\psi}^n_{\sqrt{dt}}}\,,
  \end{split}
\end{align}
where we have defined
\begin{align}
  \begin{split}
    \nket{\unnorm{\psi}^n_{dt}}&\defined-\tfrac{1}{2}\abs{\xi_t}^2\,
    \Big([n(2\sinhr^2+1)+\sinhr^2]\nket{n_{\gamma,\overline{t}}}
    \vphantom{\tightsqrt{(n+1)}}\\
    &
    {}-\coshr \sinhr\Big[e^{-2i\phi}\tightsqrt{n(n-1)}\tightket{(n-2)_{\gamma,\overline{t}}}
    \\
    &
    +e^{2i\phi}\tightsqrt{(n+1)(n+2)}\tightket{(n+2)_{\gamma,\overline{t}}}\Big]\Big)
  \end{split}
  \\
  \nket{\unnorm{\psi}^n_{\sqrt{dt}}}&\defined
  \xi_t\left(\coshr \sqrt{n}\tightket{(n-1)_{\gamma,\overline{t}}}
  -e^{2i\phi}\sinhr \tightsqrt{n+1}\tightket{(n+1)_{\gamma,\overline{t}}}\right)\,.
\end{align}
This is the promised temporal decomposition of the wave-packet state! We could immediately proceed with master equation derivations, but this would burden us with unnecessary trouble since the state contains some irrelevant details.
We discard these details of the state by recognizing that the only way the state finds its way into expressions is by interacting with the system via \cref{eq:inf-time-evol-op}. 
\begin{widetext}
\begin{align}
  \begin{split}
    U_t\nket{n_{\gamma,\xi}}&=
    \Id\system\otimes\Big[\nket{0_t}\otimes\big(\nket{n_{\gamma,\overline{t}}}
    +dt\,\nket{\unnorm{\psi}^n_{dt}}\big)
    +\sqrt{dt}\,\nket{1_t}\otimes\nket{\unnorm{\psi}^n_{\sqrt{dt}}}\Big]
    -dt\,L^\dagger S\otimes\nket{0_t}\otimes\nket{\unnorm{\psi}^n_{\sqrt{dt}}}
    +\sqrt{dt}\,L\otimes\nket{1_t}\otimes\nket{n_{\gamma,\overline{t}}}
    \\
    &\phantrel{=}{}
    +\sqrt{dt}\,(S-\Id\system)\otimes\nket{1_t}\otimes
    \nket{\unnorm{\psi}^n_{\sqrt{dt}}}
    -dt(iH\system+\tfrac{1}{2}L^\dagger L)\otimes\nket{0_t}\otimes
    \nket{n_{\gamma,\overline{t}}}\,.
  \end{split}
\end{align}
After this interaction they are subsequently projected onto either $\nbra{0_t}$ or $\nbra{1_t}$.
Since it is inconvenient to have the size of the Hilbert space change every time a measurement is made, we reattach a $\nket{0_t}$ for convenience.
While this choice makes no difference for the system dynamics---since this part of the field never interacts again and is ultimately traced out---it reveals a convenient way to represent the post interaction/projection state-of-affairs in terms of the original wave-packet state.
First, rewrite \cref{eq:squeezed-temp-decomp} to express $\nket{0_t}\otimes\nket{n_{\gamma,\overline{t}}}$ as
\begin{align}
  \label{eq:not-t-temp-decomp}
  \nket{0_t}\otimes\nket{n_{\gamma,\overline{t}}}&=\nket{n_{\gamma,\xi}}
  -dt\,\nket{0_t}\otimes\nket{\unnorm{\psi}^n_{dt}}
  -\sqrt{dt}\,\nket{1_t}\otimes\nket{\unnorm{\psi}^n_{\sqrt{dt}}}\,.
\end{align}
Use this identity to express the post interaction/projection state-of-affairs as
\begin{align}
  \begin{split}
    \label{eq:sqz-wavepacket-0-sand}
    \noprod{0_t}{0_t}U_t\nket{n_{\gamma,\xi}}&=\Id\system\otimes\nket{0_t}\otimes
    \big(\nket{n_{\gamma,\overline{t}}}+dt\,\nket{\unnorm{\psi}^n_{dt}}\big)
    -dt\,L^\dagger S\otimes\nket{0_t}\otimes\nket{\unnorm{\psi}^n_{\sqrt{dt}}}
    -dt(iH\system+\tfrac{1}{2}L^\dagger L)\otimes\nket{0_t}\otimes
    \nket{n_{\gamma,\overline{t}}}
    \\
    &=\Id\system\otimes\big(\nket{n_{\gamma,\xi}}
    -\sqrt{dt}\,\nket{1_t}\otimes\nket{\unnorm{\psi}^n_{\sqrt{dt}}}\big)
    -dt\,L^\dagger S\otimes\nket{0_t}\otimes\nket{\unnorm{\psi}^n_{\sqrt{dt}}}
    -dt(iH\system+\tfrac{1}{2}L^\dagger L)\otimes\nket{n_{\gamma,\xi}}
  \end{split}
  \\
  \begin{split}
    \label{eq:sqz-wavepacket-1-sand}
    \noprod{0_t}{1_t}U_t\nket{n_{\gamma,\xi}}&=
    \sqrt{dt}\Big[\Id\system\otimes\nket{0_t}
    \otimes\nket{\unnorm{\psi}^n_{\sqrt{dt}}}+L\otimes\nket{0_t}
    \otimes\nket{n_{\gamma,\overline{t}}}
    +(S-\Id\system)\otimes\nket{0_t}\otimes\nket{\unnorm{\psi}^n_{\sqrt{dt}}}\Big]
    \\
    &=\sqrt{dt}S\otimes\nket{0_t}
    \otimes\nket{\unnorm{\psi}^n_{\sqrt{dt}}}
    +\sqrt{dt}L\otimes\nket{n_{\gamma,\xi}}
    +dt\,L\otimes\nket{1_t}\otimes\nket{\unnorm{\psi}^n_{\sqrt{dt}}}\,.
  \end{split}
\end{align}
We find that, by utilizing \cref{eq:not-t-temp-decomp} and discarding terms of irrelevant order in $dt$, we can always fold the $dt$ correction to the $\ket{0_t}$ coefficient back into $\nket{n_{\gamma,\xi}}$, removing all dependence on $\nket{\unnorm{\psi}^n_{dt}}$ and saving us a lot of trouble.
Recall though, from the analysis in \cref{app:temp_decomp}, that these term $\nket{\unnorm{\psi}^n_{dt}}$ has physical meaning in terms of squeezing reduction conditioned on not detecting a photon, as seen in \cref{eq:infinitesimal-squeeze}.

\subsection{State update}\label{sec:state_update_sme}
Armed with the temporal decomposition of the squeezed wave-packet state, we now aim to describe the dynamics of a system illuminated by a field in such a state.
Though it is possible to directly derive the unconditional dynamics for the case where the outgoing field is beyond our reach, as was done in \cref{app:mastereqnderivation}, here we adopt the approach of solving for the dynamics conditioned upon a particular measurement record.
We can then arrive at the master equation via an alternate route, since the unconditional dynamics emerge from averaging over potential measurement records.

Given a measurement record $\mathbf{R}$ on the interval $[0,t)$ the conditional reduced state of the system at time $t$ is given by
\begin{align}
  \rho_{t|\mathbf{R}}&=\frac{\tr_{[t,\infty)}\Big[C_{\mathbf{R}}\big(\rho_0
  \otimes\oprod{0_{\gamma,\xi}}{0_{\gamma,\xi}}\big)
  C_{\mathbf{R}}^\dagger\Big]}{\Pr(\mathbf{R})}\,,
  \label{eq:normalized-cond-state}
\end{align}
where we have traced out the future field that has yet to interact with the system and made use of the composite Kraus operators $C_{\Rec}$ to capture the joint evolution of the field and system in the unitary $U(0,t)$ and the measurement record in the inner product with $\nbra{\Rec}$---the tensor product of temporal field eigenstates corresponding to the measurement record $\Rec$:
\begin{align}
  C_{\mathbf{R}}&=\bra{\mathbf{R}}U(t,0)\otimes \Id_{[t,\infty)}\,.
  \label{eq:composite-kraus-op}
\end{align}
We will again have need to reference the following more general ``bookkeeping'' operators:
\begin{align}
  \rhonorm{t}{m,n}&=\frac{\tr_{[t,\infty)}\Big[C_{\mathbf{R}}
  \big(\rho_0\otimes\oprod{m_{\gamma,\xi}}{n_{\gamma,\xi}}\big)
  C_{\mathbf{R}}^\dagger\Big]}{\Pr(\mathbf{R})}
  ={\rhonorm{t}{n,m}}\dg\,.
  \label{eq:normalized-bookkeeping-cond-state}
\end{align}

To derive the stochastic differential equation governing the evolution of the system we seek to express the update to the state given a measurement result at time $t$ in terms of the measurement result and our current description of the system given by $\rhonorm{t}{m,n}$.
Since temporal correlations in the wave-packet result in non-Markovian dynamics, we know that the reduced system state $\rhonorm{t}{0,0}$ alone will generally be insufficient, so we expect the bookkeeping operators $\rhonorm{t}{m,n}$ will play a role in capturing the necessary memory effects, just as they did in the master-equation derivation.

To proceed, we must decide on a particular field measurement to perform. Photon counting presents itself as a reasonable choice, especially considering that the temporal decomposition we have performed privileges the measurement eigenbases, so we turn our attention to this measurement in the following section.

\subsection{Photon counting}\label{sec:app_photon_sme}
We now follow the derivation procedure outline by Baragiola et al.~\cite{baragiola_quantum_2017}. 
The potential results for an infinitesimal photon-counting measurement are $\emptyset$ (no detection) and $J$ (detection).
The unnormalized updated states corresponding to these results are
\begin{align}
  \label{eq:no-detection-part-trace-update}
  \begin{split}
    \rhonorm{t+dt|\Rec,\emptyset}{m,n}
    &=
    \frac{\tr_{[t+dt,\infty)}\left[\bra{0_t}U_tC_{\mathbf{R}}
    \left(\rho_0\otimes\oprod{m_{\gamma,\xi}}{n_{\gamma,\xi}}\right)
    C_{\mathbf{R}}^\dagger U_t^\dagger\ket{0_t}\right]}
    {\Pr(\mathbf{R})\Pr(\emptyset|\Rec)} \\
    &=\frac{\tr_{[t,\infty)}\left[\oprod{0_t}{0_t}U_tC_{\mathbf{R}}
    \left(\rho_0\otimes\oprod{m_{\gamma,\xi}}{n_{\gamma,\xi}}\right)
    C_{\mathbf{R}}^\dagger U_t^\dagger\oprod{0_t}{0_t}\right]}
    {\Pr(\mathbf{R})\Pr(\emptyset|\Rec)}
  \end{split}
  \\
  \label{eq:detection-part-trace-update}
  \begin{split}
    \rhonorm{t+dt|\Rec,J}{m,n}
    &=
    \frac{\tr_{[t+dt,\infty)}\left[\bra{1_t}U_tC_{\mathbf{R}}
    \left(\rho_0\otimes\oprod{m_{\gamma,\xi}}{n_{\gamma,\xi}}\right)
    C_{\mathbf{R}}^\dagger U_t^\dagger\ket{1_t}\right]}{\Pr(\mathbf{R})\Pr(J|\Rec)} \\
    &=\frac{\tr_{[t,\infty)}\left[\oprod{0_t}{1_t}U_tC_{\mathbf{R}}
    \left(\rho_0\otimes\oprod{m_{\gamma,\xi}}{n_{\gamma,\xi}}\right)
    C_{\mathbf{R}}^\dagger U_t^\dagger\oprod{1_t}{0_t}\right]}
    {\Pr(\mathbf{R})\Pr(J|\Rec)}\,.
  \end{split}
\end{align}
We arrive at the above expressions by noting that $U_t$ acts trivially on all the past field, allowing us to express
\begin{align}
  \begin{split}
    C_{\Rec,\emptyset}&=\nbra{\Rec,\emptyset}\,U(\tight{t+dt},0)
    =\tight{\nbra{0_t}\otimes\nbra{\Rec}}\,U_t\,U(t,0)
    \\
    &=\nbra{0_t}\,U_t\nbra{\Rec}\,U(t,0)=\nbra{0_t}\,U_t\,C_\Rec
  \end{split}
  \\
  \begin{split}
    C_{\Rec,J}&=\nbra{\Rec,J}\,U(\tight{t+dt},0)
    =\nbra{1_t}\tight{\otimes}\nbra{\Rec}\,U_t\,U(t,0)
  \\
  &=\nbra{1_t}\,U_t\nbra{\Rec}\,U(t,0)=\nbra{1_t}\,U_t\,C_\Rec\,,
\end{split}
\end{align}
where we have left the tensor product with the identity on the future field implicit.
$\bra{0_t}$, $\bra{1_t}$, and the field part of $U_t$ all commute with $C_{\mathbf{R}}$, meaning we can pull the outer products and unitaries in to join with the squeezed wave-packet states and perform the substitutions in \cref{eq:sqz-wavepacket-0-sand,eq:sqz-wavepacket-1-sand}, keeping in mind that, since the system part of $U_t$ and the $\overline{t}$ part of $\ket{m_{\gamma,\xi}}$ do not commute with $C_{\mathbf{R}}$, the substitutions we perform straddle $C_{\mathbf{R}}$ and all system operators will therefore appear outside the composite Kraus operators.
First, the no detection update:
\begin{align}
  \label{eq:no-detection-part-trace-update-expanded}
  \begin{split}
    \rhonorm{t+dt|\Rec,\emptyset}{m,n}
    &=
    \Id\system\frac{\tr_{[t,\infty)}\left[C_{\mathbf{R}}
    \left(\rho_0\otimes
    \nket{m_{\gamma,\xi}}
    \nbra{n_{\gamma,\xi}}
    \right)
    C_{\mathbf{R}}^\dagger\right]}{\Pr(\mathbf{R})\Pr(\emptyset|\Rec)}\Id\system \\
    &\phantrel{=}{}
    -\sqrt{dt}\Id\system\frac{\tr_{[t,\infty)}\left[C_{\mathbf{R}}
    \left(\rho_0\otimes
    \big[\nket{1_t}\otimes\nket{\unnorm{\psi}^m_{\sqrt{dt}}}\big]
    \nbra{n_{\gamma,\xi}}
    \right)
    C_{\mathbf{R}}^\dagger\right]}{\Pr(\mathbf{R})\Pr(\emptyset|\Rec)}\Id\system \\
    &\phantrel{=}{}
    -\sqrt{dt}\Id\system\frac{\tr_{[t,\infty)}\left[C_{\mathbf{R}}
    \left(\rho_0\otimes
    \nket{m_{\gamma,\xi}}
    \big[\nbra{\unnorm{\psi}^n_{\sqrt{dt}}}\otimes\nbra{1_t}\big]
    \right)
    C_{\mathbf{R}}^\dagger\right]}{\Pr(\mathbf{R})\Pr(\emptyset|\Rec)}\Id\system \\
    &\phantrel{=}{}
    +dt\,\Id\system\frac{\tr_{[t,\infty)}\left[C_{\mathbf{R}}
    \left(\rho_0\otimes
    \big[\nket{1_t}\otimes\nket{\unnorm{\psi}^m_{\sqrt{dt}}}\big]
    \big[\nbra{\unnorm{\psi}^n_{\sqrt{dt}}}\otimes\nbra{1_t}\big]
    \right)
    C_{\mathbf{R}}^\dagger\right]}{\Pr(\mathbf{R})\Pr(\emptyset|\Rec)}\Id\system \\
    &\phantrel{=}{}
    -dt\,L^\dagger S\frac{\tr_{[t,\infty)}\left[C_{\mathbf{R}}
    \left(\rho_0\otimes
    \big[\nket{0_t}\otimes\nket{\unnorm{\psi}^m_{\sqrt{dt}}}\big]
    \nbra{n_{\gamma,\xi}}
    \right)
    C_{\mathbf{R}}^\dagger\right]}{\Pr(\mathbf{R})\Pr(\emptyset|\Rec)}\Id\system \\
    &\phantrel{=}{}
    -dt\,\Id\system\frac{\tr_{[t,\infty)}\left[C_{\mathbf{R}}
    \left(\rho_0\otimes
    \nket{m_{\gamma,\xi}}
    \big[\nbra{\unnorm{\psi}^n_{\sqrt{dt}}}\otimes\nbra{0_t}\big]
    \right)
    C_{\mathbf{R}}^\dagger\right]}{\Pr(\mathbf{R})\Pr(\emptyset|\Rec)}S^\dagger L \\
    &\phantrel{=}{}
    -dt\,(iH\system+\tfrac{1}{2}L^\dagger L)
    \frac{\tr_{[t,\infty)}\left[C_{\mathbf{R}}
    \left(\rho_0\otimes
    \nket{m_{\gamma,\xi}}
    \nbra{n_{\gamma,\xi}}
    \right)
    C_{\mathbf{R}}^\dagger\right]}{\Pr(\mathbf{R})\Pr(\emptyset|\Rec)}\Id\system \\
    &\phantrel{=}{}
    -dt\,\Id\system\frac{\tr_{[t,\infty)}\left[C_{\mathbf{R}}
    \left(\rho_0\otimes
    \nket{m_{\gamma,\xi}}
    \nbra{n_{\gamma,\xi}}
    \right)
    C_{\mathbf{R}}^\dagger\right]}{\Pr(\mathbf{R})\Pr(\emptyset|\Rec)}
    (-iH\system+\tfrac{1}{2}L^\dagger L)\,.
  \end{split}
\end{align}
Second, the detection update:
\begin{align}
  \label{eq:detection-part-trace-update-expanded}
  \begin{split}
    \rhonorm{t+dt|\Rec,J}{m,n}
    &=
    dt\,S\frac{\tr_{[t,\infty)}\left[C_{\mathbf{R}}\left(\rho_0\otimes
    \big[\nket{0_t}\otimes\nket{\unnorm{\psi}^m_{\sqrt{dt}}}\big]
    \big[\nbra{\unnorm{\psi}^n_{\sqrt{dt}}}\otimes\nbra{0_t}\big]
    \right)C_{\mathbf{R}}^\dagger\right]}{\Pr(\mathbf{R})\Pr(J|\Rec)}S\dg
    \\
    &\phantrel{=}{}
    +dt\,L\frac{\tr_{[t,\infty)}\left[C_{\mathbf{R}}\left(\rho_0\otimes
    \nket{m_{\gamma,\xi}}
    \big[\nbra{\unnorm{\psi}^n_{\sqrt{dt}}}\otimes\nbra{0_t}\big]
    \right)C_{\mathbf{R}}^\dagger\right]}{\Pr(\mathbf{R})\Pr(J|\Rec)}S\dg
    \\
    &\phantrel{=}{}
    +dt\,S\frac{\tr_{[t,\infty)}\left[C_{\mathbf{R}}\left(\rho_0\otimes
    \big[\nket{0_t}\otimes\nket{\unnorm{\psi}^m_{\sqrt{dt}}}\big]
    \nbra{n_{\gamma,\xi}}
    \right)C_{\mathbf{R}}^\dagger\right]}{\Pr(\mathbf{R})\Pr(J|\Rec)}L\dg
    \\
    &\phantrel{=}{}
    +dt\,L\frac{\tr_{[t,\infty)}\left[C_{\mathbf{R}}\left(\rho_0\otimes
    \nket{m_{\gamma,\xi}}
    \nbra{n_{\gamma,\xi}}
    \right)C_{\mathbf{R}}^\dagger\right]}{\Pr(\mathbf{R})\Pr(J|\Rec)}L\dg\,.
  \end{split}
\end{align}
These expressions aren't the simplest, but already we identify a plausible way forward, since some of the terms in the updated state are proportional to bookkeeping operators at the current time.
The terms that still need some attention are those that contain $\nket{\unnorm{\psi}^m_{\sqrt{dt}}}$.

To express the manifest partial traces in terms of the bookkeeping operators \cref{eq:normalized-bookkeeping-cond-state}, we use the temporal decomposition to establish identities up to relevant order in $dt$ and play around with the freedom we have to use the trace to turn outer products into inner products:
\begin{align}\hspace{-2em}
    \tr_{[t,\infty)}\big(\nket{\phi_t}\otimes\noprod{\psi_{[t+dt,\infty)}}
      {\psi_{[t+dt,\infty)}}
    \otimes\nbra{\phi_t}\big)
    &=
    \niprod{\phi_t}{\phi_t}
    \tr_{[t+dt,\infty)}\big(\noprod{\psi_{[t+dt,\infty)}}{\psi_{[t+dt,\infty)}}\big)
    \nonumber
    \\
    &=
    \niprod{\phi^\prime_t}{\phi^\prime_t}
    \tr_{[t+dt,\infty)}\big(\noprod{\psi_{[t+dt,\infty)}}
    {\psi_{[t+dt,\infty)}}\big)
    \nonumber
    \\
    &=
    \tr_{[t,\infty)}\big(\nket{\phi^\prime_t}\otimes\noprod{\psi_{[t+dt,\infty)}}
      {\psi_{[t+dt,\infty)}}
    \otimes\nbra{\phi^\prime_t}\big)
    \\
    \tr_{[t,\infty)}\big(\nket{\phi_t}\otimes\noprod{\psi_{[t+dt,\infty)}}
    {\psi^\prime_{[t,\infty)}}\big)
    &=
    \tr_{[t+dt,\infty)}\big(\nket{\psi_{[t+dt,\infty)}}
    \niprod{\psi^\prime_{[t,\infty)}}{\phi_t}\big)\,.
\end{align}
Performing these manipulations allows us to show
\begin{align}
  \label{eq:part-trace-one-psi-n}
  \begin{split}
    &\sqrt{dt}\,\frac{\tr_{[t,\infty)}\left[C_{\mathbf{R}}
    \left(\rho_0\otimes\big[\nket{1_t}\otimes\nket{\unnorm{\psi}^m_{\sqrt{dt}}}\big]
    \nbra{n_{\gamma,\xi}}\right)
    C_{\mathbf{R}}^\dagger\right]}{\Pr(\mathbf{R})}
    \\
    &=\sqrt{dt}\,\frac{\tr_{[t,\infty)}\left[C_{\mathbf{R}}
    \left(\rho_0\otimes
    \nket{m_{\gamma,\xi}}
    \big[\nbra{\unnorm{\psi}^n_{\sqrt{dt}}}\otimes\nbra{1_t}\big]
    \right)
    C_{\mathbf{R}}^\dagger\right]}{\Pr(\mathbf{R})}
    \\
    &=dt\,\frac{\tr_{[t,\infty)}\left[C_{\mathbf{R}}
    \left(\rho_0\otimes
    \big[\nket{1_t}\otimes\nket{\unnorm{\psi}^m_{\sqrt{dt}}}\big]
    \big[\nbra{\unnorm{\psi}^n_{\sqrt{dt}}}\otimes\nbra{1_t}\big]
    \right)
    C_{\mathbf{R}}^\dagger\right]}{\Pr(\mathbf{R})}
    \\
    &=dt\,\frac{\tr_{[t+dt,\infty)}\left[C_{\mathbf{R}}
    \left(\rho_0\otimes\nket{\unnorm{\psi}^m_{\sqrt{dt}}}
    \nbra{\unnorm{\psi}^n_{\sqrt{dt}}}\right)
    C_{\mathbf{R}}^\dagger\right]}{\Pr(\mathbf{R})}
    \\
    &=dt\,\frac{\tr_{[t,\infty)}\left[C_{\mathbf{R}}
    \left(\rho_0\otimes\big[\nket{0_t}\otimes\nket{\unnorm{\psi}^m_{\sqrt{dt}}}\big]
    \big[\nbra{\unnorm{\psi}^n_{\sqrt{dt}}}\otimes\nbra{0_t}\big]\right)
    C_{\mathbf{R}}^\dagger\right]}{\Pr(\mathbf{R})}
    \\
    &=\abs{\xi_t}^2dt\,\frac{\tr_{[t,\infty)}
    \begin{bmatrix}
      C_{\mathbf{R}}
      \Big(\rho_0\otimes
      \big[\coshr \sqrt{m}\tightnket{(m-1)_{\gamma,\xi}}
      -e^{2i\phi} \sinhr\tightsqrt{m+1}\tightnket{(m+1)_{\gamma,\xi}}\big]
      \\
      \times\big[\coshr \sqrt{n}\tightnbra{(n-1)_{\gamma,\xi}}
      -e^{-2i\phi} \sinhr \tightsqrt{n+1}\tightnbra{(n+1)_{\gamma,\xi}}\big]
      \Big)
      C_{\mathbf{R}}^\dagger
    \end{bmatrix}
    }{\Pr(\mathbf{R})}
    \\
    &=\abs{\xi_t}^2dt\big(
    \coshr^2\sqrt{mn}\rhonorm{t}{m-1,n-1}
    -\coshr \sinhr e^{-2i\phi}\tightsqrt{m(n+1)}\rhonorm{t}{m-1,n+1}
    \\
    &\phantrel{=}\hphantom{\abs{\xi_t}^2dt\big(}{}
    -\coshr \sinhr e^{2i\phi}\tightsqrt{(m+1)n}\rhonorm{t}{m+1,n-1}
    +\sinhr^2\tightsqrt{(m+1)(n+1)}\rhonorm{t}{m+1,n+1}
    \big)
  \end{split}
\end{align}
and
\begin{align}
  \begin{split}
    &dt\,\frac{\tr_{[t,\infty)}\left[C_{\mathbf{R}}
    \left(\rho_0\otimes
    \big[\nket{0_t}\otimes\nket{\unnorm{\psi}^m_{\sqrt{dt}}}\big]
    \nbra{n_{\gamma,\xi}}
    \right)
    C_{\mathbf{R}}^\dagger\right]}{\Pr(\mathbf{R})}
    =\xi_tdt\big(\coshr \sqrt{m}\rhonorm{t}{m-1,n}
    -e^{2i\phi} \sinhr \tightsqrt{m+1}\rhonorm{t}{m+1,n}\big)\,.
    \label{eq:part-trace-zero-psi-n}
  \end{split}
\end{align}

We get the expression for the remaining unaccounted terms by taking the Hermitian conjugate of \cref{eq:part-trace-zero-psi-n} and exchanging $m$ and $n$, giving us
\begin{align}
  \label{eq:part-trace-zero-psi-n-conj}
  \begin{split}
    &dt\,\frac{\tr_{[t,\infty)}\left[C_{\mathbf{R}}
    \left(\rho_0\otimes
    \nket{m_{\gamma,\xi}}
    \big[\nbra{\unnorm{\psi}^n_{\sqrt{dt}}}\otimes\nbra{0_t}\big]
    \right)
    C_{\mathbf{R}}^\dagger\right]}{\Pr(\mathbf{R})}
=\xi_t^*dt\big(\coshr \sqrt{n}\rhonorm{t}{m,n-1}
    -e^{-2i\phi} \sinhr \tightsqrt{n+1}\rhonorm{t}{m,n+1}\big)\,.
  \end{split}
\end{align}

Now we have all the partial trace terms expressed using the bookkeeping operators, so we substitute the fruit of our labors into \cref{eq:no-detection-part-trace-update-expanded,eq:detection-part-trace-update-expanded}, yielding
\begin{align}
  \label{eq:no-detection-unnorm-update}
  \begin{split}
    \Pr(\emptyset|\Rec)\rhonorm{t+dt|\Rec,\emptyset}{m,n}
    &=
    \rhonorm{t}{m,n}-idt[H\system,\rhonorm{t}{m,n}]
    -\tfrac{1}{2}dt[L\dg L,\rhonorm{t}{m,n}]_+
    \\
    &\phantrel{=}{}
    -\abs{\xi_t}^2dt\Big(
    \coshr^2\sqrt{mn}\rhonorm{t}{m-1,n-1}
    -\coshr \sinhr e^{-2i\phi}\tightsqrt{m(n+1)}\rhonorm{t}{m-1,n+1}
    \\
    &\phantrel{=}\hphantom{{}-\abs{\xi_t}^2dt\Big(}{}
    -\coshr \sinhr e^{2i\phi}\tightsqrt{(m+1)n}\rhonorm{t}{m+1,n-1}
    +\sinhr^2\tightsqrt{(m+1)(n+1)}\rhonorm{t}{m+1,n+1}
    \Big)
    \\
    &\phantrel{=}{}
    -\xi_tdt\,L\dg S\big(\coshr \sqrt{m}\rhonorm{t}{m-1,n}
    -e^{2i\phi} \sinhr \tightsqrt{m+1}\rhonorm{t}{m+1,n}\big)
    \\
    &\phantrel{=}{}
    -\xi_t^*dt\big(\coshr \sqrt{n}\rhonorm{t}{m,n-1}
    -e^{-2i\phi} \sinhr \tightsqrt{n+1}\rhonorm{t}{m,n+1}\big)S\dg L
  \end{split}
\end{align}  
and
\begin{align}
  \label{eq:detection-unnorm-update}
  \begin{split}
    \Pr(J|\Rec)\rhonorm{t+dt|\Rec,J}{m,n}
    &=
    \abs{\xi_t}^2dt\,S\big(
    \coshr^2\sqrt{mn}\rhonorm{t}{m-1,n-1}
    -\coshr \sinhr e^{-2i\phi}\tightsqrt{m(n+1)}\rhonorm{t}{m-1,n+1}
    \\
    &\phantrel{=}\hphantom{\abs{\xi_t}^2dt\,S\big(}{}
    -\coshr \sinhr e^{2i\phi}\tightsqrt{(m+1)n}\rhonorm{t}{m+1,n-1}
    +\sinhr^2\tightsqrt{(m+1)(n+1)}\rhonorm{t}{m+1,n+1}
    \big)S\dg
    \\
    &\phantrel{=}{}
    +\xi_tdt\,S\big(\coshr \sqrt{m}\rhonorm{t}{m-1,n}
    -e^{2i\phi}\sinhr \tightsqrt{m+1}\rhonorm{t}{m+1,n}\big)L\dg
    +\xi_t^*dtL\big(\coshr \sqrt{n}\rhonorm{t}{m,n-1}
    -e^{-2i\phi}\sinhr \tightsqrt{n+1}\rhonorm{t}{m,n+1}\big)S\dg
    \\
    &\phantrel{=}{}
    +dt\,L\rhonorm{t}{m,n}L\dg\,.
  \end{split}
\end{align}
We easily solve for the jump probability $\Pr(J|\Rec)$ in the case of a wave packet in squeezed vacuum by taking the trace of \cref{eq:detection-unnorm-update} when $m=n=0$, making use of the cyclic property of the trace and recalling $S\dg S=\Id\system$:
\begin{align}
    \Pr(J|\Rec)&=dt\,\tr\big(L\rhonorm{t}{0,0}L\dg
      -e^{-2i\phi}\xi_t^* \sinhr L\rhonorm{t}{0,1}S\dg
      -e^{2i\phi}S\xi_t \sinhr \rhonorm{t}{1,0}L\dg
      +\abs{\xi_t}^2\sinhr^2\rhonorm{t}{1,1}\big)\,.
\end{align}
Since there are only two possible detection results for a particular time $t$ the probability of no jump is $\Pr(\emptyset|\Rec)=1-\Pr(J|\Rec)$.
We substitute this expression into the no-jump updated state, discarding irrelevant terms by recognizing $\Pr(J|\Rec)\in\BigO(dt)$, and derive the differential state updates
\begin{align}
  \begin{split}
    d\rhonorm{t|\Rec,\emptyset}{m,n}&=\rhonorm{t|\Rec,\emptyset}{m,n}
    -\rhonorm{t}{m,n} \\
    &=dt\Big[\big(\Pr(J|\Rec)/dt\big)\rhonorm{t}{m,n}
    -i[H\system,\rhonorm{t}{m,n}]
    -\tfrac{1}{2}[L\dg L,\rhonorm{t}{m,n}]_+
    \\
    &\phantrel{=}\hphantom{dt\Big[}{}
    -\abs{\xi_t}^2\big(
    \coshr^2\sqrt{mn}\rhonorm{t}{m-1,n-1}
    -\coshr \sinhr e^{-2i\phi}\tightsqrt{m(n+1)}\rhonorm{t}{m-1,n+1}
    \\
    &\phantrel{=}\hphantom{\hphantom{dt\Big[}{}-\abs{\xi_t}^2\big(}{}
    -\coshr \sinhr e^{2i\phi}\tightsqrt{(m+1)n}\rhonorm{t}{m+1,n-1}
    +\sinhr^2\tightsqrt{(m+1)(n+1)}\rhonorm{t}{m+1,n+1}
    \big)
    \\
    &\phantrel{=}\hphantom{dt\Big[}{}
    -\xi_tL\dg S\big(\coshr \sqrt{m}\rhonorm{t}{m-1,n}
    -e^{2i\phi} \sinhr \tightsqrt{m+1}\rhonorm{t}{m+1,n}\big)
    -\xi_t^*\big(\coshr \sqrt{n}\rhonorm{t}{m,n-1}
    -e^{-2i\phi} \sinhr \tightsqrt{n+1}\rhonorm{t}{m,n+1}\big)S\dg L\Big]
  \end{split}
\end{align}
\begin{align}
  \begin{split}
    d\rhonorm{t|\Rec,J}{m,n}&=\rhonorm{t|\Rec,J}{m,n}
    -\rhonorm{t}{m,n} \\
    &=\Big[
    \big(
    L\rhonorm{t}{m,n}
    -\xit \sinhr\,e^{2i\phi}\tightsqrt{m+1}\,
    S\rhonorm{t}{m+1,n}
    +\xit \coshr \sqrt{m}\,
    S\rhonorm{t}{m-1,n}
    \big)L^\dagger
    \\
    &\phantrel{=}\hphantom{\Big[}{}
    -\big(
    L\rhonorm{t}{m,n+1}
    -\xit \sinhr \,e^{2i\phi}\tightsqrt{m+1}\,S\rhonorm{t}{m+1,n+1}
    +\xi_t \coshr \sqrt{m}\,S\rhonorm{t}{m-1,n+1}
    \big)\xi^*_t \sinhr \,e^{-2i\phi}\tightsqrt{n+1}\,S^\dagger
    \\
    &\phantrel{=}\hphantom{\Big[}{}
    +\big(
    L\rhonorm{t}{m,n-1}
    -\xi_t \sinhr \,e^{2i\phi}\tightsqrt{m+1}\,
    S\rhonorm{t}{m+1,n-1}
    +\xi_t \coshr \sqrt{m}\,
    S\rhonorm{t}{m-1,n-1}
    \big)\xi^*_t \coshr \sqrt{n}\,S^\dagger
    \Big]
    \Big/
    \big(\Pr(J|\mathbf{R})\big/dt\big)
    -\rhonorm{t}{m,n}\,.
  \end{split}
\end{align}
\end{widetext}

As explained in the main text we combine these two expressions into a single stochastic differential equation by employing the Poisson process $dN_{t|\mathbf{R}}:\emptyset\mapsto0,J\mapsto1$ to encode the measurement outcome.
\begin{align}
  \begin{split}
    d\rhonorm{t}{m,n}&=(1-dN_{t|\Rec})d\rhonorm{t|\Rec,\emptyset}{m,n}
    +dN_{t|\Rec}d\rhonorm{t|\Rec,J}{m,n}
    \\
    &=d\rhonorm{t|\Rec,\emptyset}{m,n}
    +dN_{t|\Rec}d\rhonorm{t|\Rec,J}{m,n}\,.
  \end{split}
\end{align}
The simplification in the last line follows from the observation that $\mathbb{E}[dN_{t|\Rec}]\in\BigO(dt)$ and $d\rhonorm{t|\Rec,\emptyset}{m,n}\in\BigO(dt)$, meaning their product is of irrelevant order in $dt$.

\subsection{Homodyne}\label{sec:app_homodyne_sme}

To handle homodyne detection of the quadrature $dQ_t^\phi\defined e^{-i\phi}dB_t+e^{i\phi}dB_t^\dagger$ we treat the measurement as a measurement of $\sigma_x$ on the infinitesimal field modes.
This means we need to calculate sandwiches like \cref{eq:sqz-wavepacket-0-sand,eq:sqz-wavepacket-1-sand} where the infinitesimal number states $\ket{0_t}$ and $\ket{1_t}$ (eigenstates of $d\Lambda_t$) are replaced by the infinitesimal quadrature eigenstates $ |\pm_t^\phi\rangle$
The derivation of this eigenvalue equation requires that the action of $\dBdag$ on $\nket{1_t}$ may be discarded, which is justified since this action yields no terms that survive in the state-update equations.

The new sandwiches are simply (tasty) linear combinations of the old sandwiches.
\begin{widetext}
\begin{align}
  \begin{split}
    \noprod{0_t}{\pm_t^\phi}U_t\nket{n_{\gamma,\xi}}&=
    \tfrac{1}{\sqrt{2}}\Big(\!\noprod{0_t}{0_t}U_t\nket{n_{\gamma,\xi}}
    \pm e^{-i\phi}\noprod{0_t}{1_t}U_t\nket{n_{\gamma,\xi}}\!\Big)
    \\
    &=\tfrac{1}{\sqrt{2}}\Big[\Id\system\otimes\big(\nket{n_{\gamma,\xi}}
    -\sqrt{dt}\,\nket{1_t}\otimes\nket{\unnorm{\psi}^n_{\sqrt{dt}}}\big)
    -dt\,L^\dagger S\otimes\nket{0_t}\otimes\nket{\unnorm{\psi}^n_{\sqrt{dt}}}
    -dt(iH\system+\tfrac{1}{2}L^\dagger L)\otimes\nket{n_{\gamma,\xi}}
    \\
    &\phantrel{=}\hphantom{\tfrac{1}{\sqrt{2}}\Big[}{}
    \pm e^{-i\phi}\sqrt{dt}S\otimes\nket{0_t}
    \otimes\nket{\unnorm{\psi}^n_{\sqrt{dt}}}
    \pm e^{-i\phi}\sqrt{dt}L\otimes\nket{n_{\gamma,\xi}}
    \pm e^{-i\phi}dt\,L\otimes\nket{1_t}\otimes\nket{\unnorm{\psi}^n_{\sqrt{dt}}}
    \Big]\,.
  \end{split}
\end{align}
Now write the state updates:
\begin{align}
  \begin{split}
    \rhonorm{t+dt|\Rec,\pm}{m,n}
    &=\frac{\tr_{[t,\infty)}\left[\noprod{0_t}{\pm^\phi_t}U_tC_{\mathbf{R}}
    \left(\rho_0\otimes\oprod{m_{\gamma,\xi}}{n_{\gamma,\xi}}\right)
    C_{\mathbf{R}}^\dagger U_t^\dagger\noprod{\pm^\phi_t}{0_t}\right]}
    {\Pr(\mathbf{R})\Pr(\tight{R_t=\pm})}
    \\
    &=
    \frac{\tr_{[t,\infty)}\left[\noprod{0_t}{0_t}U_tC_{\mathbf{R}}
    \left(\rho_0\otimes\oprod{m_{\gamma,\xi}}{n_{\gamma,\xi}}\right)
    C_{\mathbf{R}}^\dagger U_t^\dagger\noprod{0_t}{0_t}\right]}
    {2\Pr(\mathbf{R})\Pr(\tight{R_t=\pm})}
    \\
    &\phantrel{=}{}
    \pm e^{-i\phi}\frac{\tr_{[t,\infty)}\left[\noprod{0_t}{1_t}U_tC_{\mathbf{R}}
    \left(\rho_0\otimes\oprod{m_{\gamma,\xi}}{n_{\gamma,\xi}}\right)
    C_{\mathbf{R}}^\dagger U_t^\dagger\noprod{0_t}{0_t}\right]}
    {2\Pr(\mathbf{R})\Pr(\tight{R_t=\pm})}
    \\
    &\phantrel{=}{}
    \pm e^{i\phi}\frac{\tr_{[t,\infty)}\left[\noprod{0_t}{0_t}U_tC_{\mathbf{R}}
    \left(\rho_0\otimes\oprod{m_{\gamma,\xi}}{n_{\gamma,\xi}}\right)
    C_{\mathbf{R}}^\dagger U_t^\dagger\noprod{1_t}{0_t}\right]}
    {2\Pr(\mathbf{R})\Pr(\tight{R_t=\pm})}
    \\
    &\phantrel{=}{}
    +\frac{\tr_{[t,\infty)}\left[\noprod{0_t}{1_t}U_tC_{\mathbf{R}}
    \left(\rho_0\otimes\oprod{m_{\gamma,\xi}}{n_{\gamma,\xi}}\right)
    C_{\mathbf{R}}^\dagger U_t^\dagger\noprod{1_t}{0_t}\right]}
    {2\Pr(\mathbf{R})\Pr(\tight{R_t=\pm})}\,.
  \end{split}
\end{align}
Some of these terms are proportional to the jump updates we previously worked out, so we identify those and expand the additional cross terms:
\begin{multline}
    2\Pr(\tight{R_t=\pm})\rhonorm{t+dt|\Rec,\pm}{m,n}
    =
    \Pr(\emptyset|\Rec)\rhonorm{t+dt|\Rec,\emptyset}{m,n}
    +\Pr(J|\Rec)\rhonorm{t+dt|\Rec,J}{m,n}
    \\
    \pm e^{-i\phi}\Bigg [
    \sqrt{dt}\,S
    \frac{\tr_{[t,\infty)}\left[C_{\mathbf{R}}
    \left(\rho_0\otimes
    \big[\nket{0_t}\otimes\nket{\unnorm{\psi}^m_{\sqrt{dt}}}\big]
    \nbra{n_{\gamma,\xi}}
    \right)
    C_{\mathbf{R}}^\dagger\right]}
    {\Pr(\mathbf{R})}
    +\sqrt{dt}\,L
    \frac{\tr_{[t,\infty)}\left[C_{\mathbf{R}}
    \left(\rho_0\otimes
    \nket{m_{\gamma,\xi}}
    \nbra{n_{\gamma,\xi}}
    \right)
    C_{\mathbf{R}}^\dagger\right]}
    {\Pr(\mathbf{R})}
    \\
    -dt\,L
    \frac{\tr_{[t,\infty)}\left[C_{\mathbf{R}}
    \left(\rho_0\otimes
    \nket{m_{\gamma,\xi}}
    \big[\nbra{\unnorm{\psi}^n_{\sqrt{dt}}}\otimes\nbra{1_t}\big]
    \right)
    C_{\mathbf{R}}^\dagger\right]}
    {\Pr(\mathbf{R})}
    +dt\,L
    \frac{\tr_{[t,\infty)}\left[C_{\mathbf{R}}
    \left(\rho_0\otimes
    \big[\nket{1_t}\otimes\nket{\unnorm{\psi}^m_{\sqrt{dt}}}\big]
    \nbra{n_{\gamma,\xi}}
    \right)
    C_{\mathbf{R}}^\dagger\right]}
    {\Pr(\mathbf{R})}
    \Bigg]
    \\
    \pm e^{i\phi}\Bigg[
    \sqrt{dt}
    \frac{\tr_{[t,\infty)}\left[C_{\mathbf{R}}
    \left(\rho_0\otimes
    \nket{m_{\gamma,\xi}}
    \big[\nbra{\unnorm{\psi}^n_{\sqrt{dt}}}\otimes\nbra{0_t}\big]
    \right)
    C_{\mathbf{R}}^\dagger\right]}
    {\Pr(\mathbf{R})}
    S\dg
    +\sqrt{dt}
    \frac{\tr_{[t,\infty)}\left[C_{\mathbf{R}}
    \left(\rho_0\otimes
    \nket{m_{\gamma,\xi}}
    \nbra{n_{\gamma,\xi}}
    \right)
    C_{\mathbf{R}}^\dagger\right]}
    {\Pr(\mathbf{R})}
    L\dg
    \\
    -dt
    \frac{\tr_{[t,\infty)}\left[C_{\mathbf{R}}
    \left(\rho_0\otimes
    \big[\nket{1_t}\otimes\nket{\unnorm{\psi}^m_{\sqrt{dt}}}\big]
    \nbra{n_{\gamma,\xi}}
    \right)
    C_{\mathbf{R}}^\dagger\right]}
    {\Pr(\mathbf{R})}
    L\dg
    +dt
    \frac{\tr_{[t,\infty)}\left[C_{\mathbf{R}}
    \left(\rho_0\otimes
    \nket{m_{\gamma,\xi}}
    \big[\nbra{\unnorm{\psi}^n_{\sqrt{dt}}}\otimes\nbra{1_t}\big]
    \right)
    C_{\mathbf{R}}^\dagger\right]}
    {\Pr(\mathbf{R})}
    L\dg
    \Bigg ]
    \,.
\end{multline}
We rework these remaining partial traces in terms of the bookkeeping operators by once again by making the substitutions worked out in \cref{eq:part-trace-one-psi-n,eq:part-trace-zero-psi-n,eq:part-trace-zero-psi-n-conj}.
Even though we have a $\sqrt{dt}$ prefactor in the terms analogous to \cref{eq:part-trace-zero-psi-n,eq:part-trace-zero-psi-n-conj} instead of a $dt$ prefactor, the trace itself consists of an order unity term with corrections of order $dt$, which we may still discard with a $\sqrt{dt}$ prefactor.
\begin{align}
  \begin{split}
    2\Pr(\tight{R_t=\pm})\rhonorm{t+dt|\Rec,\pm}{m,n}&=
    \Expt{\rhonorm{t+dt|\Rec}{m,n}}
    \\
    &\phantrel{=}{}
    \pm\sqrt{dt}\Big[
    e^{-i\phi}\Big(S\xi_t\big[\coshr \sqrt{m}\rhonorm{t}{m-1,n}
    -e^{2i\phi} \sinhr \tightsqrt{m+1}\rhonorm{t}{m+1,n}\big]
    +L\rhonorm{t}{m,n}
    \Big)
    \\
    &\phantrel{=}\hphantom{{}\pm\sqrt{dt}\Big[}{}
    +e^{i\phi}\Big(\xi_t^*\big[\coshr \sqrt{n}\rhonorm{t}{m,n-1}
    -e^{-2i\phi}\sinhr \tightsqrt{n+1}\rhonorm{t}{m,n+1}\big]S\dg
    +\rhonorm{t}{m,n}L\dg
    \Big)
    \Big]\,.
  \end{split}
\end{align}

\end{widetext}

\bibliography{references}

\begin{thebibliography}{100}%
\makeatletter
\providecommand \@ifxundefined [1]{%
 \@ifx{#1\undefined}
}%
\providecommand \@ifnum [1]{%
 \ifnum #1\expandafter \@firstoftwo
 \else \expandafter \@secondoftwo
 \fi
}%
\providecommand \@ifx [1]{%
 \ifx #1\expandafter \@firstoftwo
 \else \expandafter \@secondoftwo
 \fi
}%
\providecommand \natexlab [1]{#1}%
\providecommand \enquote  [1]{``#1''}%
\providecommand \bibnamefont  [1]{#1}%
\providecommand \bibfnamefont [1]{#1}%
\providecommand \citenamefont [1]{#1}%
\providecommand \href@noop [0]{\@secondoftwo}%
\providecommand \href [0]{\begingroup \@sanitize@url \@href}%
\providecommand \@href[1]{\@@startlink{#1}\@@href}%
\providecommand \@@href[1]{\endgroup#1\@@endlink}%
\providecommand \@sanitize@url [0]{\catcode `\\12\catcode `\$12\catcode
  `\&12\catcode `\#12\catcode `\^12\catcode `\_12\catcode `\%12\relax}%
\providecommand \@@startlink[1]{}%
\providecommand \@@endlink[0]{}%
\providecommand \url  [0]{\begingroup\@sanitize@url \@url }%
\providecommand \@url [1]{\endgroup\@href {#1}{\urlprefix }}%
\providecommand \urlprefix  [0]{URL }%
\providecommand \Eprint [0]{\href }%
\providecommand \doibase [0]{https://doi.org/}%
\providecommand \selectlanguage [0]{\@gobble}%
\providecommand \bibinfo  [0]{\@secondoftwo}%
\providecommand \bibfield  [0]{\@secondoftwo}%
\providecommand \translation [1]{[#1]}%
\providecommand \BibitemOpen [0]{}%
\providecommand \bibitemStop [0]{}%
\providecommand \bibitemNoStop [0]{.\EOS\space}%
\providecommand \EOS [0]{\spacefactor3000\relax}%
\providecommand \BibitemShut  [1]{\csname bibitem#1\endcsname}%
\let\auto@bib@innerbib\@empty
\bibitem [{\citenamefont {Yuen}(1976)}]{yuen_twophoton_1976}%
  \BibitemOpen
  \bibfield  {author} {\bibinfo {author} {\bibfnamefont {H.~P.}\ \bibnamefont
  {Yuen}},\ }\bibfield  {title} {\bibinfo {title} {Two-photon coherent states
  of the radiation field},\ }\href {https://doi.org/10.1103/PhysRevA.13.2226}
  {\bibfield  {journal} {\bibinfo  {journal} {Phys. Rev. A}\ }\textbf {\bibinfo
  {volume} {13}},\ \bibinfo {pages} {2226} (\bibinfo {year}
  {1976})}\BibitemShut {NoStop}%
\bibitem [{\citenamefont {Caves}\ and\ \citenamefont
  {Schumaker}(1985)}]{caves_twophoton_1985}%
  \BibitemOpen
  \bibfield  {author} {\bibinfo {author} {\bibfnamefont {C.~M.}\ \bibnamefont
  {Caves}}\ and\ \bibinfo {author} {\bibfnamefont {B.~L.}\ \bibnamefont
  {Schumaker}},\ }\bibfield  {title} {\bibinfo {title} {New formalism for
  two-photon quantum optics. i. quadrature phases and squeezed states},\ }\href
  {https://doi.org/10.1103/PhysRevA.31.3068} {\bibfield  {journal} {\bibinfo
  {journal} {Phys. Rev. A}\ }\textbf {\bibinfo {volume} {31}},\ \bibinfo
  {pages} {3068} (\bibinfo {year} {1985})}\BibitemShut {NoStop}%
\bibitem [{\citenamefont {Schumaker}\ and\ \citenamefont
  {Caves}(1985)}]{schumaker_twophoton_1985}%
  \BibitemOpen
  \bibfield  {author} {\bibinfo {author} {\bibfnamefont {B.~L.}\ \bibnamefont
  {Schumaker}}\ and\ \bibinfo {author} {\bibfnamefont {C.~M.}\ \bibnamefont
  {Caves}},\ }\bibfield  {title} {\bibinfo {title} {New formalism for
  two-photon quantum optics. ii. mathematical foundation and compact
  notation},\ }\href {https://doi.org/10.1103/PhysRevA.31.3093} {\bibfield
  {journal} {\bibinfo  {journal} {Phys. Rev. A}\ }\textbf {\bibinfo {volume}
  {31}},\ \bibinfo {pages} {3093} (\bibinfo {year} {1985})}\BibitemShut
  {NoStop}%
\bibitem [{\citenamefont {Lawrie}\ \emph {et~al.}(2019)\citenamefont {Lawrie},
  \citenamefont {Lett}, \citenamefont {Marino},\ and\ \citenamefont
  {Pooser}}]{lawrie2019quantum}%
  \BibitemOpen
  \bibfield  {author} {\bibinfo {author} {\bibfnamefont {B.~J.}\ \bibnamefont
  {Lawrie}}, \bibinfo {author} {\bibfnamefont {P.~D.}\ \bibnamefont {Lett}},
  \bibinfo {author} {\bibfnamefont {A.~M.}\ \bibnamefont {Marino}},\ and\
  \bibinfo {author} {\bibfnamefont {R.~C.}\ \bibnamefont {Pooser}},\ }\bibfield
   {title} {\bibinfo {title} {Quantum sensing with squeezed light},\ }\href
  {https://doi.org/10.1021/acsphotonics.9b00250} {\bibfield  {journal}
  {\bibinfo  {journal} {ACS Photonics}\ }\textbf {\bibinfo {volume} {6}},\
  \bibinfo {pages} {1307} (\bibinfo {year} {2019})}\BibitemShut {NoStop}%
\bibitem [{\citenamefont {Hamilton}\ \emph {et~al.}(2017)\citenamefont
  {Hamilton}, \citenamefont {Kruse}, \citenamefont {Sansoni}, \citenamefont
  {Barkhofen}, \citenamefont {Silberhorn},\ and\ \citenamefont
  {Jex}}]{hamilton2017gaussian}%
  \BibitemOpen
  \bibfield  {author} {\bibinfo {author} {\bibfnamefont {C.~S.}\ \bibnamefont
  {Hamilton}}, \bibinfo {author} {\bibfnamefont {R.}~\bibnamefont {Kruse}},
  \bibinfo {author} {\bibfnamefont {L.}~\bibnamefont {Sansoni}}, \bibinfo
  {author} {\bibfnamefont {S.}~\bibnamefont {Barkhofen}}, \bibinfo {author}
  {\bibfnamefont {C.}~\bibnamefont {Silberhorn}},\ and\ \bibinfo {author}
  {\bibfnamefont {I.}~\bibnamefont {Jex}},\ }\bibfield  {title} {\bibinfo
  {title} {Gaussian boson sampling},\ }\href
  {https://doi.org/10.1103/PhysRevLett.119.170501} {\bibfield  {journal}
  {\bibinfo  {journal} {Phys. Rev. Lett.}\ }\textbf {\bibinfo {volume} {119}},\
  \bibinfo {pages} {170501} (\bibinfo {year} {2017})}\BibitemShut {NoStop}%
\bibitem [{\citenamefont {{M. Tse {\em et al.}}}(2019)}]{aligo_squeezing_2019}%
  \BibitemOpen
  \bibfield  {author} {\bibinfo {author} {\bibnamefont {{M. Tse {\em et
  al.}}}},\ }\bibfield  {title} {\bibinfo {title} {Quantum-enhanced advanced
  ligo detectors in the era of gravitational-wave astronomy},\ }\href
  {https://doi.org/10.1103/PhysRevLett.123.231107} {\bibfield  {journal}
  {\bibinfo  {journal} {Phys. Rev. Lett.}\ }\textbf {\bibinfo {volume} {123}},\
  \bibinfo {pages} {231107} (\bibinfo {year} {2019})}\BibitemShut {NoStop}%
\bibitem [{\citenamefont {{Virgo Collaboration}}(2019)}]{virgo_squeezing_2019}%
  \BibitemOpen
  \bibfield  {author} {\bibinfo {author} {\bibnamefont {{Virgo
  Collaboration}}},\ }\bibfield  {title} {\bibinfo {title} {Increasing the
  astrophysical reach of the advanced virgo detector via the application of
  squeezed vacuum states of light},\ }\href
  {https://doi.org/10.1103/PhysRevLett.123.231108} {\bibfield  {journal}
  {\bibinfo  {journal} {Phys. Rev. Lett.}\ }\textbf {\bibinfo {volume} {123}},\
  \bibinfo {pages} {231108} (\bibinfo {year} {2019})}\BibitemShut {NoStop}%
\bibitem [{\citenamefont {Caves}(1980)}]{caves_prl_1980}%
  \BibitemOpen
  \bibfield  {author} {\bibinfo {author} {\bibfnamefont {C.~M.}\ \bibnamefont
  {Caves}},\ }\bibfield  {title} {\bibinfo {title} {Quantum-mechanical
  radiation-pressure fluctuations in an interferometer},\ }\href
  {https://doi.org/10.1103/PhysRevLett.45.75} {\bibfield  {journal} {\bibinfo
  {journal} {Phys. Rev. Lett.}\ }\textbf {\bibinfo {volume} {45}},\ \bibinfo
  {pages} {75} (\bibinfo {year} {1980})}\BibitemShut {NoStop}%
\bibitem [{\citenamefont {Caves}(1981)}]{caves_prd_1981}%
  \BibitemOpen
  \bibfield  {author} {\bibinfo {author} {\bibfnamefont {C.~M.}\ \bibnamefont
  {Caves}},\ }\bibfield  {title} {\bibinfo {title} {Quantum-mechanical noise in
  an interferometer},\ }\href {https://doi.org/10.1103/PhysRevD.23.1693}
  {\bibfield  {journal} {\bibinfo  {journal} {Phys. Rev. D}\ }\textbf {\bibinfo
  {volume} {23}},\ \bibinfo {pages} {1693} (\bibinfo {year}
  {1981})}\BibitemShut {NoStop}%
\bibitem [{\citenamefont {Couteau}(2018)}]{couteau2018spontaneous}%
  \BibitemOpen
  \bibfield  {author} {\bibinfo {author} {\bibfnamefont {C.}~\bibnamefont
  {Couteau}},\ }\bibfield  {title} {\bibinfo {title} {Spontaneous parametric
  down-conversion},\ }\href {https://doi.org/10.1080/00107514.2018.1488463}
  {\bibfield  {journal} {\bibinfo  {journal} {Contemporary Physics}\ }\textbf
  {\bibinfo {volume} {59}},\ \bibinfo {pages} {291} (\bibinfo {year}
  {2018})}\BibitemShut {NoStop}%
\bibitem [{\citenamefont {Knill}\ \emph {et~al.}(2001)\citenamefont {Knill},
  \citenamefont {Laflamme},\ and\ \citenamefont {Milburn}}]{knill2001scheme}%
  \BibitemOpen
  \bibfield  {author} {\bibinfo {author} {\bibfnamefont {E.}~\bibnamefont
  {Knill}}, \bibinfo {author} {\bibfnamefont {R.}~\bibnamefont {Laflamme}},\
  and\ \bibinfo {author} {\bibfnamefont {G.~J.}\ \bibnamefont {Milburn}},\
  }\bibfield  {title} {\bibinfo {title} {A scheme for efficient quantum
  computation with linear optics},\ }\href@noop {} {\bibfield  {journal}
  {\bibinfo  {journal} {nature}\ }\textbf {\bibinfo {volume} {409}},\ \bibinfo
  {pages} {46} (\bibinfo {year} {2001})}\BibitemShut {NoStop}%
\bibitem [{\citenamefont {Nielsen}(2004)}]{nielsen2004optical}%
  \BibitemOpen
  \bibfield  {author} {\bibinfo {author} {\bibfnamefont {M.~A.}\ \bibnamefont
  {Nielsen}},\ }\bibfield  {title} {\bibinfo {title} {Optical quantum
  computation using cluster states},\ }\href
  {https://doi.org/10.1103/PhysRevLett.93.040503} {\bibfield  {journal}
  {\bibinfo  {journal} {Phys. Rev. Lett.}\ }\textbf {\bibinfo {volume} {93}},\
  \bibinfo {pages} {040503} (\bibinfo {year} {2004})}\BibitemShut {NoStop}%
\bibitem [{\citenamefont {Gimeno-Segovia}\ \emph {et~al.}(2015)\citenamefont
  {Gimeno-Segovia}, \citenamefont {Shadbolt}, \citenamefont {Browne},\ and\
  \citenamefont {Rudolph}}]{gimeno2015from}%
  \BibitemOpen
  \bibfield  {author} {\bibinfo {author} {\bibfnamefont {M.}~\bibnamefont
  {Gimeno-Segovia}}, \bibinfo {author} {\bibfnamefont {P.}~\bibnamefont
  {Shadbolt}}, \bibinfo {author} {\bibfnamefont {D.~E.}\ \bibnamefont
  {Browne}},\ and\ \bibinfo {author} {\bibfnamefont {T.}~\bibnamefont
  {Rudolph}},\ }\bibfield  {title} {\bibinfo {title} {From three-photon
  greenberger-horne-zeilinger states to ballistic universal quantum
  computation},\ }\href {https://doi.org/10.1103/PhysRevLett.115.020502}
  {\bibfield  {journal} {\bibinfo  {journal} {Phys. Rev. Lett.}\ }\textbf
  {\bibinfo {volume} {115}},\ \bibinfo {pages} {020502} (\bibinfo {year}
  {2015})}\BibitemShut {NoStop}%
\bibitem [{\citenamefont {Asavanant}\ \emph {et~al.}(2019)\citenamefont
  {Asavanant}, \citenamefont {Shiozawa}, \citenamefont {Yokoyama},
  \citenamefont {Charoensombutamon}, \citenamefont {Emura}, \citenamefont
  {Alexander}, \citenamefont {Takeda}, \citenamefont {Yoshikawa}, \citenamefont
  {Menicucci}, \citenamefont {Yonezawa} \emph
  {et~al.}}]{asavanant2019generation}%
  \BibitemOpen
  \bibfield  {author} {\bibinfo {author} {\bibfnamefont {W.}~\bibnamefont
  {Asavanant}}, \bibinfo {author} {\bibfnamefont {Y.}~\bibnamefont {Shiozawa}},
  \bibinfo {author} {\bibfnamefont {S.}~\bibnamefont {Yokoyama}}, \bibinfo
  {author} {\bibfnamefont {B.}~\bibnamefont {Charoensombutamon}}, \bibinfo
  {author} {\bibfnamefont {H.}~\bibnamefont {Emura}}, \bibinfo {author}
  {\bibfnamefont {R.~N.}\ \bibnamefont {Alexander}}, \bibinfo {author}
  {\bibfnamefont {S.}~\bibnamefont {Takeda}}, \bibinfo {author} {\bibfnamefont
  {J.-i.}\ \bibnamefont {Yoshikawa}}, \bibinfo {author} {\bibfnamefont {N.~C.}\
  \bibnamefont {Menicucci}}, \bibinfo {author} {\bibfnamefont {H.}~\bibnamefont
  {Yonezawa}}, \emph {et~al.},\ }\bibfield  {title} {\bibinfo {title}
  {Generation of time-domain-multiplexed two-dimensional cluster state},\
  }\href@noop {} {\bibfield  {journal} {\bibinfo  {journal} {Science}\ }\textbf
  {\bibinfo {volume} {366}},\ \bibinfo {pages} {373} (\bibinfo {year}
  {2019})}\BibitemShut {NoStop}%
\bibitem [{\citenamefont {Larsen}\ \emph {et~al.}(2019)\citenamefont {Larsen},
  \citenamefont {Guo}, \citenamefont {Breum}, \citenamefont
  {Neergaard-Nielsen},\ and\ \citenamefont
  {Andersen}}]{larsen2019deterministic}%
  \BibitemOpen
  \bibfield  {author} {\bibinfo {author} {\bibfnamefont {M.~V.}\ \bibnamefont
  {Larsen}}, \bibinfo {author} {\bibfnamefont {X.}~\bibnamefont {Guo}},
  \bibinfo {author} {\bibfnamefont {C.~R.}\ \bibnamefont {Breum}}, \bibinfo
  {author} {\bibfnamefont {J.~S.}\ \bibnamefont {Neergaard-Nielsen}},\ and\
  \bibinfo {author} {\bibfnamefont {U.~L.}\ \bibnamefont {Andersen}},\
  }\bibfield  {title} {\bibinfo {title} {Deterministic generation of a
  two-dimensional cluster state},\ }\href@noop {} {\bibfield  {journal}
  {\bibinfo  {journal} {Science}\ }\textbf {\bibinfo {volume} {366}},\ \bibinfo
  {pages} {369} (\bibinfo {year} {2019})}\BibitemShut {NoStop}%
\bibitem [{\citenamefont {Menicucci}\ \emph {et~al.}(2006)\citenamefont
  {Menicucci}, \citenamefont {Van~Loock}, \citenamefont {Gu}, \citenamefont
  {Weedbrook}, \citenamefont {Ralph},\ and\ \citenamefont
  {Nielsen}}]{menicucci2006universal}%
  \BibitemOpen
  \bibfield  {author} {\bibinfo {author} {\bibfnamefont {N.~C.}\ \bibnamefont
  {Menicucci}}, \bibinfo {author} {\bibfnamefont {P.}~\bibnamefont
  {Van~Loock}}, \bibinfo {author} {\bibfnamefont {M.}~\bibnamefont {Gu}},
  \bibinfo {author} {\bibfnamefont {C.}~\bibnamefont {Weedbrook}}, \bibinfo
  {author} {\bibfnamefont {T.~C.}\ \bibnamefont {Ralph}},\ and\ \bibinfo
  {author} {\bibfnamefont {M.~A.}\ \bibnamefont {Nielsen}},\ }\bibfield
  {title} {\bibinfo {title} {Universal quantum computation with
  continuous-variable cluster states},\ }\href@noop {} {\bibfield  {journal}
  {\bibinfo  {journal} {Physical review letters}\ }\textbf {\bibinfo {volume}
  {97}},\ \bibinfo {pages} {110501} (\bibinfo {year} {2006})}\BibitemShut
  {NoStop}%
\bibitem [{\citenamefont {Braunstein}(2005)}]{braunstein2005squeezing}%
  \BibitemOpen
  \bibfield  {author} {\bibinfo {author} {\bibfnamefont {S.~L.}\ \bibnamefont
  {Braunstein}},\ }\bibfield  {title} {\bibinfo {title} {Squeezing as an
  irreducible resource},\ }\href {https://doi.org/10.1103/PhysRevA.71.055801}
  {\bibfield  {journal} {\bibinfo  {journal} {Phys. Rev. A}\ }\textbf {\bibinfo
  {volume} {71}},\ \bibinfo {pages} {055801} (\bibinfo {year}
  {2005})}\BibitemShut {NoStop}%
\bibitem [{\citenamefont {Caves}\ and\ \citenamefont
  {Schumaker}(1986)}]{caves1986broadband}%
  \BibitemOpen
  \bibfield  {author} {\bibinfo {author} {\bibfnamefont {C.}~\bibnamefont
  {Caves}}\ and\ \bibinfo {author} {\bibfnamefont {B.}~\bibnamefont
  {Schumaker}},\ }\bibfield  {title} {\bibinfo {title} {Broadband squeezing},\
  }in\ \href@noop {} {\emph {\bibinfo {booktitle} {Quantum Optics IV}}}\
  (\bibinfo  {publisher} {Springer},\ \bibinfo {year} {1986})\ pp.\ \bibinfo
  {pages} {20--30}\BibitemShut {NoStop}%
\bibitem [{\citenamefont {Gardiner}\ and\ \citenamefont
  {Collett}(1985)}]{InputOutput85}%
  \BibitemOpen
  \bibfield  {author} {\bibinfo {author} {\bibfnamefont {C.~W.}\ \bibnamefont
  {Gardiner}}\ and\ \bibinfo {author} {\bibfnamefont {M.~J.}\ \bibnamefont
  {Collett}},\ }\bibfield  {title} {\bibinfo {title} {Input and output in
  damped quantum systems: Quantum stochastic differential equations and the
  master equation},\ }\href {https://doi.org/10.1103/PhysRevA.31.3761}
  {\bibfield  {journal} {\bibinfo  {journal} {Phys. Rev. A}\ }\textbf {\bibinfo
  {volume} {31}},\ \bibinfo {pages} {3761} (\bibinfo {year}
  {1985})}\BibitemShut {NoStop}%
\bibitem [{\citenamefont {Gardiner}(1986)}]{Gardiner86}%
  \BibitemOpen
  \bibfield  {author} {\bibinfo {author} {\bibfnamefont {C.~W.}\ \bibnamefont
  {Gardiner}},\ }\bibfield  {title} {\bibinfo {title} {Inhibition of atomic
  phase decays by squeezed light: A direct effect of squeezing},\ }\href
  {https://doi.org/10.1103/PhysRevLett.56.1917} {\bibfield  {journal} {\bibinfo
   {journal} {Phys. Rev. Lett.}\ }\textbf {\bibinfo {volume} {56}},\ \bibinfo
  {pages} {1917} (\bibinfo {year} {1986})}\BibitemShut {NoStop}%
\bibitem [{\citenamefont {Wiseman}(1994)}]{WisemanPhD94}%
  \BibitemOpen
  \bibfield  {author} {\bibinfo {author} {\bibfnamefont {H.~M.}\ \bibnamefont
  {Wiseman}},\ }\emph {\bibinfo {title} {Quantum trajectories and feedback}},\
  \href {http://www.ict.griffith.edu.au/wiseman/PhDThesis.ps.z} {Ph.D.
  thesis},\ \bibinfo  {school} {University of Queensland} (\bibinfo {year}
  {1994})\BibitemShut {NoStop}%
\bibitem [{\citenamefont {Honegger}\ and\ \citenamefont
  {Rieckers}(1997)}]{honegger1997}%
  \BibitemOpen
  \bibfield  {author} {\bibinfo {author} {\bibfnamefont {R.}~\bibnamefont
  {Honegger}}\ and\ \bibinfo {author} {\bibfnamefont {A.}~\bibnamefont
  {Rieckers}},\ }\bibfield  {title} {\bibinfo {title} {Squeezed variances of
  smeared boson fields},\ }\href@noop {} {\bibfield  {journal} {\bibinfo
  {journal} {Helvetica Physica Acta}\ }\textbf {\bibinfo {volume} {70}},\
  \bibinfo {pages} {507} (\bibinfo {year} {1997})}\BibitemShut {NoStop}%
\bibitem [{\citenamefont {Hellmich}\ \emph
  {et~al.}(2002{\natexlab{a}})\citenamefont {Hellmich}, \citenamefont
  {Honegger}, \citenamefont {K\"{o}stler}, \citenamefont {K\"{u}mmerer},\ and\
  \citenamefont {Rieckers}}]{HellHone02}%
  \BibitemOpen
  \bibfield  {author} {\bibinfo {author} {\bibfnamefont {J.}~\bibnamefont
  {Hellmich}}, \bibinfo {author} {\bibfnamefont {R.}~\bibnamefont {Honegger}},
  \bibinfo {author} {\bibfnamefont {C.}~\bibnamefont {K\"{o}stler}}, \bibinfo
  {author} {\bibfnamefont {B.}~\bibnamefont {K\"{u}mmerer}},\ and\ \bibinfo
  {author} {\bibfnamefont {A.}~\bibnamefont {Rieckers}},\ }\bibfield  {title}
  {\bibinfo {title} {Couplings to classical and non-classical squeezed white
  noise as stationary markov processes},\ }\href
  {https://doi.org/10.2977/prims/1145476415} {\bibfield  {journal} {\bibinfo
  {journal} {Publ. Res. Inst. Math. Sci.}\ }\textbf {\bibinfo {volume} {38}},\
  \bibinfo {pages} {1–31} (\bibinfo {year} {2002}{\natexlab{a}})}\BibitemShut
  {NoStop}%
\bibitem [{\citenamefont {Gough}(2003)}]{Gough:2003aa}%
  \BibitemOpen
  \bibfield  {author} {\bibinfo {author} {\bibfnamefont {J.}~\bibnamefont
  {Gough}},\ }\bibfield  {title} {\bibinfo {title} {Quantum white noises and
  the master equation for gaussian reference states},\ }\href
  {https://arxiv.org/abs/quant-ph/0309103} {\bibfield  {journal} {\bibinfo
  {journal} {Russian Journal of Mathematical Physics}\ }\textbf {\bibinfo
  {volume} {10}},\ \bibinfo {pages} {142} (\bibinfo {year} {2003})}\BibitemShut
  {NoStop}%
\bibitem [{\citenamefont {Barchielli}\ \emph {et~al.}(2009)\citenamefont
  {Barchielli}, \citenamefont {Gregoratti},\ and\ \citenamefont
  {Licciardo}}]{Barchielli_2009}%
  \BibitemOpen
  \bibfield  {author} {\bibinfo {author} {\bibfnamefont {A.}~\bibnamefont
  {Barchielli}}, \bibinfo {author} {\bibfnamefont {M.}~\bibnamefont
  {Gregoratti}},\ and\ \bibinfo {author} {\bibfnamefont {M.}~\bibnamefont
  {Licciardo}},\ }\bibfield  {title} {\bibinfo {title} {Feedback control of the
  fluorescence light squeezing},\ }\href
  {https://doi.org/10.1209/0295-5075/85/14006} {\bibfield  {journal} {\bibinfo
  {journal} {{EPL} (Europhysics Letters)}\ }\textbf {\bibinfo {volume} {85}},\
  \bibinfo {pages} {14006} (\bibinfo {year} {2009})}\BibitemShut {NoStop}%
\bibitem [{\citenamefont
  {Parthasarathy}(2012)}]{parthasarathy2012introduction}%
  \BibitemOpen
  \bibfield  {author} {\bibinfo {author} {\bibfnamefont {K.~R.}\ \bibnamefont
  {Parthasarathy}},\ }\href@noop {} {\emph {\bibinfo {title} {An introduction
  to quantum stochastic calculus}}},\ Vol.~\bibinfo {volume} {85}\ (\bibinfo
  {publisher} {Birkh{\"a}user},\ \bibinfo {year} {2012})\BibitemShut {NoStop}%
\bibitem [{\citenamefont {Accardi}\ \emph {et~al.}(2013)\citenamefont
  {Accardi}, \citenamefont {Lu},\ and\ \citenamefont
  {Volovich}}]{accardi2013quantum}%
  \BibitemOpen
  \bibfield  {author} {\bibinfo {author} {\bibfnamefont {L.}~\bibnamefont
  {Accardi}}, \bibinfo {author} {\bibfnamefont {Y.~G.}\ \bibnamefont {Lu}},\
  and\ \bibinfo {author} {\bibfnamefont {I.}~\bibnamefont {Volovich}},\
  }\href@noop {} {\emph {\bibinfo {title} {Quantum theory and its stochastic
  limit}}}\ (\bibinfo  {publisher} {Springer Science \& Business Media},\
  \bibinfo {year} {2013})\BibitemShut {NoStop}%
\bibitem [{\citenamefont {D{\k a}browska}\ and\ \citenamefont
  {Gough}(2016{\natexlab{a}})}]{DabGou2016a}%
  \BibitemOpen
  \bibfield  {author} {\bibinfo {author} {\bibfnamefont {A.}~\bibnamefont {D{\k
  a}browska}}\ and\ \bibinfo {author} {\bibfnamefont {J.}~\bibnamefont
  {Gough}},\ }\bibfield  {title} {\bibinfo {title} {Belavkin filtering with
  squeezed light sources},\ }\href {https://doi.org/10.1134/S1061920816020035}
  {\bibfield  {journal} {\bibinfo  {journal} {Russian Journal of Mathematical
  Physics}\ }\textbf {\bibinfo {volume} {23}},\ \bibinfo {pages} {172}
  (\bibinfo {year} {2016}{\natexlab{a}})}\BibitemShut {NoStop}%
\bibitem [{\citenamefont {D{\k a}browska}\ and\ \citenamefont
  {Gough}(2016{\natexlab{b}})}]{DabrGough2016b}%
  \BibitemOpen
  \bibfield  {author} {\bibinfo {author} {\bibfnamefont {A.}~\bibnamefont {D{\k
  a}browska}}\ and\ \bibinfo {author} {\bibfnamefont {J.}~\bibnamefont
  {Gough}},\ }\bibfield  {title} {\bibinfo {title} {Quantum trajectories for
  squeezed input processes: Explicit solutions},\ }\href
  {https://doi.org/10.1142/S1230161216500049} {\bibfield  {journal} {\bibinfo
  {journal} {Open Systems \& Information Dynamics}\ }\textbf {\bibinfo {volume}
  {23}},\ \bibinfo {pages} {1650004} (\bibinfo {year}
  {2016}{\natexlab{b}})}\BibitemShut {NoStop}%
\bibitem [{\citenamefont {Wiseman}\ and\ \citenamefont
  {Milburn}(2010)}]{wiseman_quantum_2010}%
  \BibitemOpen
  \bibfield  {author} {\bibinfo {author} {\bibfnamefont {H.~M.}\ \bibnamefont
  {Wiseman}}\ and\ \bibinfo {author} {\bibfnamefont {G.~J.}\ \bibnamefont
  {Milburn}},\ }\href@noop {} {\emph {\bibinfo {title} {Quantum measurement and
  control}}}\ (\bibinfo  {publisher} {Cambridge University Press},\ \bibinfo
  {year} {2010})\ \bibinfo {note} {google-Books-ID: ZNjvHaH8qA4C}\BibitemShut
  {NoStop}%
\bibitem [{\citenamefont {Gardiner}\ \emph {et~al.}(1987)\citenamefont
  {Gardiner}, \citenamefont {Parkins},\ and\ \citenamefont
  {Collett}}]{Gard_io_III_1987}%
  \BibitemOpen
  \bibfield  {author} {\bibinfo {author} {\bibfnamefont {C.~W.}\ \bibnamefont
  {Gardiner}}, \bibinfo {author} {\bibfnamefont {A.~S.}\ \bibnamefont
  {Parkins}},\ and\ \bibinfo {author} {\bibfnamefont {M.~J.}\ \bibnamefont
  {Collett}},\ }\bibfield  {title} {\bibinfo {title} {Input and output in
  damped quantum systems. ii. methods in non-white-noise situations and
  application to inhibition of atomic phase decays},\ }\bibfield  {booktitle}
  {\emph {\bibinfo {booktitle} {Journal of the Optical Society of America B}},\
  }\href {https://doi.org/10.1364/JOSAB.4.001683} {\bibfield  {journal}
  {\bibinfo  {journal} {Journal of the Optical Society of America B}\ }\textbf
  {\bibinfo {volume} {4}},\ \bibinfo {pages} {1683} (\bibinfo {year}
  {1987})}\BibitemShut {NoStop}%
\bibitem [{\citenamefont {Toyli}\ \emph {et~al.}(2016)\citenamefont {Toyli},
  \citenamefont {Eddins}, \citenamefont {Boutin}, \citenamefont {Puri},
  \citenamefont {Hover}, \citenamefont {Bolkhovsky}, \citenamefont {Oliver},
  \citenamefont {Blais},\ and\ \citenamefont {Siddiqi}}]{toyli_resonance_2016}%
  \BibitemOpen
  \bibfield  {author} {\bibinfo {author} {\bibfnamefont {D.~M.}\ \bibnamefont
  {Toyli}}, \bibinfo {author} {\bibfnamefont {A.~W.}\ \bibnamefont {Eddins}},
  \bibinfo {author} {\bibfnamefont {S.}~\bibnamefont {Boutin}}, \bibinfo
  {author} {\bibfnamefont {S.}~\bibnamefont {Puri}}, \bibinfo {author}
  {\bibfnamefont {D.}~\bibnamefont {Hover}}, \bibinfo {author} {\bibfnamefont
  {V.}~\bibnamefont {Bolkhovsky}}, \bibinfo {author} {\bibfnamefont {W.~D.}\
  \bibnamefont {Oliver}}, \bibinfo {author} {\bibfnamefont {A.}~\bibnamefont
  {Blais}},\ and\ \bibinfo {author} {\bibfnamefont {I.}~\bibnamefont
  {Siddiqi}},\ }\bibfield  {title} {\bibinfo {title} {Resonance fluorescence
  from an artificial atom in squeezed vacuum},\ }\bibfield  {journal} {\bibinfo
   {journal} {Physical Review X}\ }\textbf {\bibinfo {volume} {6}},\ \href
  {https://doi.org/10.1103/PhysRevX.6.031004} {10.1103/PhysRevX.6.031004}
  (\bibinfo {year} {2016})\BibitemShut {NoStop}%
\bibitem [{\citenamefont {Andersen}\ \emph {et~al.}(2016)\citenamefont
  {Andersen}, \citenamefont {Gehring}, \citenamefont {Marquardt},\ and\
  \citenamefont {Leuchs}}]{30years_of_squeezing}%
  \BibitemOpen
  \bibfield  {author} {\bibinfo {author} {\bibfnamefont {U.~L.}\ \bibnamefont
  {Andersen}}, \bibinfo {author} {\bibfnamefont {T.}~\bibnamefont {Gehring}},
  \bibinfo {author} {\bibfnamefont {C.}~\bibnamefont {Marquardt}},\ and\
  \bibinfo {author} {\bibfnamefont {G.}~\bibnamefont {Leuchs}},\ }\bibfield
  {title} {\bibinfo {title} {30 years of squeezed light generation},\ }\href
  {https://doi.org/10.1088/0031-8949/91/5/053001} {\bibfield  {journal}
  {\bibinfo  {journal} {Physica Scripta}\ }\textbf {\bibinfo {volume} {91}},\
  \bibinfo {pages} {053001} (\bibinfo {year} {2016})}\BibitemShut {NoStop}%
\bibitem [{\citenamefont {Vahlbruch}\ \emph {et~al.}(2016)\citenamefont
  {Vahlbruch}, \citenamefont {Mehmet}, \citenamefont {Danzmann},\ and\
  \citenamefont {Schnabel}}]{vahlbruch2016detection}%
  \BibitemOpen
  \bibfield  {author} {\bibinfo {author} {\bibfnamefont {H.}~\bibnamefont
  {Vahlbruch}}, \bibinfo {author} {\bibfnamefont {M.}~\bibnamefont {Mehmet}},
  \bibinfo {author} {\bibfnamefont {K.}~\bibnamefont {Danzmann}},\ and\
  \bibinfo {author} {\bibfnamefont {R.}~\bibnamefont {Schnabel}},\ }\bibfield
  {title} {\bibinfo {title} {Detection of 15 db squeezed states of light and
  their application for the absolute calibration of photoelectric quantum
  efficiency},\ }\href {https://doi.org/10.1103/PhysRevLett.117.110801}
  {\bibfield  {journal} {\bibinfo  {journal} {Phys. Rev. Lett.}\ }\textbf
  {\bibinfo {volume} {117}},\ \bibinfo {pages} {110801} (\bibinfo {year}
  {2016})}\BibitemShut {NoStop}%
\bibitem [{\citenamefont {Wakui}\ \emph {et~al.}(2014)\citenamefont {Wakui},
  \citenamefont {Eto}, \citenamefont {Benichi}, \citenamefont {Izumi},
  \citenamefont {Yanagida}, \citenamefont {Ema}, \citenamefont {Numata},
  \citenamefont {Fukuda}, \citenamefont {Takeoka},\ and\ \citenamefont
  {Sasaki}}]{wakui2014ultrabroadband}%
  \BibitemOpen
  \bibfield  {author} {\bibinfo {author} {\bibfnamefont {K.}~\bibnamefont
  {Wakui}}, \bibinfo {author} {\bibfnamefont {Y.}~\bibnamefont {Eto}}, \bibinfo
  {author} {\bibfnamefont {H.}~\bibnamefont {Benichi}}, \bibinfo {author}
  {\bibfnamefont {S.}~\bibnamefont {Izumi}}, \bibinfo {author} {\bibfnamefont
  {T.}~\bibnamefont {Yanagida}}, \bibinfo {author} {\bibfnamefont
  {K.}~\bibnamefont {Ema}}, \bibinfo {author} {\bibfnamefont {T.}~\bibnamefont
  {Numata}}, \bibinfo {author} {\bibfnamefont {D.}~\bibnamefont {Fukuda}},
  \bibinfo {author} {\bibfnamefont {M.}~\bibnamefont {Takeoka}},\ and\ \bibinfo
  {author} {\bibfnamefont {M.}~\bibnamefont {Sasaki}},\ }\bibfield  {title}
  {\bibinfo {title} {Ultrabroadband direct detection of nonclassical photon
  statistics at telecom wavelength},\ }\href
  {https://doi.org/10.1038/srep04535} {\bibfield  {journal} {\bibinfo
  {journal} {Scientific reports}\ }\textbf {\bibinfo {volume} {4}},\ \bibinfo
  {pages} {1} (\bibinfo {year} {2014})}\BibitemShut {NoStop}%
\bibitem [{\citenamefont {Kashiwazaki}\ \emph {et~al.}(2020)\citenamefont
  {Kashiwazaki}, \citenamefont {Takanashi}, \citenamefont {Yamashima},
  \citenamefont {Kazama}, \citenamefont {Enbutsu}, \citenamefont {Kasahara},
  \citenamefont {Umeki},\ and\ \citenamefont
  {Furusawa}}]{kashiwazaki2020continuous}%
  \BibitemOpen
  \bibfield  {author} {\bibinfo {author} {\bibfnamefont {T.}~\bibnamefont
  {Kashiwazaki}}, \bibinfo {author} {\bibfnamefont {N.}~\bibnamefont
  {Takanashi}}, \bibinfo {author} {\bibfnamefont {T.}~\bibnamefont
  {Yamashima}}, \bibinfo {author} {\bibfnamefont {T.}~\bibnamefont {Kazama}},
  \bibinfo {author} {\bibfnamefont {K.}~\bibnamefont {Enbutsu}}, \bibinfo
  {author} {\bibfnamefont {R.}~\bibnamefont {Kasahara}}, \bibinfo {author}
  {\bibfnamefont {T.}~\bibnamefont {Umeki}},\ and\ \bibinfo {author}
  {\bibfnamefont {A.}~\bibnamefont {Furusawa}},\ }\bibfield  {title} {\bibinfo
  {title} {Continuous-wave 6-{dB}-squeezed light with 2.5-{THz}-bandwidth from
  single-mode {PPLN} waveguide},\ }\href {https://doi.org/10.1063/1.5142437}
  {\bibfield  {journal} {\bibinfo  {journal} {APL Photonics}\ }\textbf
  {\bibinfo {volume} {5}},\ \bibinfo {pages} {036104} (\bibinfo {year}
  {2020})}\BibitemShut {NoStop}%
\bibitem [{\citenamefont {Vaidya}\ \emph {et~al.}(2020)\citenamefont {Vaidya},
  \citenamefont {Morrison}, \citenamefont {Helt}, \citenamefont {Shahrokshahi},
  \citenamefont {Mahler}, \citenamefont {Collins}, \citenamefont {Tan},
  \citenamefont {Lavoie}, \citenamefont {Repingon}, \citenamefont {Menotti}
  \emph {et~al.}}]{vaidya2020broadband}%
  \BibitemOpen
  \bibfield  {author} {\bibinfo {author} {\bibfnamefont {V.~D.}\ \bibnamefont
  {Vaidya}}, \bibinfo {author} {\bibfnamefont {B.}~\bibnamefont {Morrison}},
  \bibinfo {author} {\bibfnamefont {L.}~\bibnamefont {Helt}}, \bibinfo {author}
  {\bibfnamefont {R.}~\bibnamefont {Shahrokshahi}}, \bibinfo {author}
  {\bibfnamefont {D.}~\bibnamefont {Mahler}}, \bibinfo {author} {\bibfnamefont
  {M.}~\bibnamefont {Collins}}, \bibinfo {author} {\bibfnamefont
  {K.}~\bibnamefont {Tan}}, \bibinfo {author} {\bibfnamefont {J.}~\bibnamefont
  {Lavoie}}, \bibinfo {author} {\bibfnamefont {A.}~\bibnamefont {Repingon}},
  \bibinfo {author} {\bibfnamefont {M.}~\bibnamefont {Menotti}}, \emph
  {et~al.},\ }\bibfield  {title} {\bibinfo {title} {Broadband
  quadrature-squeezed vacuum and nonclassical photon number correlations from a
  nanophotonic device},\ }\href {https://doi.org/10.1126/sciadv.aba9186}
  {\bibfield  {journal} {\bibinfo  {journal} {Science advances}\ }\textbf
  {\bibinfo {volume} {6}},\ \bibinfo {pages} {eaba9186} (\bibinfo {year}
  {2020})}\BibitemShut {NoStop}%
\bibitem [{\citenamefont {Chen}\ \emph {et~al.}(2021)\citenamefont {Chen},
  \citenamefont {Briggs}, \citenamefont {Hou},\ and\ \citenamefont
  {Fan}}]{pao-kang2021}%
  \BibitemOpen
  \bibfield  {author} {\bibinfo {author} {\bibfnamefont {P.-K.}\ \bibnamefont
  {Chen}}, \bibinfo {author} {\bibfnamefont {I.}~\bibnamefont {Briggs}},
  \bibinfo {author} {\bibfnamefont {S.}~\bibnamefont {Hou}},\ and\ \bibinfo
  {author} {\bibfnamefont {L.}~\bibnamefont {Fan}},\ }\bibfield  {title}
  {\bibinfo {title} {Ultra-broadband quadrature squeezing with thin-film
  lithium niobate nanophotonics},\ }\href {http://arxiv.org/abs/2107.02250}
  {\bibfield  {journal} {\bibinfo  {journal} {arXiv:2107.02250 [quant-ph]}\ }
  (\bibinfo {year} {2021})},\ \bibinfo {note} {arXiv:2107.02250}\BibitemShut
  {NoStop}%
\bibitem [{\citenamefont {Carmichael}\ and\ \citenamefont
  {Walls}(1973)}]{Carmichael_1973}%
  \BibitemOpen
  \bibfield  {author} {\bibinfo {author} {\bibfnamefont {H.~J.}\ \bibnamefont
  {Carmichael}}\ and\ \bibinfo {author} {\bibfnamefont {D.~F.}\ \bibnamefont
  {Walls}},\ }\bibfield  {title} {\bibinfo {title} {Master equation for
  strongly interacting systems},\ }\href
  {https://doi.org/10.1088/0305-4470/6/10/014} {\bibfield  {journal} {\bibinfo
  {journal} {Journal of Physics A: Mathematical, Nuclear and General}\ }\textbf
  {\bibinfo {volume} {6}},\ \bibinfo {pages} {1552} (\bibinfo {year}
  {1973})}\BibitemShut {NoStop}%
\bibitem [{\citenamefont {Yeoman}\ and\ \citenamefont
  {Barnett}(1996)}]{yeoman_influence_1996}%
  \BibitemOpen
  \bibfield  {author} {\bibinfo {author} {\bibfnamefont {G.}~\bibnamefont
  {Yeoman}}\ and\ \bibinfo {author} {\bibfnamefont {S.~M.}\ \bibnamefont
  {Barnett}},\ }\bibfield  {title} {\bibinfo {title} {Influence of squeezing
  bandwidths on resonance fluorescence},\ }\href
  {https://doi.org/10.1080/09500349608232870} {\bibfield  {journal} {\bibinfo
  {journal} {Journal of Modern Optics}\ }\textbf {\bibinfo {volume} {43}},\
  \bibinfo {pages} {2037} (\bibinfo {year} {1996})}\BibitemShut {NoStop}%
\bibitem [{\citenamefont {Ficek}\ \emph {et~al.}(1997)\citenamefont {Ficek},
  \citenamefont {Dalton},\ and\ \citenamefont {Wahiddin}}]{Ficek97}%
  \BibitemOpen
  \bibfield  {author} {\bibinfo {author} {\bibfnamefont {Z.}~\bibnamefont
  {Ficek}}, \bibinfo {author} {\bibfnamefont {B.~J.}\ \bibnamefont {Dalton}},\
  and\ \bibinfo {author} {\bibfnamefont {M.~R.~B.}\ \bibnamefont {Wahiddin}},\
  }\bibfield  {title} {\bibinfo {title} {Spectral linewidth narrowing by a
  narrow bandwidth squeezed vacuum in a cavity},\ }\href
  {https://doi.org/10.1080/09500349708230713} {\bibfield  {journal} {\bibinfo
  {journal} {Journal of Modern Optics}\ }\textbf {\bibinfo {volume} {44}},\
  \bibinfo {pages} {1005} (\bibinfo {year} {1997})}\BibitemShut {NoStop}%
\bibitem [{\citenamefont {Tanas}(1999)}]{Tanas1999}%
  \BibitemOpen
  \bibfield  {author} {\bibinfo {author} {\bibfnamefont {R.}~\bibnamefont
  {Tanas}},\ }\bibfield  {title} {\bibinfo {title} {Atoms in a narrow-bandwidth
  squeezed vacuum},\ }\href
  {http://www.physics.sk/aps/pubs/1999/aps-1999-49-4-595.ps} {\bibfield
  {journal} {\bibinfo  {journal} {Acta Physica Slovaca}\ }\textbf {\bibinfo
  {volume} {49}},\ \bibinfo {pages} {595} (\bibinfo {year} {1999})}\BibitemShut
  {NoStop}%
\bibitem [{\citenamefont {Kowalewska-kudlaszyk}\ and\ \citenamefont
  {Tana{\'s}}(2001)}]{Kowalewska-kudlaszyk:2001aa}%
  \BibitemOpen
  \bibfield  {author} {\bibinfo {author} {\bibfnamefont {A.}~\bibnamefont
  {Kowalewska-kudlaszyk}}\ and\ \bibinfo {author} {\bibfnamefont
  {R.}~\bibnamefont {Tana{\'s}}},\ }\bibfield  {title} {\bibinfo {title}
  {Generalized master equation for a two-level atom in a strong field and
  tailored reservoirs},\ }\bibfield  {booktitle} {\emph {\bibinfo {booktitle}
  {Journal of Modern Optics}},\ }\href
  {https://doi.org/10.1080/09500340108232462} {\bibfield  {journal} {\bibinfo
  {journal} {Journal of Modern Optics}\ }\textbf {\bibinfo {volume} {48}},\
  \bibinfo {pages} {347} (\bibinfo {year} {2001})}\BibitemShut {NoStop}%
\bibitem [{\citenamefont {Parkins}(1990)}]{parkins_rabi_1990}%
  \BibitemOpen
  \bibfield  {author} {\bibinfo {author} {\bibfnamefont {A.~S.}\ \bibnamefont
  {Parkins}},\ }\bibfield  {title} {\bibinfo {title} {Rabi sideband narrowing
  via strongly driven resonance fluorescence in a narrow-bandwidth squeezed
  vacuum},\ }\href {https://doi.org/10.1103/PhysRevA.42.4352} {\bibfield
  {journal} {\bibinfo  {journal} {Physical Review A}\ }\textbf {\bibinfo
  {volume} {42}},\ \bibinfo {pages} {4352} (\bibinfo {year}
  {1990})}\BibitemShut {NoStop}%
\bibitem [{\citenamefont {Cirac}(1992)}]{Cirac1992}%
  \BibitemOpen
  \bibfield  {author} {\bibinfo {author} {\bibfnamefont {J.~I.}\ \bibnamefont
  {Cirac}},\ }\bibfield  {title} {\bibinfo {title} {Interaction of a two-level
  atom with a cavity mode in the bad-cavity limit},\ }\href
  {https://doi.org/10.1103/PhysRevA.46.4354} {\bibfield  {journal} {\bibinfo
  {journal} {Phys. Rev. A}\ }\textbf {\bibinfo {volume} {46}},\ \bibinfo
  {pages} {4354} (\bibinfo {year} {1992})}\BibitemShut {NoStop}%
\bibitem [{\citenamefont {Gardiner}\ and\ \citenamefont
  {Parkins}(1994)}]{GardPar94}%
  \BibitemOpen
  \bibfield  {author} {\bibinfo {author} {\bibfnamefont {C.~W.}\ \bibnamefont
  {Gardiner}}\ and\ \bibinfo {author} {\bibfnamefont {A.~S.}\ \bibnamefont
  {Parkins}},\ }\bibfield  {title} {\bibinfo {title} {Driving atoms with light
  of arbitrary statistics},\ }\href {https://doi.org/10.1103/PhysRevA.50.1792}
  {\bibfield  {journal} {\bibinfo  {journal} {Phys. Rev. A}\ }\textbf {\bibinfo
  {volume} {50}},\ \bibinfo {pages} {1792} (\bibinfo {year}
  {1994})}\BibitemShut {NoStop}%
\bibitem [{\citenamefont {Kochan}\ and\ \citenamefont
  {Carmichael}(1994)}]{KochCarm94}%
  \BibitemOpen
  \bibfield  {author} {\bibinfo {author} {\bibfnamefont {P.}~\bibnamefont
  {Kochan}}\ and\ \bibinfo {author} {\bibfnamefont {H.~J.}\ \bibnamefont
  {Carmichael}},\ }\bibfield  {title} {\bibinfo {title} {Photon-statistics
  dependence of single-atom absorption},\ }\href
  {https://doi.org/10.1103/PhysRevA.50.1700} {\bibfield  {journal} {\bibinfo
  {journal} {Phys. Rev. A}\ }\textbf {\bibinfo {volume} {50}},\ \bibinfo
  {pages} {1700} (\bibinfo {year} {1994})}\BibitemShut {NoStop}%
\bibitem [{\citenamefont {Smyth}\ and\ \citenamefont
  {Swain}(1999)}]{SmythSwain1999}%
  \BibitemOpen
  \bibfield  {author} {\bibinfo {author} {\bibfnamefont {W.~S.}\ \bibnamefont
  {Smyth}}\ and\ \bibinfo {author} {\bibfnamefont {S.}~\bibnamefont {Swain}},\
  }\bibfield  {title} {\bibinfo {title} {Complete quenching of fluorescence
  from a two-level atom driven by a weak, narrow-band, nonclassical light
  field},\ }\href {https://doi.org/10.1103/PhysRevA.59.R2579} {\bibfield
  {journal} {\bibinfo  {journal} {Phys. Rev. A}\ }\textbf {\bibinfo {volume}
  {59}},\ \bibinfo {pages} {R2579} (\bibinfo {year} {1999})}\BibitemShut
  {NoStop}%
\bibitem [{\citenamefont {Messikh}\ \emph {et~al.}(2000)\citenamefont
  {Messikh}, \citenamefont {Tana\ifmmode~\acute{s}\else \'{s}\fi{}},\ and\
  \citenamefont {Ficek}}]{Messikh2000}%
  \BibitemOpen
  \bibfield  {author} {\bibinfo {author} {\bibfnamefont {A.}~\bibnamefont
  {Messikh}}, \bibinfo {author} {\bibfnamefont {R.}~\bibnamefont
  {Tana\ifmmode~\acute{s}\else \'{s}\fi{}}},\ and\ \bibinfo {author}
  {\bibfnamefont {Z.}~\bibnamefont {Ficek}},\ }\bibfield  {title} {\bibinfo
  {title} {Response of a two-level atom to a narrow-bandwidth squeezed-vacuum
  excitation},\ }\href {https://doi.org/10.1103/PhysRevA.61.033811} {\bibfield
  {journal} {\bibinfo  {journal} {Phys. Rev. A}\ }\textbf {\bibinfo {volume}
  {61}},\ \bibinfo {pages} {033811} (\bibinfo {year} {2000})}\BibitemShut
  {NoStop}%
\bibitem [{\citenamefont {Ritsch}\ and\ \citenamefont
  {Zoller}(1988{\natexlab{a}})}]{Ritsch:1988aa}%
  \BibitemOpen
  \bibfield  {author} {\bibinfo {author} {\bibfnamefont {H.}~\bibnamefont
  {Ritsch}}\ and\ \bibinfo {author} {\bibfnamefont {P.}~\bibnamefont
  {Zoller}},\ }\bibfield  {title} {\bibinfo {title} {Atomic transitions in
  finite-bandwidth squeezed light},\ }\href
  {https://doi.org/10.1103/PhysRevLett.61.1097} {\bibfield  {journal} {\bibinfo
   {journal} {Phys. Rev. Lett.}\ }\textbf {\bibinfo {volume} {61}},\ \bibinfo
  {pages} {1097} (\bibinfo {year} {1988}{\natexlab{a}})}\BibitemShut {NoStop}%
\bibitem [{\citenamefont {Ritsch}\ and\ \citenamefont
  {Zoller}(1988{\natexlab{b}})}]{Ritsch:1988ab}%
  \BibitemOpen
  \bibfield  {author} {\bibinfo {author} {\bibfnamefont {H.}~\bibnamefont
  {Ritsch}}\ and\ \bibinfo {author} {\bibfnamefont {P.}~\bibnamefont
  {Zoller}},\ }\bibfield  {title} {\bibinfo {title} {Systems driven by colored
  squeezed noise: The atomic absorption spectrum},\ }\href
  {https://doi.org/10.1103/PhysRevA.38.4657} {\bibfield  {journal} {\bibinfo
  {journal} {Phys. Rev. A}\ }\textbf {\bibinfo {volume} {38}},\ \bibinfo
  {pages} {4657} (\bibinfo {year} {1988}{\natexlab{b}})}\BibitemShut {NoStop}%
\bibitem [{\citenamefont {Gough}\ \emph {et~al.}(2012)\citenamefont {Gough},
  \citenamefont {James}, \citenamefont {Nurdin},\ and\ \citenamefont
  {Combes}}]{gough_quantum_2012}%
  \BibitemOpen
  \bibfield  {author} {\bibinfo {author} {\bibfnamefont {J.~E.}\ \bibnamefont
  {Gough}}, \bibinfo {author} {\bibfnamefont {M.~R.}\ \bibnamefont {James}},
  \bibinfo {author} {\bibfnamefont {H.~I.}\ \bibnamefont {Nurdin}},\ and\
  \bibinfo {author} {\bibfnamefont {J.}~\bibnamefont {Combes}},\ }\bibfield
  {title} {\bibinfo {title} {Quantum filtering for systems driven by fields in
  single-photon states or superposition of coherent states},\ }\href
  {https://doi.org/10.1103/PhysRevA.86.043819} {\bibfield  {journal} {\bibinfo
  {journal} {Physical Review A}\ }\textbf {\bibinfo {volume} {86}},\ \bibinfo
  {pages} {043819} (\bibinfo {year} {2012})}\BibitemShut {NoStop}%
\bibitem [{\citenamefont {Baragiola}\ \emph {et~al.}(2012)\citenamefont
  {Baragiola}, \citenamefont {Cook}, \citenamefont {Bra{\'n}czyk},\ and\
  \citenamefont {Combes}}]{baragiola_n-photon_2012}%
  \BibitemOpen
  \bibfield  {author} {\bibinfo {author} {\bibfnamefont {B.~Q.}\ \bibnamefont
  {Baragiola}}, \bibinfo {author} {\bibfnamefont {R.~L.}\ \bibnamefont {Cook}},
  \bibinfo {author} {\bibfnamefont {A.~M.}\ \bibnamefont {Bra{\'n}czyk}},\ and\
  \bibinfo {author} {\bibfnamefont {J.}~\bibnamefont {Combes}},\ }\bibfield
  {title} {\bibinfo {title} {N-photon wave packets interacting with an
  arbitrary quantum system},\ }\href
  {https://doi.org/10.1103/PhysRevA.86.013811} {\bibfield  {journal} {\bibinfo
  {journal} {Physical Review A}\ }\textbf {\bibinfo {volume} {86}},\ \bibinfo
  {pages} {013811} (\bibinfo {year} {2012})}\BibitemShut {NoStop}%
\bibitem [{\citenamefont {Baragiola}\ and\ \citenamefont
  {Combes}(2017)}]{baragiola_quantum_2017}%
  \BibitemOpen
  \bibfield  {author} {\bibinfo {author} {\bibfnamefont {B.~Q.}\ \bibnamefont
  {Baragiola}}\ and\ \bibinfo {author} {\bibfnamefont {J.}~\bibnamefont
  {Combes}},\ }\bibfield  {title} {\bibinfo {title} {Quantum trajectories for
  propagating {Fock} states},\ }\href
  {https://doi.org/10.1103/PhysRevA.96.023819} {\bibfield  {journal} {\bibinfo
  {journal} {Physical Review A}\ }\textbf {\bibinfo {volume} {96}},\ \bibinfo
  {pages} {023819} (\bibinfo {year} {2017})}\BibitemShut {NoStop}%
\bibitem [{\citenamefont {D\c{a}browska}\ \emph {et~al.}(2017)\citenamefont
  {D\c{a}browska}, \citenamefont {Sarbicki},\ and\ \citenamefont
  {Chru\'{s}ci\'{n}ski}}]{dabrowska_quantum_2017}%
  \BibitemOpen
  \bibfield  {author} {\bibinfo {author} {\bibfnamefont {A.}~\bibnamefont
  {D\c{a}browska}}, \bibinfo {author} {\bibfnamefont {G.}~\bibnamefont
  {Sarbicki}},\ and\ \bibinfo {author} {\bibfnamefont {D.}~\bibnamefont
  {Chru\'{s}ci\'{n}ski}},\ }\bibfield  {title} {\bibinfo {title} {Quantum
  trajectories for a system interacting with environment in a single-photon
  state: {Counting} and diffusive processes},\ }\href
  {https://doi.org/10.1103/PhysRevA.96.053819} {\bibfield  {journal} {\bibinfo
  {journal} {Physical Review A}\ }\textbf {\bibinfo {volume} {96}},\ \bibinfo
  {pages} {053819} (\bibinfo {year} {2017})},\ \bibinfo {note} {publisher:
  American Physical Society}\BibitemShut {NoStop}%
\bibitem [{\citenamefont {D\c{a}browska}(2019)}]{dabrowska_quantum_2019}%
  \BibitemOpen
  \bibfield  {author} {\bibinfo {author} {\bibfnamefont {A.~M.}\ \bibnamefont
  {D\c{a}browska}},\ }\bibfield  {title} {\bibinfo {title} {Quantum
  trajectories for environment in superposition of coherent states},\ }\href
  {https://doi.org/10.1007/s11128-019-2340-4} {\bibfield  {journal} {\bibinfo
  {journal} {Quantum Information Processing}\ }\textbf {\bibinfo {volume}
  {18}},\ \bibinfo {pages} {224} (\bibinfo {year} {2019})}\BibitemShut
  {NoStop}%
\bibitem [{\citenamefont {D{\k{a}}browska}\ \emph {et~al.}(2019)\citenamefont
  {D{\k{a}}browska}, \citenamefont {Sarbicki},\ and\ \citenamefont
  {Chru{\'{s}}ci{\'{n}}ski}}]{Dabrowska_2019_nphoton}%
  \BibitemOpen
  \bibfield  {author} {\bibinfo {author} {\bibfnamefont {A.}~\bibnamefont
  {D{\k{a}}browska}}, \bibinfo {author} {\bibfnamefont {G.}~\bibnamefont
  {Sarbicki}},\ and\ \bibinfo {author} {\bibfnamefont {D.}~\bibnamefont
  {Chru{\'{s}}ci{\'{n}}ski}},\ }\bibfield  {title} {\bibinfo {title} {Quantum
  trajectories for a system interacting with environment {inN}-photon state},\
  }\href {https://doi.org/10.1088/1751-8121/ab01ac} {\bibfield  {journal}
  {\bibinfo  {journal} {Journal of Physics A: Mathematical and Theoretical}\
  }\textbf {\bibinfo {volume} {52}},\ \bibinfo {pages} {105303} (\bibinfo
  {year} {2019})}\BibitemShut {NoStop}%
\bibitem [{\citenamefont {Kiilerich}\ and\ \citenamefont
  {M{\o}lmer}(2019)}]{kiilerich_input-output_2019}%
  \BibitemOpen
  \bibfield  {author} {\bibinfo {author} {\bibfnamefont {A.~H.}\ \bibnamefont
  {Kiilerich}}\ and\ \bibinfo {author} {\bibfnamefont {K.}~\bibnamefont
  {M{\o}lmer}},\ }\bibfield  {title} {\bibinfo {title} {Input-output theory
  with quantum pulses},\ }\href
  {https://doi.org/10.1103/PhysRevLett.123.123604} {\bibfield  {journal}
  {\bibinfo  {journal} {Physical Review Letters}\ }\textbf {\bibinfo {volume}
  {123}},\ \bibinfo {pages} {123604} (\bibinfo {year} {2019})},\ \bibinfo
  {note} {publisher: American Physical Society}\BibitemShut {NoStop}%
\bibitem [{\citenamefont {Kiilerich}\ and\ \citenamefont
  {M\o{}lmer}(2020)}]{kiilerich_quantum_2020}%
  \BibitemOpen
  \bibfield  {author} {\bibinfo {author} {\bibfnamefont {A.~H.}\ \bibnamefont
  {Kiilerich}}\ and\ \bibinfo {author} {\bibfnamefont {K.}~\bibnamefont
  {M\o{}lmer}},\ }\bibfield  {title} {\bibinfo {title} {Quantum interactions
  with pulses of radiation},\ }\href
  {https://doi.org/10.1103/PhysRevA.102.023717} {\bibfield  {journal} {\bibinfo
   {journal} {Phys. Rev. A}\ }\textbf {\bibinfo {volume} {102}},\ \bibinfo
  {pages} {023717} (\bibinfo {year} {2020})}\BibitemShut {NoStop}%
\bibitem [{\citenamefont {\ifmmode~\mbox{\c{C}}\else \c{C}\fi{}akmak}\ \emph
  {et~al.}(2017)\citenamefont {\ifmmode~\mbox{\c{C}}\else \c{C}\fi{}akmak},
  \citenamefont {Pezzutto}, \citenamefont {Paternostro},\ and\ \citenamefont
  {M\"ustecapl\ifmmode \imath \else \i \fi{}o\ifmmode~\breve{g}\else
  \u{g}\fi{}lu}}]{Cakmak2017}%
  \BibitemOpen
  \bibfield  {author} {\bibinfo {author} {\bibfnamefont {B.}~\bibnamefont
  {\ifmmode~\mbox{\c{C}}\else \c{C}\fi{}akmak}}, \bibinfo {author}
  {\bibfnamefont {M.}~\bibnamefont {Pezzutto}}, \bibinfo {author}
  {\bibfnamefont {M.}~\bibnamefont {Paternostro}},\ and\ \bibinfo {author}
  {\bibfnamefont {O.~E.}\ \bibnamefont {M\"ustecapl\ifmmode \imath \else \i
  \fi{}o\ifmmode~\breve{g}\else \u{g}\fi{}lu}},\ }\bibfield  {title} {\bibinfo
  {title} {Non-markovianity, coherence, and system-environment correlations in
  a long-range collision model},\ }\href
  {https://doi.org/10.1103/PhysRevA.96.022109} {\bibfield  {journal} {\bibinfo
  {journal} {Phys. Rev. A}\ }\textbf {\bibinfo {volume} {96}},\ \bibinfo
  {pages} {022109} (\bibinfo {year} {2017})}\BibitemShut {NoStop}%
\bibitem [{\citenamefont {Daryanoosh}\ \emph {et~al.}(2018)\citenamefont
  {Daryanoosh}, \citenamefont {Baragiola}, \citenamefont {Guff},\ and\
  \citenamefont {Gilchrist}}]{Daryanoosh:2018aa}%
  \BibitemOpen
  \bibfield  {author} {\bibinfo {author} {\bibfnamefont {S.}~\bibnamefont
  {Daryanoosh}}, \bibinfo {author} {\bibfnamefont {B.~Q.}\ \bibnamefont
  {Baragiola}}, \bibinfo {author} {\bibfnamefont {T.}~\bibnamefont {Guff}},\
  and\ \bibinfo {author} {\bibfnamefont {A.}~\bibnamefont {Gilchrist}},\
  }\bibfield  {title} {\bibinfo {title} {Quantum master equations for entangled
  qubit environments},\ }\href {https://doi.org/10.1103/PhysRevA.98.062104}
  {\bibfield  {journal} {\bibinfo  {journal} {Phys. Rev. A}\ }\textbf {\bibinfo
  {volume} {98}},\ \bibinfo {pages} {062104} (\bibinfo {year}
  {2018})}\BibitemShut {NoStop}%
\bibitem [{\citenamefont {Whalen}(2019)}]{Whalen:2019aa}%
  \BibitemOpen
  \bibfield  {author} {\bibinfo {author} {\bibfnamefont {S.~J.}\ \bibnamefont
  {Whalen}},\ }\bibfield  {title} {\bibinfo {title} {Collision model for
  non-markovian quantum trajectories},\ }\href
  {https://doi.org/10.1103/PhysRevA.100.052113} {\bibfield  {journal} {\bibinfo
   {journal} {Phys. Rev. A}\ }\textbf {\bibinfo {volume} {100}},\ \bibinfo
  {pages} {052113} (\bibinfo {year} {2019})}\BibitemShut {NoStop}%
\bibitem [{\citenamefont {Taranto}\ \emph {et~al.}(2019)\citenamefont
  {Taranto}, \citenamefont {Milz}, \citenamefont {Pollock},\ and\ \citenamefont
  {Modi}}]{Taranto:2019aa}%
  \BibitemOpen
  \bibfield  {author} {\bibinfo {author} {\bibfnamefont {P.}~\bibnamefont
  {Taranto}}, \bibinfo {author} {\bibfnamefont {S.}~\bibnamefont {Milz}},
  \bibinfo {author} {\bibfnamefont {F.~A.}\ \bibnamefont {Pollock}},\ and\
  \bibinfo {author} {\bibfnamefont {K.}~\bibnamefont {Modi}},\ }\bibfield
  {title} {\bibinfo {title} {Structure of quantum stochastic processes with
  finite markov order},\ }\href {https://doi.org/10.1103/PhysRevA.99.042108}
  {\bibfield  {journal} {\bibinfo  {journal} {Phys. Rev. A}\ }\textbf {\bibinfo
  {volume} {99}},\ \bibinfo {pages} {042108} (\bibinfo {year}
  {2019})}\BibitemShut {NoStop}%
\bibitem [{\citenamefont {Rodrigues}\ \emph {et~al.}(2019)\citenamefont
  {Rodrigues}, \citenamefont {De~Chiara}, \citenamefont {Paternostro},\ and\
  \citenamefont {Landi}}]{Rodrigues:2019aa}%
  \BibitemOpen
  \bibfield  {author} {\bibinfo {author} {\bibfnamefont {F.~L.~S.}\
  \bibnamefont {Rodrigues}}, \bibinfo {author} {\bibfnamefont {G.}~\bibnamefont
  {De~Chiara}}, \bibinfo {author} {\bibfnamefont {M.}~\bibnamefont
  {Paternostro}},\ and\ \bibinfo {author} {\bibfnamefont {G.~T.}\ \bibnamefont
  {Landi}},\ }\bibfield  {title} {\bibinfo {title} {Thermodynamics of weakly
  coherent collisional models},\ }\href
  {https://doi.org/10.1103/PhysRevLett.123.140601} {\bibfield  {journal}
  {\bibinfo  {journal} {Phys. Rev. Lett.}\ }\textbf {\bibinfo {volume} {123}},\
  \bibinfo {pages} {140601} (\bibinfo {year} {2019})}\BibitemShut {NoStop}%
\bibitem [{\citenamefont {Cilluffo}\ \emph {et~al.}(2020)\citenamefont
  {Cilluffo}, \citenamefont {Carollo}, \citenamefont {Lorenzo}, \citenamefont
  {Gross}, \citenamefont {Palma},\ and\ \citenamefont
  {Ciccarello}}]{cilluffo_collisional_2020}%
  \BibitemOpen
  \bibfield  {author} {\bibinfo {author} {\bibfnamefont {D.}~\bibnamefont
  {Cilluffo}}, \bibinfo {author} {\bibfnamefont {A.}~\bibnamefont {Carollo}},
  \bibinfo {author} {\bibfnamefont {S.}~\bibnamefont {Lorenzo}}, \bibinfo
  {author} {\bibfnamefont {J.~A.}\ \bibnamefont {Gross}}, \bibinfo {author}
  {\bibfnamefont {G.~M.}\ \bibnamefont {Palma}},\ and\ \bibinfo {author}
  {\bibfnamefont {F.}~\bibnamefont {Ciccarello}},\ }\bibfield  {title}
  {\bibinfo {title} {Collisional picture of quantum optics with giant
  emitters},\ }\href {https://doi.org/10.1103/PhysRevResearch.2.043070}
  {\bibfield  {journal} {\bibinfo  {journal} {Physical Review Research}\
  }\textbf {\bibinfo {volume} {2}},\ \bibinfo {pages} {043070} (\bibinfo {year}
  {2020})},\ \bibinfo {note} {arXiv: 2006.08631}\BibitemShut {NoStop}%
\bibitem [{\citenamefont {Carollo}\ \emph {et~al.}(2020)\citenamefont
  {Carollo}, \citenamefont {Cilluffo},\ and\ \citenamefont
  {Ciccarello}}]{carollo_mechanism_2020}%
  \BibitemOpen
  \bibfield  {author} {\bibinfo {author} {\bibfnamefont {A.}~\bibnamefont
  {Carollo}}, \bibinfo {author} {\bibfnamefont {D.}~\bibnamefont {Cilluffo}},\
  and\ \bibinfo {author} {\bibfnamefont {F.}~\bibnamefont {Ciccarello}},\
  }\bibfield  {title} {\bibinfo {title} {Mechanism of decoherence-free coupling
  between giant atoms},\ }\href
  {https://doi.org/10.1103/PhysRevResearch.2.043184} {\bibfield  {journal}
  {\bibinfo  {journal} {Physical Review Research}\ }\textbf {\bibinfo {volume}
  {2}},\ \bibinfo {pages} {043184} (\bibinfo {year} {2020})},\ \bibinfo {note}
  {arXiv: 2006.13940}\BibitemShut {NoStop}%
\bibitem [{\citenamefont {Cattaneo}\ \emph {et~al.}(2020)\citenamefont
  {Cattaneo}, \citenamefont {De~Chiara}, \citenamefont {Maniscalco},
  \citenamefont {Zambrini},\ and\ \citenamefont
  {Giorgi}}]{Cattaneo_collisional_2020}%
  \BibitemOpen
  \bibfield  {author} {\bibinfo {author} {\bibfnamefont {M.}~\bibnamefont
  {Cattaneo}}, \bibinfo {author} {\bibfnamefont {G.}~\bibnamefont {De~Chiara}},
  \bibinfo {author} {\bibfnamefont {S.}~\bibnamefont {Maniscalco}}, \bibinfo
  {author} {\bibfnamefont {R.}~\bibnamefont {Zambrini}},\ and\ \bibinfo
  {author} {\bibfnamefont {G.~L.}\ \bibnamefont {Giorgi}},\ }\bibfield  {title}
  {\bibinfo {title} {Collision models can efficiently simulate any multipartite
  {Markovian} quantum dynamics},\ }\href {http://arxiv.org/abs/2010.13910}
  {\bibfield  {journal} {\bibinfo  {journal} {arXiv:2010.13910 [quant-ph]}\ }
  (\bibinfo {year} {2020})},\ \bibinfo {note} {arXiv: 2010.13910}\BibitemShut
  {NoStop}%
\bibitem [{\citenamefont {Ciccarello}\ \emph {et~al.}(2021)\citenamefont
  {Ciccarello}, \citenamefont {Lorenzo}, \citenamefont {Giovannetti},\ and\
  \citenamefont {Palma}}]{ciccarello_CM_review_2021}%
  \BibitemOpen
  \bibfield  {author} {\bibinfo {author} {\bibfnamefont {F.}~\bibnamefont
  {Ciccarello}}, \bibinfo {author} {\bibfnamefont {S.}~\bibnamefont {Lorenzo}},
  \bibinfo {author} {\bibfnamefont {V.}~\bibnamefont {Giovannetti}},\ and\
  \bibinfo {author} {\bibfnamefont {G.~M.}\ \bibnamefont {Palma}},\ }\bibfield
  {title} {\bibinfo {title} {Quantum collision models: open system dynamics
  from repeated interactions},\ }\href {http://arxiv.org/abs/2106.11974}
  {\bibfield  {journal} {\bibinfo  {journal} {arXiv:2106.11974 [quant-ph]}\ }
  (\bibinfo {year} {2021})},\ \bibinfo {note} {arXiv: 2106.11974}\BibitemShut
  {NoStop}%
\bibitem [{\citenamefont {Gardiner}\ \emph {et~al.}(2004)\citenamefont
  {Gardiner}, \citenamefont {Zoller},\ and\ \citenamefont
  {Zoller}}]{q_noise_gard2004}%
  \BibitemOpen
  \bibfield  {author} {\bibinfo {author} {\bibfnamefont {C.}~\bibnamefont
  {Gardiner}}, \bibinfo {author} {\bibfnamefont {P.}~\bibnamefont {Zoller}},\
  and\ \bibinfo {author} {\bibfnamefont {P.}~\bibnamefont {Zoller}},\
  }\href@noop {} {\emph {\bibinfo {title} {Quantum noise: a handbook of
  Markovian and non-Markovian quantum stochastic methods with applications to
  quantum optics}}}\ (\bibinfo  {publisher} {Springer Science \& Business
  Media},\ \bibinfo {year} {2004})\BibitemShut {NoStop}%
\bibitem [{\citenamefont {Gross}\ \emph {et~al.}(2017)\citenamefont {Gross},
  \citenamefont {Caves}, \citenamefont {Milburn},\ and\ \citenamefont
  {Combes}}]{gross_qubit_2017}%
  \BibitemOpen
  \bibfield  {author} {\bibinfo {author} {\bibfnamefont {J.~A.}\ \bibnamefont
  {Gross}}, \bibinfo {author} {\bibfnamefont {C.~M.}\ \bibnamefont {Caves}},
  \bibinfo {author} {\bibfnamefont {G.~J.}\ \bibnamefont {Milburn}},\ and\
  \bibinfo {author} {\bibfnamefont {J.}~\bibnamefont {Combes}},\ }\bibfield
  {title} {\bibinfo {title} {Qubit models of weak continuous measurements:
  {Markovian} conditional and open-system dynamics},\ }\bibfield  {journal}
  {\bibinfo  {journal} {Quantum Science and Technology}\ }\href
  {https://doi.org/10.1088/2058-9565/aaa39f} {10.1088/2058-9565/aaa39f}
  (\bibinfo {year} {2017})\BibitemShut {NoStop}%
\bibitem [{\citenamefont {Combes}\ \emph {et~al.}(2017)\citenamefont {Combes},
  \citenamefont {Kerckhoff},\ and\ \citenamefont {Sarovar}}]{combes_slh_2017}%
  \BibitemOpen
  \bibfield  {author} {\bibinfo {author} {\bibfnamefont {J.}~\bibnamefont
  {Combes}}, \bibinfo {author} {\bibfnamefont {J.}~\bibnamefont {Kerckhoff}},\
  and\ \bibinfo {author} {\bibfnamefont {M.}~\bibnamefont {Sarovar}},\
  }\bibfield  {title} {\bibinfo {title} {The {SLH} framework for modeling
  quantum input-output networks},\ }\href
  {https://doi.org/10.1080/23746149.2017.1343097} {\bibfield  {journal}
  {\bibinfo  {journal} {Advances in Physics: X}\ }\textbf {\bibinfo {volume}
  {2}},\ \bibinfo {pages} {784} (\bibinfo {year} {2017})},\ \bibinfo {note}
  {arXiv: 1611.00375}\BibitemShut {NoStop}%
\bibitem [{\citenamefont {Trivedi}\ \emph {et~al.}(2019)\citenamefont
  {Trivedi}, \citenamefont {Fischer}, \citenamefont {Mishra},\ and\
  \citenamefont {Vu\ifmmode \check{c}\else
  \v{c}\fi{}kovi\ifmmode~\acute{c}\else \'{c}\fi{}}}]{TrivFisc2019}%
  \BibitemOpen
  \bibfield  {author} {\bibinfo {author} {\bibfnamefont {R.}~\bibnamefont
  {Trivedi}}, \bibinfo {author} {\bibfnamefont {K.}~\bibnamefont {Fischer}},
  \bibinfo {author} {\bibfnamefont {S.~D.}\ \bibnamefont {Mishra}},\ and\
  \bibinfo {author} {\bibfnamefont {J.}~\bibnamefont {Vu\ifmmode \check{c}\else
  \v{c}\fi{}kovi\ifmmode~\acute{c}\else \'{c}\fi{}}},\ }\bibfield  {title}
  {\bibinfo {title} {Point-coupling hamiltonian for frequency-independent
  linear optical devices},\ }\href
  {https://doi.org/10.1103/PhysRevA.100.043827} {\bibfield  {journal} {\bibinfo
   {journal} {Phys. Rev. A}\ }\textbf {\bibinfo {volume} {100}},\ \bibinfo
  {pages} {043827} (\bibinfo {year} {2019})}\BibitemShut {NoStop}%
\bibitem [{\citenamefont {Hudson}\ and\ \citenamefont
  {Parthasarathy}(1984)}]{HudsonPartha1984}%
  \BibitemOpen
  \bibfield  {author} {\bibinfo {author} {\bibfnamefont {R.~L.}\ \bibnamefont
  {Hudson}}\ and\ \bibinfo {author} {\bibfnamefont {K.~R.}\ \bibnamefont
  {Parthasarathy}},\ }\bibfield  {title} {\bibinfo {title} {Quantum {I}to's
  formula and stochastic evolutions},\ }\href
  {https://doi.org/10.1007/BF01258530} {\bibfield  {journal} {\bibinfo
  {journal} {Communications in Mathematical Physics}\ }\textbf {\bibinfo
  {volume} {93}},\ \bibinfo {pages} {301} (\bibinfo {year} {1984})}\BibitemShut
  {NoStop}%
\bibitem [{\citenamefont {Yurke}\ and\ \citenamefont
  {Denker}(1984)}]{YurkeDenker84}%
  \BibitemOpen
  \bibfield  {author} {\bibinfo {author} {\bibfnamefont {B.}~\bibnamefont
  {Yurke}}\ and\ \bibinfo {author} {\bibfnamefont {J.~S.}\ \bibnamefont
  {Denker}},\ }\bibfield  {title} {\bibinfo {title} {Quantum network theory},\
  }\href {https://doi.org/10.1103/PhysRevA.29.1419} {\bibfield  {journal}
  {\bibinfo  {journal} {Phys. Rev. A}\ }\textbf {\bibinfo {volume} {29}},\
  \bibinfo {pages} {1419} (\bibinfo {year} {1984})}\BibitemShut {NoStop}%
\bibitem [{\citenamefont {Blow}\ \emph {et~al.}(1990)\citenamefont {Blow},
  \citenamefont {Loudon}, \citenamefont {Phoenix},\ and\ \citenamefont
  {Shepherd}}]{Blow1990}%
  \BibitemOpen
  \bibfield  {author} {\bibinfo {author} {\bibfnamefont {K.~J.}\ \bibnamefont
  {Blow}}, \bibinfo {author} {\bibfnamefont {R.}~\bibnamefont {Loudon}},
  \bibinfo {author} {\bibfnamefont {S.~J.~D.}\ \bibnamefont {Phoenix}},\ and\
  \bibinfo {author} {\bibfnamefont {T.~J.}\ \bibnamefont {Shepherd}},\
  }\bibfield  {title} {\bibinfo {title} {Continuum fields in quantum optics},\
  }\href {https://doi.org/10.1103/PhysRevA.42.4102} {\bibfield  {journal}
  {\bibinfo  {journal} {Phys. Rev. A}\ }\textbf {\bibinfo {volume} {42}},\
  \bibinfo {pages} {4102} (\bibinfo {year} {1990})}\BibitemShut {NoStop}%
\bibitem [{\citenamefont {Fan}\ \emph {et~al.}(2010)\citenamefont {Fan},
  \citenamefont {Kocaba\ifmmode~\mbox{\c{s}}\else \c{s}\fi{}},\ and\
  \citenamefont {Shen}}]{ScatInputOutput2010}%
  \BibitemOpen
  \bibfield  {author} {\bibinfo {author} {\bibfnamefont {S.}~\bibnamefont
  {Fan}}, \bibinfo {author} {\bibfnamefont {i.~m. c.~E.}\ \bibnamefont
  {Kocaba\ifmmode~\mbox{\c{s}}\else \c{s}\fi{}}},\ and\ \bibinfo {author}
  {\bibfnamefont {J.-T.}\ \bibnamefont {Shen}},\ }\bibfield  {title} {\bibinfo
  {title} {Input-output formalism for few-photon transport in one-dimensional
  nanophotonic waveguides coupled to a qubit},\ }\href
  {https://doi.org/10.1103/PhysRevA.82.063821} {\bibfield  {journal} {\bibinfo
  {journal} {Phys. Rev. A}\ }\textbf {\bibinfo {volume} {82}},\ \bibinfo
  {pages} {063821} (\bibinfo {year} {2010})}\BibitemShut {NoStop}%
\bibitem [{\citenamefont {van Handel}\ \emph {et~al.}(2005)\citenamefont {van
  Handel}, \citenamefont {Stockton},\ and\ \citenamefont
  {Mabuchi}}]{Handel_2005}%
  \BibitemOpen
  \bibfield  {author} {\bibinfo {author} {\bibfnamefont {R.}~\bibnamefont {van
  Handel}}, \bibinfo {author} {\bibfnamefont {J.~K.}\ \bibnamefont
  {Stockton}},\ and\ \bibinfo {author} {\bibfnamefont {H.}~\bibnamefont
  {Mabuchi}},\ }\bibfield  {title} {\bibinfo {title} {Modelling and feedback
  control design for quantum state preparation},\ }\href
  {https://doi.org/10.1088/1464-4266/7/10/001} {\bibfield  {journal} {\bibinfo
  {journal} {Journal of Optics B: Quantum and Semiclassical Optics}\ }\textbf
  {\bibinfo {volume} {7}},\ \bibinfo {pages} {S179} (\bibinfo {year}
  {2005})}\BibitemShut {NoStop}%
\bibitem [{\citenamefont {Xu}\ and\ \citenamefont
  {Fan}(2016)}]{XuFan_fano2016}%
  \BibitemOpen
  \bibfield  {author} {\bibinfo {author} {\bibfnamefont {S.}~\bibnamefont
  {Xu}}\ and\ \bibinfo {author} {\bibfnamefont {S.}~\bibnamefont {Fan}},\
  }\bibfield  {title} {\bibinfo {title} {Fano interference in two-photon
  transport},\ }\href {https://doi.org/10.1103/PhysRevA.94.043826} {\bibfield
  {journal} {\bibinfo  {journal} {Phys. Rev. A}\ }\textbf {\bibinfo {volume}
  {94}},\ \bibinfo {pages} {043826} (\bibinfo {year} {2016})}\BibitemShut
  {NoStop}%
\bibitem [{\citenamefont {Gardiner}\ \emph {et~al.}(1992)\citenamefont
  {Gardiner}, \citenamefont {Parkins},\ and\ \citenamefont {Zoller}}]{Gard92}%
  \BibitemOpen
  \bibfield  {author} {\bibinfo {author} {\bibfnamefont {C.~W.}\ \bibnamefont
  {Gardiner}}, \bibinfo {author} {\bibfnamefont {A.~S.}\ \bibnamefont
  {Parkins}},\ and\ \bibinfo {author} {\bibfnamefont {P.}~\bibnamefont
  {Zoller}},\ }\bibfield  {title} {\bibinfo {title} {Wave-function quantum
  stochastic differential equations and quantum-jump simulation methods},\
  }\href {https://doi.org/10.1103/PhysRevA.46.4363} {\bibfield  {journal}
  {\bibinfo  {journal} {Phys. Rev. A}\ }\textbf {\bibinfo {volume} {46}},\
  \bibinfo {pages} {4363} (\bibinfo {year} {1992})}\BibitemShut {NoStop}%
\bibitem [{\citenamefont {Hellmich}\ \emph
  {et~al.}(2002{\natexlab{b}})\citenamefont {Hellmich}, \citenamefont
  {Honegger}, \citenamefont {K{\"o}stler}, \citenamefont {K{\"u}mmerer},\ and\
  \citenamefont {Rieckers}}]{Hellmich02}%
  \BibitemOpen
  \bibfield  {author} {\bibinfo {author} {\bibfnamefont {J.}~\bibnamefont
  {Hellmich}}, \bibinfo {author} {\bibfnamefont {R.}~\bibnamefont {Honegger}},
  \bibinfo {author} {\bibfnamefont {C.}~\bibnamefont {K{\"o}stler}}, \bibinfo
  {author} {\bibfnamefont {B.}~\bibnamefont {K{\"u}mmerer}},\ and\ \bibinfo
  {author} {\bibfnamefont {A.}~\bibnamefont {Rieckers}},\ }\bibfield  {title}
  {\bibinfo {title} {Couplings to classical and non-classical squeezed white
  noise as stationary markov processes},\ }\href
  {https://dx.doi.org/10.2977/prims/1145476415} {\bibfield  {journal} {\bibinfo
   {journal} {Publ. Res. Inst. Math. Sci.}\ }\textbf {\bibinfo {volume} {38}},\
  \bibinfo {pages} {1} (\bibinfo {year} {2002}{\natexlab{b}})}\BibitemShut
  {NoStop}%
\bibitem [{\citenamefont {Mollow}(1969)}]{Mollow1969}%
  \BibitemOpen
  \bibfield  {author} {\bibinfo {author} {\bibfnamefont {B.~R.}\ \bibnamefont
  {Mollow}},\ }\bibfield  {title} {\bibinfo {title} {Power spectrum of light
  scattered by two-level systems},\ }\href
  {https://doi.org/10.1103/PhysRev.188.1969} {\bibfield  {journal} {\bibinfo
  {journal} {Physical Review}\ }\textbf {\bibinfo {volume} {188}},\ \bibinfo
  {pages} {1969} (\bibinfo {year} {1969})}\BibitemShut {NoStop}%
\bibitem [{\citenamefont {Garrison}\ and\ \citenamefont
  {Chiao}(2008)}]{garrison_quantum_2008}%
  \BibitemOpen
  \bibfield  {author} {\bibinfo {author} {\bibfnamefont {J.}~\bibnamefont
  {Garrison}}\ and\ \bibinfo {author} {\bibfnamefont {R.}~\bibnamefont
  {Chiao}},\ }\href@noop {} {\emph {\bibinfo {title} {Quantum {Optics}}}}\
  (\bibinfo  {publisher} {OUP Oxford},\ \bibinfo {year} {2008})\BibitemShut
  {NoStop}%
\bibitem [{\citenamefont {Jack}\ and\ \citenamefont
  {Collett}(2000)}]{JackColl00}%
  \BibitemOpen
  \bibfield  {author} {\bibinfo {author} {\bibfnamefont {M.~W.}\ \bibnamefont
  {Jack}}\ and\ \bibinfo {author} {\bibfnamefont {M.~J.}\ \bibnamefont
  {Collett}},\ }\bibfield  {title} {\bibinfo {title} {Continuous measurement
  and non-markovian quantum trajectories},\ }\href
  {https://doi.org/10.1103/PhysRevA.61.062106} {\bibfield  {journal} {\bibinfo
  {journal} {Phys. Rev. A}\ }\textbf {\bibinfo {volume} {61}},\ \bibinfo
  {pages} {062106} (\bibinfo {year} {2000})}\BibitemShut {NoStop}%
\bibitem [{\citenamefont {Str{\"u}mpfer}\ and\ \citenamefont
  {Schulten}(2012)}]{Strumpfer2012}%
  \BibitemOpen
  \bibfield  {author} {\bibinfo {author} {\bibfnamefont {J.}~\bibnamefont
  {Str{\"u}mpfer}}\ and\ \bibinfo {author} {\bibfnamefont {K.}~\bibnamefont
  {Schulten}},\ }\bibfield  {title} {\bibinfo {title} {Open quantum dynamics
  calculations with the hierarchy equations of motion on parallel computers},\
  }\bibfield  {booktitle} {\emph {\bibinfo {booktitle} {Journal of Chemical
  Theory and Computation}},\ }\href {https://doi.org/10.1021/ct3003833}
  {\bibfield  {journal} {\bibinfo  {journal} {Journal of Chemical Theory and
  Computation}\ }\textbf {\bibinfo {volume} {8}},\ \bibinfo {pages} {2808}
  (\bibinfo {year} {2012})}\BibitemShut {NoStop}%
\bibitem [{\citenamefont {Neu}(2015)}]{neu2015singular}%
  \BibitemOpen
  \bibfield  {author} {\bibinfo {author} {\bibfnamefont {J.~C.}\ \bibnamefont
  {Neu}},\ }\href@noop {} {\emph {\bibinfo {title} {Singular perturbation in
  the physical sciences}}},\ Vol.\ \bibinfo {volume} {167}\ (\bibinfo
  {publisher} {American Mathematical Soc.},\ \bibinfo {year}
  {2015})\BibitemShut {NoStop}%
\bibitem [{\citenamefont {Baragiola}(2014)}]{Baragiola:2014aa}%
  \BibitemOpen
  \bibfield  {author} {\bibinfo {author} {\bibfnamefont {B.~Q.}\ \bibnamefont
  {Baragiola}},\ }\emph {\bibinfo {title} {Open systems dynamics for
  propagating quantum fields}},\ \href {https://arxiv.org/abs/1408.4447} {Ph.D.
  thesis},\ \bibinfo  {school} {University of New Mexico} (\bibinfo {year}
  {2014})\BibitemShut {NoStop}%
\bibitem [{\citenamefont {Lax}(1963)}]{Lax:1963aa}%
  \BibitemOpen
  \bibfield  {author} {\bibinfo {author} {\bibfnamefont {M.}~\bibnamefont
  {Lax}},\ }\bibfield  {title} {\bibinfo {title} {Formal theory of quantum
  fluctuations from a driven state},\ }\href
  {https://doi.org/10.1103/PhysRev.129.2342} {\bibfield  {journal} {\bibinfo
  {journal} {Phys. Rev.}\ }\textbf {\bibinfo {volume} {129}},\ \bibinfo {pages}
  {2342} (\bibinfo {year} {1963})}\BibitemShut {NoStop}%
\bibitem [{\citenamefont {Sathyamoorthy}\ \emph {et~al.}(2014)\citenamefont
  {Sathyamoorthy}, \citenamefont {Tornberg}, \citenamefont {Kockum},
  \citenamefont {Baragiola}, \citenamefont {Combes}, \citenamefont {Wilson},
  \citenamefont {Stace},\ and\ \citenamefont
  {Johansson}}]{sathyamoorthy_quantum_2014}%
  \BibitemOpen
  \bibfield  {author} {\bibinfo {author} {\bibfnamefont {S.~R.}\ \bibnamefont
  {Sathyamoorthy}}, \bibinfo {author} {\bibfnamefont {L.}~\bibnamefont
  {Tornberg}}, \bibinfo {author} {\bibfnamefont {A.~F.}\ \bibnamefont
  {Kockum}}, \bibinfo {author} {\bibfnamefont {B.~Q.}\ \bibnamefont
  {Baragiola}}, \bibinfo {author} {\bibfnamefont {J.}~\bibnamefont {Combes}},
  \bibinfo {author} {\bibfnamefont {C.}~\bibnamefont {Wilson}}, \bibinfo
  {author} {\bibfnamefont {T.~M.}\ \bibnamefont {Stace}},\ and\ \bibinfo
  {author} {\bibfnamefont {G.}~\bibnamefont {Johansson}},\ }\bibfield  {title}
  {\bibinfo {title} {Quantum nondemolition detection of a propagating microwave
  photon},\ }\href {https://doi.org/10.1103/PhysRevLett.112.093601} {\bibfield
  {journal} {\bibinfo  {journal} {Physical Review Letters}\ }\textbf {\bibinfo
  {volume} {112}},\ \bibinfo {pages} {093601} (\bibinfo {year}
  {2014})}\BibitemShut {NoStop}%
\bibitem [{\citenamefont {Carmichael}\ \emph
  {et~al.}(1987{\natexlab{a}})\citenamefont {Carmichael}, \citenamefont
  {Lane},\ and\ \citenamefont {Walls}}]{carmichael_resonance_1987}%
  \BibitemOpen
  \bibfield  {author} {\bibinfo {author} {\bibfnamefont {H.~J.}\ \bibnamefont
  {Carmichael}}, \bibinfo {author} {\bibfnamefont {A.~S.}\ \bibnamefont
  {Lane}},\ and\ \bibinfo {author} {\bibfnamefont {D.~F.}\ \bibnamefont
  {Walls}},\ }\bibfield  {title} {\bibinfo {title} {Resonance fluorescence from
  an atom in a squeezed vacuum},\ }\href
  {https://doi.org/10.1103/PhysRevLett.58.2539} {\bibfield  {journal} {\bibinfo
   {journal} {Physical Review Letters}\ }\textbf {\bibinfo {volume} {58}},\
  \bibinfo {pages} {2539} (\bibinfo {year} {1987}{\natexlab{a}})}\BibitemShut
  {NoStop}%
\bibitem [{\citenamefont {Carmichael}\ \emph
  {et~al.}(1987{\natexlab{b}})\citenamefont {Carmichael}, \citenamefont
  {Lane},\ and\ \citenamefont {Walls}}]{Carmichael:1987aa}%
  \BibitemOpen
  \bibfield  {author} {\bibinfo {author} {\bibfnamefont {H.~J.}\ \bibnamefont
  {Carmichael}}, \bibinfo {author} {\bibfnamefont {A.~S.}\ \bibnamefont
  {Lane}},\ and\ \bibinfo {author} {\bibfnamefont {D.~F.}\ \bibnamefont
  {Walls}},\ }\bibfield  {title} {\bibinfo {title} {Resonance fluorescence in a
  squeezed vacuum},\ }\bibfield  {booktitle} {\emph {\bibinfo {booktitle}
  {Journal of Modern Optics}},\ }\href
  {https://doi.org/10.1080/09500348714550771} {\bibfield  {journal} {\bibinfo
  {journal} {Journal of Modern Optics}\ }\textbf {\bibinfo {volume} {34}},\
  \bibinfo {pages} {821} (\bibinfo {year} {1987}{\natexlab{b}})}\BibitemShut
  {NoStop}%
\bibitem [{\citenamefont {Swain}(2004)}]{Swain2004}%
  \BibitemOpen
  \bibfield  {author} {\bibinfo {author} {\bibfnamefont {S.}~\bibnamefont
  {Swain}},\ }\bibinfo {title} {Novel spectroscopy with two-level atoms in
  squeezed fields},\ in\ \href {https://doi.org/10.1007/978-3-662-09645-1_8}
  {\emph {\bibinfo {booktitle} {Quantum Squeezing}}},\ \bibinfo {editor}
  {edited by\ \bibinfo {editor} {\bibfnamefont {P.~D.}\ \bibnamefont
  {Drummond}}\ and\ \bibinfo {editor} {\bibfnamefont {Z.}~\bibnamefont
  {Ficek}}}\ (\bibinfo  {publisher} {Springer Berlin Heidelberg},\ \bibinfo
  {address} {Berlin, Heidelberg},\ \bibinfo {year} {2004})\ pp.\ \bibinfo
  {pages} {263--309}\BibitemShut {NoStop}%
\bibitem [{\citenamefont {Ficek}(2004)}]{Ficek2004}%
  \BibitemOpen
  \bibfield  {author} {\bibinfo {author} {\bibfnamefont {Z.}~\bibnamefont
  {Ficek}},\ }\bibinfo {title} {Spectroscopy with three-level atoms in a
  squeezed field},\ in\ \href {https://doi.org/10.1007/978-3-662-09645-1_9}
  {\emph {\bibinfo {booktitle} {Quantum Squeezing}}},\ \bibinfo {editor}
  {edited by\ \bibinfo {editor} {\bibfnamefont {P.~D.}\ \bibnamefont
  {Drummond}}\ and\ \bibinfo {editor} {\bibfnamefont {Z.}~\bibnamefont
  {Ficek}}}\ (\bibinfo  {publisher} {Springer Berlin Heidelberg},\ \bibinfo
  {address} {Berlin, Heidelberg},\ \bibinfo {year} {2004})\ pp.\ \bibinfo
  {pages} {311--335}\BibitemShut {NoStop}%
\bibitem [{\citenamefont {Joshi}\ and\ \citenamefont
  {Puri}(1994)}]{JoshiPuri1994}%
  \BibitemOpen
  \bibfield  {author} {\bibinfo {author} {\bibfnamefont {A.}~\bibnamefont
  {Joshi}}\ and\ \bibinfo {author} {\bibfnamefont {R.~R.}\ \bibnamefont
  {Puri}},\ }\bibfield  {title} {\bibinfo {title} {On the nature of fluorescent
  spectrum of a two atom dicke model in a narrow bandwidth squeezed bath},\
  }\href {https://doi.org/10.1142/S0217979294000051} {\bibfield  {journal}
  {\bibinfo  {journal} {International Journal of Modern Physics B}\ }\textbf
  {\bibinfo {volume} {08}},\ \bibinfo {pages} {121} (\bibinfo {year}
  {1994})}\BibitemShut {NoStop}%
\bibitem [{\citenamefont {Steck}(2007)}]{steck2007_QO_textbook}%
  \BibitemOpen
  \bibfield  {author} {\bibinfo {author} {\bibfnamefont {D.~A.}\ \bibnamefont
  {Steck}},\ }\href
  {http://atomoptics.uoregon.edu/~dsteck/teaching/quantum-optics/} {\bibinfo
  {title} {Quantum and atom optics}} (\bibinfo {year} {2007})\BibitemShut
  {NoStop}%
\bibitem [{\citenamefont {Goetsch}\ and\ \citenamefont
  {Graham}(1994)}]{GoetGrah94}%
  \BibitemOpen
  \bibfield  {author} {\bibinfo {author} {\bibfnamefont {P.}~\bibnamefont
  {Goetsch}}\ and\ \bibinfo {author} {\bibfnamefont {R.}~\bibnamefont
  {Graham}},\ }\bibfield  {title} {\bibinfo {title} {Linear stochastic wave
  equations for continuously measured quantum systems},\ }\href
  {https://doi.org/10.1103/PhysRevA.50.5242} {\bibfield  {journal} {\bibinfo
  {journal} {Phys. Rev. A}\ }\textbf {\bibinfo {volume} {50}},\ \bibinfo
  {pages} {5242} (\bibinfo {year} {1994})}\BibitemShut {NoStop}%
\bibitem [{\citenamefont {Gheri}\ \emph {et~al.}(1998)\citenamefont {Gheri},
  \citenamefont {Ellinger}, \citenamefont {Pellizzari},\ and\ \citenamefont
  {Zoller}}]{Gheri:1998aa}%
  \BibitemOpen
  \bibfield  {author} {\bibinfo {author} {\bibfnamefont {K.~M.}\ \bibnamefont
  {Gheri}}, \bibinfo {author} {\bibfnamefont {K.}~\bibnamefont {Ellinger}},
  \bibinfo {author} {\bibfnamefont {T.}~\bibnamefont {Pellizzari}},\ and\
  \bibinfo {author} {\bibfnamefont {P.}~\bibnamefont {Zoller}},\ }\bibfield
  {title} {\bibinfo {title} {Photon-wavepackets as flying quantum bits},\
  }\bibfield  {booktitle} {\emph {\bibinfo {booktitle} {Fortschritte der
  Physik}},\ }\href
  {https://doi.org/10.1002/(SICI)1521-3978(199806)46:4/5<401::AID-PROP401>3.0.CO;2-W}
  {\bibfield  {journal} {\bibinfo  {journal} {Fortschritte der Physik}\
  }\textbf {\bibinfo {volume} {46}},\ \bibinfo {pages} {401} (\bibinfo {year}
  {1998})}\BibitemShut {NoStop}%
\bibitem [{\citenamefont {Di\'osi}\ and\ \citenamefont
  {Ferialdi}(2014)}]{DiosiFerialdi2014}%
  \BibitemOpen
  \bibfield  {author} {\bibinfo {author} {\bibfnamefont {L.}~\bibnamefont
  {Di\'osi}}\ and\ \bibinfo {author} {\bibfnamefont {L.}~\bibnamefont
  {Ferialdi}},\ }\bibfield  {title} {\bibinfo {title} {General non-markovian
  structure of gaussian master and stochastic schr\"odinger equations},\ }\href
  {https://doi.org/10.1103/PhysRevLett.113.200403} {\bibfield  {journal}
  {\bibinfo  {journal} {Phys. Rev. Lett.}\ }\textbf {\bibinfo {volume} {113}},\
  \bibinfo {pages} {200403} (\bibinfo {year} {2014})}\BibitemShut {NoStop}%
\bibitem [{\citenamefont {Ferialdi}(2016)}]{Ferialdi2016}%
  \BibitemOpen
  \bibfield  {author} {\bibinfo {author} {\bibfnamefont {L.}~\bibnamefont
  {Ferialdi}},\ }\bibfield  {title} {\bibinfo {title} {Exact closed master
  equation for gaussian non-markovian dynamics},\ }\href
  {https://doi.org/10.1103/PhysRevLett.116.120402} {\bibfield  {journal}
  {\bibinfo  {journal} {Phys. Rev. Lett.}\ }\textbf {\bibinfo {volume} {116}},\
  \bibinfo {pages} {120402} (\bibinfo {year} {2016})}\BibitemShut {NoStop}%
\bibitem [{\citenamefont {Trivedi}\ \emph {et~al.}(2021)\citenamefont
  {Trivedi}, \citenamefont {Malz},\ and\ \citenamefont
  {Cirac}}]{trivedi_Malz_Cirac_convergence_2021}%
  \BibitemOpen
  \bibfield  {author} {\bibinfo {author} {\bibfnamefont {R.}~\bibnamefont
  {Trivedi}}, \bibinfo {author} {\bibfnamefont {D.}~\bibnamefont {Malz}},\ and\
  \bibinfo {author} {\bibfnamefont {J.~I.}\ \bibnamefont {Cirac}},\ }\bibfield
  {title} {\bibinfo {title} {Convergence of numerical approximations to
  non-{Markovian} bosonic gaussian environments},\ }\href
  {http://arxiv.org/abs/2107.07196} {\bibfield  {journal} {\bibinfo  {journal}
  {arXiv:2107.07196 [quant-ph]}\ } (\bibinfo {year} {2021})},\ \bibinfo {note}
  {arXiv: 2107.07196}\BibitemShut {NoStop}%
\bibitem [{\citenamefont {Gross}\ and\ \citenamefont
  {Combes}(2021)}]{gross2021pysme}%
  \BibitemOpen
  \bibfield  {author} {\bibinfo {author} {\bibfnamefont {J.~A.}\ \bibnamefont
  {Gross}}\ and\ \bibinfo {author} {\bibfnamefont {J.}~\bibnamefont {Combes}},\
  }\bibfield  {title} {\bibinfo {title} {{PySME}: A package for simulating
  stochastic master equations in python},\ }\bibfield  {journal} {\bibinfo
  {journal} {github.com}\ }\href {https://doi.org/10.5281/zenodo.5348871}
  {10.5281/zenodo.5348871} (\bibinfo {year} {2021})\BibitemShut {NoStop}%
\end{thebibliography}%

\end{document}